\documentclass[sigconf]{acmart}

\acmSubmissionID{443}
\usepackage{booktabs} 

\usepackage{algorithm}
\usepackage{algpseudocode}
\usepackage{siunitx}
\usepackage{subcaption}

\usepackage{color}
\usepackage{amsmath}
\usepackage{tikz}

\DeclareMathOperator*{\argmin}{argmin}

\acmJournal{TOG}

\copyrightyear{2023}
\acmYear{2023}
\setcopyright{rightsretained}
\acmConference[SA Conference Papers '23]{SIGGRAPH Asia 2023 Conference Papers}{December 12--15, 2023}{Sydney, NSW, Australia}
\acmBooktitle{SIGGRAPH Asia 2023 Conference Papers (SA Conference Papers '23), December 12--15, 2023, Sydney, NSW, Australia}
\acmDOI{10.1145/3610548.3618176}
\acmISBN{979-8-4007-0315-7/23/12}

\begin{document}
\title{Discovering Fatigued Movements for Virtual Character Animation}

\author{Noshaba Cheema}
\email{noshaba.cheema@dfki.de}
\affiliation{%
 \institution{DFKI, Max Planck Institute for Informatics, Saarland Informatics Campus}
 \country{Germany}}
 
\author{Rui Xu}
\email{rui.xu@dfki.de}
\affiliation{%
 \institution{DFKI, Saarland Informatics Campus}
 \country{Germany}}

\author{Nam Hee Kim}
\email{namhee.kim@aalto.fi}
\affiliation{%
\institution{Aalto University}
\country{Finland}}

\author{Perttu Hämäläinen}
\email{perttu.hamalainen@aalto.fi}
\affiliation{%
\institution{Aalto University}
\country{Finland}}

\author{Vladislav Golyanik}
\email{golyanik@mpi-inf.mpg.de}
\affiliation{%
 \institution{Max Planck Institute for Informatics, Saarland Informatics Campus}
 \country{Germany}}

 \author{Marc Habermann}
\email{mhaberma@mpi-inf.mpg.de}
\affiliation{%
 \institution{Max Planck Institute for Informatics, Saarland Informatics Campus}
 \country{Germany}}

 \author{Christian Theobalt}
\email{theobalt@mpi-inf.mpg.de}
\affiliation{%
 \institution{Max Planck Institute for Informatics, Saarland Informatics Campus}
 \country{Germany}}

\author{Philipp Slusallek}
\email{philipp.slusallek@dfki.de}
\affiliation{%
 \institution{DFKI, Saarland Informatics Campus}
 \country{Germany}}


\renewcommand\shortauthors{N. Cheema et al.}

%
%
\begin{abstract}
Virtual character animation and movement synthesis have advanced rapidly during recent years, especially through a combination of extensive motion capture datasets and machine learning. A remaining challenge is interactively simulating characters that fatigue when performing extended motions, which is indispensable for the realism of generated animations. However, capturing such movements is problematic, as performing movements like backflips with fatigued variations up to exhaustion raises capture cost and risk of injury. Surprisingly, little research has been done on faithful fatigue modeling. To address this, we propose a deep reinforcement learning-based approach, which---for the first time in literature---generates control policies for full-body physically simulated agents aware of cumulative fatigue. For this, we first leverage Generative Adversarial Imitation Learning (GAIL) to learn an expert policy for the skill; Second, we learn a fatigue policy by limiting the generated constant torque bounds based on endurance time to non-linear, state- and time-dependent limits in the joint-actuation space using a Three-Compartment Controller (3CC) model. Our results demonstrate that agents can adapt to different fatigue and rest rates interactively, and discover realistic recovery strategies without the need for any captured data of fatigued movement.
\end{abstract}
%
%





\maketitle


%
%
\begin{figure}
    \includegraphics[width=\linewidth]{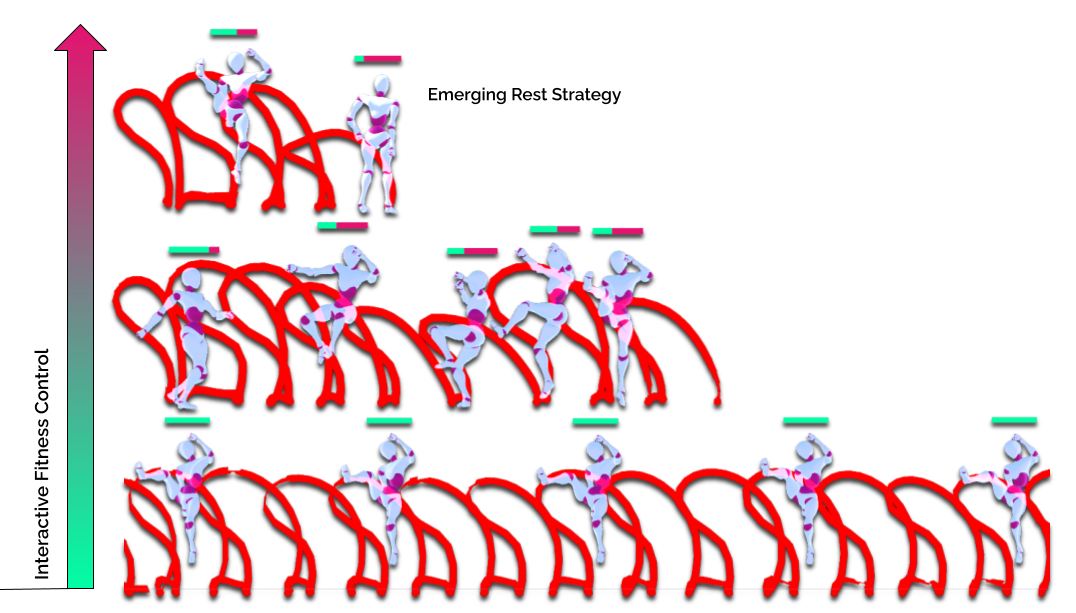}
    \vspace{-20pt}
	\caption{We propose a deep reinforcement learning approach, which explicitly accounts for a character's fatigue.We show that our agents discover novel fatigued motions and recovery strategies without requiring explicit fatigued data. Moreover, our approach enables interactive fitness control within one policy, letting characters fatigue faster or slower depending on the user}
    \vspace{-10pt}
	\label{fig:teaser}
\end{figure}
%
%
%
%
\section{Introduction} \label{sec:intro} 
%
%
For many applications, ranging from games and films to robotics, synthesizing life-like and realistic character animation and control is a crucial yet challenging element. 
A relatively unexplored yet important area in this direction is the adequate simulation of physiological changes over time.
In particular, this is true for cumulative fatigue, i.e., humans getting tired over time when performing strenuous motions. 
Simulating this effect plausibly aids to the overall realism in interactive applications, such as sports simulators, where athletes start to change their execution patterns due to muscle fatigue, and in games, when virtual characters run out of resources.
To achieve this, current games use comparably simple heuristics and pre-defined kinematic animations in state-machines such as motion graphs \cite{kovar2002motion, lee2002interactive} to model character fatigue. 
However, \emph{many} such methods do not incorporate any biomechanical realism, require \emph{prohibitively} extensive motion capture data of fatigued motions~\cite{kider2011data}.
%
%
\par 
Previous work targeting character animation can be roughly categorized into data-driven kinematic-based methods (KM)~\cite{min2012motion, levine2012continuous, buttner2015motion, holden2017phase, starke2020local} and physics-based simulation, paired with Deep Reinforcement Learning (DRL)~\cite{peng2021amp, baker2019emergent, pathak2017curiosity, yin2021discovering, jiang2019synthesis}. 
KMs provide an easy means to make use of the inherent realism that comes from captured or hand-animated motion data,  lacks the ability to generalize to novel behaviours. 
Meanwhile, characters animated via KMs lack the ability to react to dynamic stimuli such as external perturbations, unless the prior motion capture incorporates the vast amount of interaction and perturbation scenarios.
On the other hand, DRL methods provide a promising direction in this field, as even sparse rewards or space constraints already allow for the automatic generation of interactive movements and novel emergent behaviours
without the need for explicitly capturing variations of data that fulfill the given constraints.
However, what kind of rewards or constraints constitute for realistic movements and behaviours remains a fundamental challenge in animation research. 
Interestingly, none of the previous works in either of the two categories has focused on modeling character fatigue in a biologically plausible fashion while overall animation realism would benefit from such an approach.
Only a few works in the biomechanically-based simulation literature~\cite{komura2000creating, cheema2020predicting} have looked into this problem. 
However, previous works have only been applied to carefully hand-crafted single limb settings, as they require expensive musculoskeletal models~\cite{komura2000creating} or careful reward modeling and hyper-parameter tuning~\cite{cheema2020predicting} and are not able to demonstrate the emergence of fatigue and recovery behaviours to the same extent as our work. 
%
%
\par
In this paper, we propose a novel fatigue-aware policy generation framework for physically simulated characters, which allows for the emergence of realistic fatigue and recovery effects. 
For this we explore a two-step approach: 
First, we pre-train the policy to allow for a viable number of stable behaviours to emerge, and a well-behaved initial state for fatigue transfer-learning.
Second, based on the learned stable behaviours, we refine the policy by modifying the emerging torque limits to nonlinear, state- and time-dependent limits using a Three-Compartment Controller (3CC) model \cite{Xia2008ARecovery}, which we adapt from the biomechanics literature in a new context not envisioned by the original paper.
To show the effectiveness of our framework, we integrate our fatigue module into a state-of-the-art generative adversarial imitation learning (GAIL) framework \cite{ho2016generative, torabi2018generative, peng2021amp}. 
%
%
\par 
In summary, the main contributions of this work are as follows: 
\begin{itemize} 
    \item \textbf{Novel Functionality.} 
    The first approach for emergent fatigue and recovery behaviours in full-body Reinforcement Learning for 3D virtual character animation.
    \item \textbf{Efficient Torque-based Fatigue-System for Animation.} 
    The first full-body fatigue system based solely on joint actuation torques for interactive character simulation.
    \item \textbf{Interactive Fatigue and Rest Control.} 
    The use of a Three-Compartment Controller (3CC) fatigue model to limit joint actuation torques based on observed fatigue and residual strength capacity, which allows for interactive control of different fitness levels within one policy.
\end{itemize} 
Our results demonstrate several emergent fatigue behaviours during repetitive athletic tasks, such as arms bending during cartwheels and decreasing kick and jump height after athletic martial arts kicks (Fig.~\ref{fig:teaser}), as well as waiting behaviours to recover from the fatigue to be able to continue.
Our agents learn unseen motion patterns while resting in order to most effectively recover from the experienced fatigue. For example, our agent learns to effectively compensate for the momentum after a backflip or a cartwheel while ending up in a motion state where motor units start to recover.
As the risk of injury could prohibit motion capture of fatigued yet complex movements, our method brings an added benefit of bypassing this constraint.
%
%
\section{Related Work} \label{sec:related}
%
%
%
This section reviews physics-based approaches, deep reinforcement learning,  muscoskeletal methods and techniques supporting fatigue. 
Note that as fatigue modeling relies on the physical forces, purely kinematic methods cannot be applied to our problem unless one explicitly captures fatigued motions. 
%
%
\paragraph{Physics-based Methods}
Physics-based methods allow movement generation with physical realism and environmental interaction; they give direct insights on the forces required and being applied to the character for a given task by leveraging a more general knowledge of the physical equations of motion \cite{raibert1991animation, wampler2014generalizing, geijtenbeek2013flexible}. 
A fundamental challenge in physics-based approaches is the design of controllers for simulated characters. 
Task-specific controllers (e.g.~for locomotion), achieved significant success \cite{coros2010generalized, felis2016synthesis, yin2007simbicon, ye2010optimal, geijtenbeek2013flexible, lee2010data}. 
However, such manually designed controllers remain hard to generalize to diverse movements
and tasks. 
With the wide availability of motion capture data, tracking-based controllers have become a popular research domain \cite{wampler2014generalizing, ye2010synthesis, hamalainen2015online, tassa2012synthesis}, 
though they remain limited in motion quality and long-term planning. 
More recently, deep reinforcement learning-based methods have become a promising research direction to account for long-term planning as well as emergent and reactive behaviour -- three aspects necessary for emergent fatiguing behaviour over multiple repetitions. 
%
%
\paragraph{Deep Reinforcement Learning}
Deep reinforcement learning (DRL) has been successfully applied to physics-based characters animation \cite{liu2017learning,peng2016terrain,teh2017distral}. 
Here, policy gradient methods emerged for continuous control problems \cite{schulman2017proximal, schulman2015trust, sutton1998introduction}.
Imitation learning addresses the challenge of designing task-specific reward functions by learning a policy from examples by explicitly tracking the sequence of target poses in the motion clip \cite{peng2018deepmimic, liu2018learning}. 
While this technique can imitate a single motion clip, it becomes difficult to scale without including high-level motion planners \cite{bergamin2019drecon, park2019learning, won2020scalable, won2021control, lee2021learningFamily, lee2021learningTime, zhang2023simulating} or pose-based control using model-based RL \cite{fussell2021supertrack, yao2022controlvae} or behavioural cloning \cite{won2022physics}, which is prone to drift if small amounts of demonstrations are available \cite{ross2011reduction}. 
Recently, methods based on generative adversarial imitation learning (GAIL) have shown to be an appealing alternative \cite{peng2021amp, peng2022ase, lee2022deep, xu2021gan, bae2023pmp, hassan2023synthesizing}, where an adversarial discriminator is trained to serve as an objective function for training a control policy to imitate the demonstrations. 
We make use of this technique to learn highly diverse and athletic movements. 
However, while these methods are able to generate diverse and natural looking movements, they lack bio-mechanical insights which are important for movement realism. 
%
%
\paragraph{Muscoskeletal Methods and Biomechanical Cumulative Fatigue}
Several works \cite{taga1995model, anderson2001dynamic, geyer2010muscle, ackermann2012predictive, ijspeert2007swimming, maufroy2008towards, thelen2003generating} developed musculoskeletal models that use biomimetic muscles and tendons to simulate a variety of human and animal motions. 
Controlling a muscle-based virtual characters was also explored in computer animation, from upper- \cite{lee2006heads, lee2009comprehensive, lee2018dexterous, tsang2005helping, sueda2008musculotendon}, to lower- \cite{wang2012optimizing, park2022generative}, and full-body movements \cite{geijtenbeek2013flexible, lee2014locomotion, jiang2019synthesis, lee2019scalable, wang2012optimizing}. 
Such methods are computationally expensive, especially for interactive applications such as games. 
\citet{jiang2019synthesis} convert an optimal control problem in the muscle actuation space to an equal problem in the joint-actuation space. 
The generated torque limits do not take accumulated fatigue variation over time into account. Muscoskeletal approaches to predicting muscle fatigue are based on detailed muscle activation patterns  \cite{giat1993musculotendon, giat1996model,  ding2000predictive}. 
These approaches incorporate fatigue as a modifier of relatively complex muscle models. 
While they can provide realistic predictions of forces for isolated muscles, they are cumbersome for joint or whole body applications.
In contrast, \citet{liu2002dynamical} proposed a computationally efficient motor unit (MU)-based fatigue model, using three muscle activation states to estimate perceived biomechanical fatigue: resting, activated and fatigued.  
Improving upon this model, \citet{Xia2008ARecovery} introduced a Three-Compartment Controller (3CC) model for dynamic load conditions, eliminating the need for explicit modeling of muscle actuators. 
%
%
\paragraph{Cumulative Fatigue Modeling in Simulated Characters}
To the best of our knowledge little work has been done in this area.
\citet{kider2011data} captured extensive amounts of motion capture and biosignal data, including EKG, BVP, GSR, respiration, and skin temperature, to estimate fatigue of human characters. 
However, capturing data for all variances is time-consuming and expensive. 
\citet{komura2000creating} make use of a muscoskeletal model, which re-targets existing motion clips to fatigued animations automatically using the musculoskeletal fatigue model  \citet{giat1993musculotendon, giat1996model} for lower-body movements. While they achieve variance over time based on biomechanically accurate fatigue assumptions, their method needs an expensive muscoskeletal-model and cannot account for any emergent recovery behaviours. 
\citet{cheema2020predicting} use a Three-Compartment-Controller (3CC-$r$) model \cite{Xia2008ARecovery, looft2018modification} in a fatigue-related reward function, which does not require expensive modeling and simulation of muscle-tendons. 
They predict ergonomic differences of user interface configurations with a single arm model in a pointing task.
However, they do not consider acrobatic full-body movements.
Additionally, none of the mentioned works indicate the emergence of rest behaviours to the same extent as our method and have only been applied to carefully crafted single limb settings with limited movements.
%

%
%
\section{Preliminaries - 3CC Model}
\label{sec:3CC}
We first review how cumulative fatigue can be modeled efficiently using only joint actuation torques with a 3CC-model \cite{Xia2008ARecovery, looft2018modification}, which has been used for ergonomic assessment of endurance times and fatigue in biomechanics \cite{frey2012three} and HCI \cite{jang2017modeling, cheema2020predicting}.
%
%
\paragraph{Motor Units}
The 3CC model assumes motor units (MUs) to be in one of three possible states (compartments):
1) \emph{active} -- MUs contributing to the task 
2) \emph{fatigued} -- fatigued MUs without activation 
3) \emph{resting} -- inactive MUs not required for the task.

These are usually expressed as a percentage of maximum voluntary contraction (\emph{\%MVC}), which can practically be expressed as percentage of maximum voluntary force (\emph{\%MVF}) or torque (\emph{\%MVT}). Rested MUs ($M_R$) become activated ($M_A$) once a target load ($TL$) needs to be held. Active MUs are then directly contributing to the task. Once an MU is activated, its force decays over time and becomes fatigued ($M_F$). 
An initial non-fatigued state starts out with $M_{R_0} = 100\%$ and $M_{F_0} = M_{A_0} = 0\%$. The following system of equations describes the change of rate over time $t$ for each compartment:
\begin{subequations}
\vspace{-10pt}
    \begin{align}
        \frac{\partial M_A}{\partial t} &= C(t) - F \cdot M_A \label{eq:3cc:1}\\
        \frac{\partial M_R}{\partial t} &= -C(t) + R_r \cdot M_F \label{eq:3cc:2}\\
        \frac{\partial M_F}{\partial t} &= F \cdot M_A - R_r \cdot M_F \label{eq:3cc:3}
    \end{align}\label{eq:3cc}
\end{subequations}
Here, $R_r$ is defined as
\vspace{-10pt}
\begin{equation}
    R_r = \begin{cases}
        r \cdot R & \text{if $M_A \geq TL$} \\
        R & \text{else},
    \end{cases}
    \label{eq:rR}
\end{equation}
where $F$ and $R$ denote the fatigue and recovery coefficients, and $r$ as an additional rest recovery multiplier for intermittent tasks \cite{looft2018modification}. A change in $F$ denotes the change of rate in fatigue, whereas a change in $R$ indicates the overall rate of recovery, as well as an upper bound for maximal fatigue and loads that can be held indefinitely. For example, $R = F \cdot 0.2$ indicates that 20\% $MVC$ can be held indefinitely, which would be in accordance to empirical studies \cite{rohmert1960ermittlung}. In this case the limit of $M_F$ would be at 80\%. An increase of $r$ indicates an increased recovery rate during tasks with intermittent rest periods when $r > 1$.
$C(t)$ in Eq.~(\ref{eq:3cc:1}) and (\ref{eq:3cc:2}) is a bounded proportional controller, which produces the force required for the target load ($TL$) by controlling the size of $M_A$ and $M_R$. 
%
\paragraph{Motor Activation-Deactivation Drive C(t)}
To obtain behaviours matching muscle physiology -- e.g., active MUs decaying over time -- control theory is applied. Therefore, $C(t)$ is introduced as a muscle activation-deactivation drive between rested and active MUs:
\begin{equation}
 C(t)=
  \begin{cases} 
    L_R \cdot (TL-M_A) &\textrm{if}\; M_A \geq TL\\
    L_D \cdot (TL-M_A) &\textrm{if}\; M_A < TL\; \textrm{and}\; M_R > TL - M_A\\
    L_D \cdot M_R      &\textrm{if}\; M_A < TL\; \textrm{and}\; M_R \leq TL - M_A.
    \label{eq:c_t}
  \end{cases}
\end{equation}

The three cases can be described in the following way:
\begin{itemize}
    \item \emph{Case 1:} If there are more active motor units $M_A$ than required for the target load $TL$, then $M_A$ decays and $M_R$ increases in Eq.~(\ref{eq:3cc}), which makes the muscle go into a recovery state. In this case $C(t)$ becomes negative.
    \item \emph{Case 2:} When there are not enough active motor units $M_A$ compared to the required target $TL$ but the difference is smaller than the available rested MUs $M_R$, then rested MUs $M_R$ become active MUs $M_A$. In this case $C(t)$ is positive and greater or equal than $F \cdot M_A$ in Eq.~(\ref{eq:3cc:1}).
    \item \emph{Case 3:} When there are not enough active motor units $M_A$ compared to the required target $TL$ and not enough rested MUs $M_R$ to compensate the discrepancy, the muscle starts to fatigue and the target load cannot be held any longer. In this case, $C(t)$ is positive but smaller than $F \cdot M_A$ in Eq.~(\ref{eq:3cc:1}). Here, $M_A$ decays and $M_R$ becomes (near) zero.
\end{itemize}
$L_D$ and $L_R$ are muscle force development and relaxation factors, which describe the sensitivity towards the target load. Since the time course of either is negligible compared to the time course of fatigue (e.g. varying $L_D/L_R$ from 2 to 50, or a change of 2500\%, only alters the endurance time by 10\%), \citet{Xia2008ARecovery} set the same arbitrary value to each $L_D = L_R = 10$.
\paragraph{Residual Capacity}
Residual Capacity (RC) describes the remaining motor strength capabilities or stamina due to fatigue in percent. 0\% indicates no strength reserve, and 100\% indicates full non-fatigued strength:
\begin{equation}
    RC(t) = M_A + M_R = 100\% - M_F.
    \label{eq:RC}
\end{equation}
\par
While the 3CC model is a great analysis tool for fatigue and endurance time estimation~\cite{frey2010endurance, cheema2020predicting, jang2017modeling}, it does not directly lend itself to create full-body character animations modeling the described fatigue effects in Sec.~\ref{sec:intro}.  
To approach this challenge, we leverage Generative Adversarial Imitation Learning (GAIL) \cite{ho2016generative, torabi2018generative, peng2021amp, peng2022ase} 
and constrain the action space based on the computed Residual Capacity of the 3CC model, which we describe in more detail in the following section.

\begin{figure}
	\includegraphics[width=\linewidth]{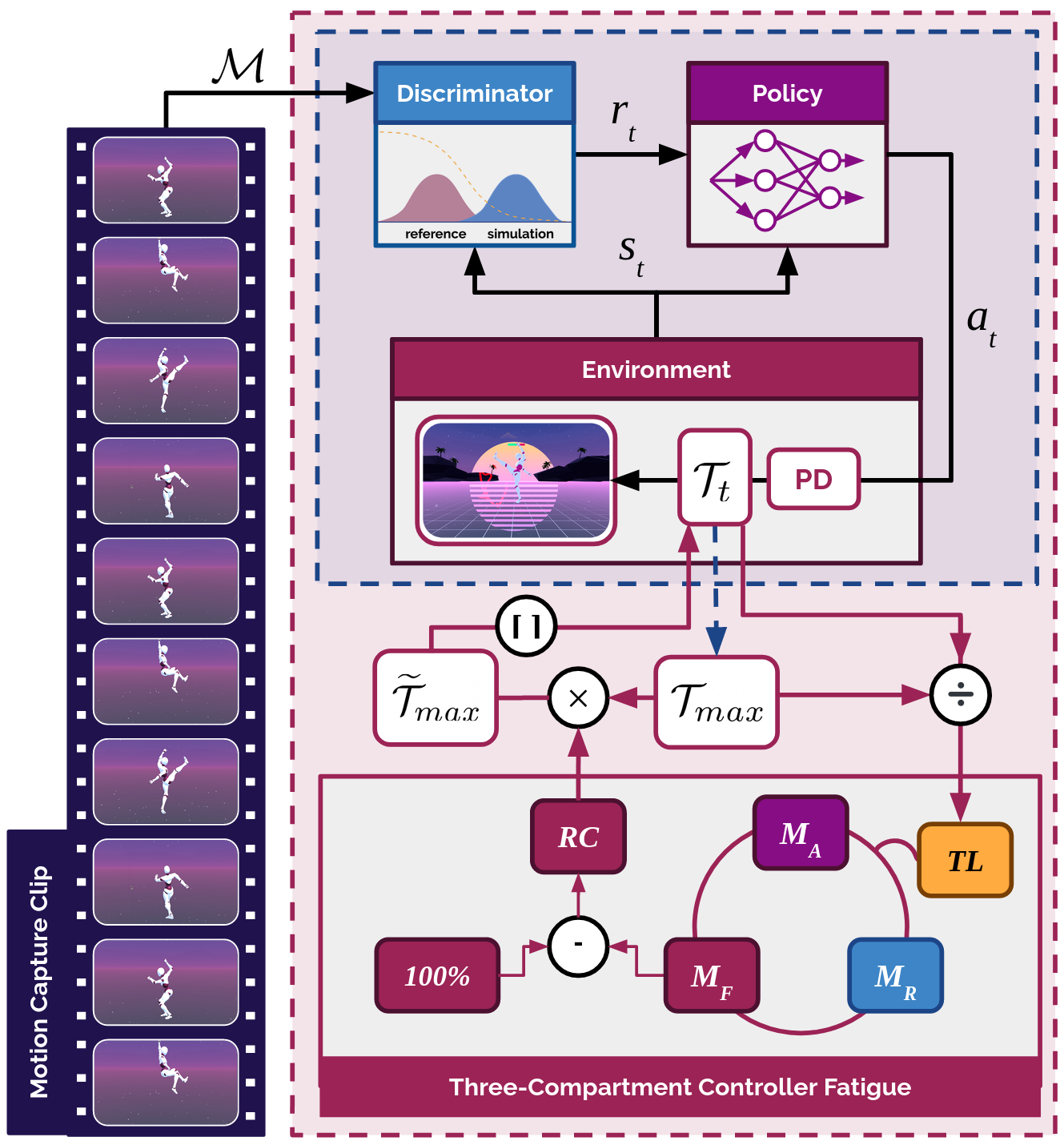}
    \vspace{-15pt}
	\caption{\textbf{Overview.} Our framework consists of two stages: A pre-training stage where the policy learns to imitate the motion clip.Then a transfer learning stage, where an adaptive fatigue policy is learned by constraining the output torques of the PD-controllers based on the computed torque bounds by the Residual Capacity ($RC$) from the 3CC-model.}
    \vspace{-10pt}
	\label{fig:overview}
\end{figure}

\section{Fatigue Modeling for Character Animation} \label{sec:system}
Our method takes as input a motion clip $\mathcal{M}$ of the full-body skeleton of the humanoid character represented by a sequence of poses $\hat q_t$.
This motion clip does \textit{not} contain varying fatigue levels.
Nonetheless, our goal is to generate an animation that mimics the original behaviour while the character fatigues over time and learns plausible recovery strategies.
At the technical core, we explore a two-step approach that effectively blends ideas from biomechanical cumulative fatigue modeling, biologically-inspired torque limit constraining, and Deep Reinforcement Learning to enable the emergence of realistic symptoms of fatigue in character animation (see Fig.~\ref{fig:overview}).
\par 
We first pre-train the policies on the reference motions to estimate the maximum constant torque bounds $\mathcal{T}_{max}$ across the tasks. 
The actions $\mathbf{a}_t := (\mathbf{u}_t, \beta_t)$ at time $t$ from the policy $\pi$ specify target positions $\mathbf{u}_t$ for PD-controllers positioned at each of the character's joints and a stiffness and damping multiplier $\beta_t$, similar to \cite{yuan2021simpoe}, which we query at the policy frequency.
Modulating stiffness and damping introduces the possibility for the character to relax and tense its whole body, which is appropriate for the context of fatigue modeling, as fatigued virtual characters using proportion-derivative (PD) controllers with fixed stiffness/damping parameters may choose overly stiff/conservative motions instead of relaxed motions.
The output torques are then applied to the character physics simulation (Sec.~\ref{sec:pre_train}). 
Once an expert policy has been learned, we use transfer learning to learn a policy, which is able to adapt to fatigue by constraining the torque-bounds over time based on RC computed by the 3CC fatigue modules resulting in $\mathcal{\widetilde{T}}_{max}$ (Sec.~\ref{sec:transfer_learning}). 
This forces the policy to handle lower torque levels and discover fatigue and rest behaviours in attempt to fulfill the task based on the given constraints. 
Importantly, the model can be trained on a single ($F$, $R$, $r$)-triplet and adapt to novel triplets during inference, which makes training more efficient. 
Once trained, the agents exhibit unseen fatigued movement patterns and unseen rest recovery strategies emerge to overcome the loss of strength. 
%
%
\subsection{Imitation Objective and Torque-Estimation} \label{sec:pre_train}
In this section, we describe the pre-training. 
We start with a general formulation of a DRL problem, then continue with the imitation objective, and finally describe the constant torque-bound estimation.
%
%
\paragraph{RL Problem Formulation}
At each time step $t$, the agent observes a state $\mathbf{s}_t$ based on its environment observations and samples an action $\mathbf{a}_t$ from a policy $\pi(\mathbf{a}_t | \mathbf{s}_t)$ in accordance to the observed state, which leads to a new state $\mathbf{s}_{t+1}$ and a reward $r_t = r(\mathbf{s}_t, \mathbf{a}_t, \mathbf{s}_{t+1})$. 
The agent's objective is to maximize its expected discounted return $J(\pi)$ \cite{sutton1998introduction}
%
%
\begin{equation}
    J(\pi) = \mathbb{E}_{p(\tau | \pi)} \left[\sum_{t=0}^{T-1} \gamma^t r_t \right],
    \label{eq:RL_return}
\end{equation}
%
%
where $p(\tau | \pi) = p(\mathbf{s}_0) \prod_{t=0}^{T-1} p(\mathbf{s}_{t+1} | \mathbf{s}_t, \mathbf{a}_t) \pi(\mathbf{a}_t | \mathbf{s}_t)$ represents the likelihood of the trajectory $\tau = \left\{(s_t, a_t, r_t)_{t=0}^{T-1}, s_T\right\}$ under a policy $\pi$. 
Here, $p(\mathbf{s}_0)$ denotes the initial state distribution, $T$ is the time horizon of a trajectory, and $\gamma \in [0, 1]$ is the discount factor. 
To design a reward objective, which can imitate diverse athletic movements, we leverage Generative Adversarial Imitation Learning (GAIL) \cite{ho2016generative, peng2021amp, peng2022ase}.
%
%
\paragraph{Imitation Objective}
In GAIL, the objective to imitate a given task is modeled as a discriminator $D(\mathbf{s}, \mathbf{a})$, which is trained to predict whether a given state $\mathbf{s}$ and action $\mathbf{a}$ is sampled from the demonstrations $\mathcal{M}$ or generated from the policy $\pi$ \cite{ho2016generative}. 
This formulation of GAIL requires access to the demonstrator's actions, which, however, are not given when only motion clips are provided as demonstrations. 
Similar to \citet{torabi2018generative}, we train the discriminator on state transitions $D(\mathbf{s}, \mathbf{s}')$ instead of state-action pairs $D(\mathbf{s}, \mathbf{a})$ to overcome this limitation:
%
%
\begin{equation}
    \argmin_D -\mathbb{E}_{p^{\mathcal{M}(\mathbf{s}, \mathbf{s}')}} \left[\log\left(D(\mathbf{s}, \mathbf{s}')\right)\right] - \mathbb{E}_{p^{\mathcal{\pi}(\mathbf{s}, \mathbf{s}')}}\left[\log\left(1 - D(\mathbf{s}, \mathbf{s}')\right)\right],
    \label{eq:disc_obj}
\end{equation}
%
%
where $p^{\mathcal{M}(s, s')}$ and $p^{\mathcal{\pi}(s, s')}$ denote the likelihoods of observing a state transition from state $\mathbf{s}$ to $\mathbf{s}'$ in the dataset $\mathcal{M}$, and following the policy $\pi$, respectively. Additionally, we incorporate the gradient penalty regularizers~\cite{peng2021amp}. The discriminator is then trained using the following objective
%
%
\begin{align}
\begin{split}
    \argmin_D & -\mathbb{E}_{p^{\mathcal{M}(\mathbf{s}, \mathbf{s}')}} \left[\log\left(D(\mathbf{s}, \mathbf{s}')\right)\right] - \mathbb{E}_{p^{\mathcal{\pi}(\mathbf{s}, \mathbf{s}')}}\left[\log\left(1 - D(\mathbf{s}, \mathbf{s}')\right)\right] \\
    & + w_{\text{gp}} \mathbb{E}_{d^{\mathcal{M}(\mathbf{s}, \mathbf{s}')}} \left[\|\nabla_{\phi} D(\phi) |_{\phi=(\mathbf{s}, \mathbf{s}')}\|^2\right],
\end{split}
\end{align}
%
%
where $w_{\text{gp}}$ is a gradient penalty coefficient. 
Akin to \citet{peng2022ase}, we use the imitation objective for an adversarial imitation policy by defining the reward objective in Eq.~(\ref{eq:RL_return}) as
%
%
\begin{equation}
    r_t = -\log(1 - D(\Phi(\mathbf{s}_t), \Phi(\mathbf{s}_{t+1}))),
    \label{eq:gail_reward}
\end{equation}
%
%
where $\Phi(\mathbf{s}_t)$ denotes a feature map based on the state space $\mathbf{s}_t$.
%
%
\paragraph{Torque Estimation}
We use the following PD-controller formulation to estimate the joint torque bounds and actuation torques \cite{featherstone2014rigid, yuan2021simpoe} at every control query:
%
%
\begin{equation}
    \mathcal{T}_d := \beta_d k_\texttt{p}^d (u_d - \hat{u}_d) - \beta_d k_\texttt{d}^d \dot{\hat u}_d
    \label{eq:pdControl}
\end{equation}
%
%
where $u_d$ denotes the target orientation and $\beta_d$ a stiffness and damping modulation parameter given from the policy's action $a_d := (u_d, \beta_d)$. $\hat{u}_d$ denotes the current orientation of the DoF $d$ and $\dot{\hat u}_t$ the current velocity. $k_\texttt{p}$ and $k_\texttt{d}$ specify constant stiffness and damping parameters.
The pre-training has the nice side-effect that we can automatically estimate maximum torques as 
%
%
\begin{equation}
    \mathcal{T}_{max_d} := \max(\mathcal{T}_d, \mathcal{T}_{max_d})
\end{equation}
%
%
with $\mathcal{T}_{max_{d_0}} = 0$, which avoids manual or grid search for this hyperparameter.
Additionally, we consider physiological symmetries by ensuring that two symmetric joints, e.g. left and right elbow, have the same value by computing the \emph{minimum} between the two, as lower energy movements tend to look more natural \cite{Yu2018LearningLocomotion}. In addition, we found we are able to greatly reduce outliers for potential $\mathcal{T}_{max_d}$ candidates.
We note that our method works for hand-crafted torque limits, as well as limits which depend on a single and multiple motions. 
In these cases, just the rate of fatigue and recovery for an $(F,R,r)$ setting changes.

%
\paragraph{Transfer Learning: Fatigue-based Torque Limits} \label{sec:transfer_learning}
Inspired by \citet{jiang2019synthesis}, we limit constant torque-bounds to nonlinear state-dependent limits in the joint actuation space. 
While their method allows for bio-mechanically enhanced torque-based actuation, it does not allow for variance over time from cumulative fatigue. 
Thus, we multiply the residual capacity for each 3CC-model with the maximum torque bounds found in the previous stage and then use transfer learning to make the policy adapt to the loss of strength as explained next.
%
%
\paragraph{Fatigued Torque Bounds}
We assume the target load $TL$ to be the incoming joint actuator torques of the respective PD-controller of each DoF, which the character requires to reach the target position. 
Each DoF $d$ is modeled by a 3CC model as
%
%
\begin{equation}
    TL_d(\mathcal{T}_d) = \frac{\mathcal{T}_d}{\mathcal{T}_{max_d}} \cdot 100\%
    \label{eq:3cc_TL}
\end{equation}
%
%
being the ratio of the incoming actuator torque $\mathcal{T}_d$ computed by the respective PD-controller and the constant maximum torque-bound found in the previous step representing the percentage of maximum voluntary contraction $\%MVC$.
$M_{A_d}$, $M_{R_d}$ and $M_{F_d}$ are computed in accordance to the incoming target load per DoF using Eq.~(\ref{eq:3cc})--(\ref{eq:c_t}). 
To estimate the fatigued torque bounds $\mathcal{\widetilde T}_{max_d}$, we leverage the residual capacity $RC$ as a time-varying multiplier to the previously found torque bound limits:
%
%
\begin{equation}
    \mathcal{\widetilde T}_{max_d} = RC_d \cdot \mathcal{T}_{max_d},
    \label{eq:rc_T_max}
\end{equation}
%
%
where $RC_d = 100\% - M_{F_d}$ (see Sec.~\ref{sec:3CC}).
The final fatigued torque $\widetilde{\mathcal{T}}_d$ applied to the environment for each DoF is computed by clipping the incoming joint actuator torque $\mathcal{T}_d$ within $RC_d \cdot \left[-\mathcal{T}_{max_d}, \mathcal{T}_{max_d}\right]$ with $\mathcal{\widetilde{T}}_d$ being defined as
%
%
\begin{equation}
    \mathcal{\widetilde{T}}_d \in \left[-\mathcal{\widetilde T}_{max_d}, \mathcal{\widetilde T}_{max_d}\right]= \left[- RC_d \cdot \mathcal{T}_{max_d}, RC_d \cdot \mathcal{T}_{max_d}\right].
    \label{eq:appl_torque}
\end{equation}

\paragraph{Transfer Learning}
Simply reducing the joint actuator torques of an agent with a policy trained with full torque-bounds will let the agent fall as the policy never learned to deal with loss of strength. 
Thus, we apply transfer learning, where the policy is trained on the time-varying torque outputs based on the Residual Capacity. 
While $R$ and $F$ values are joint specific \cite{frey2012three, looft2018modification} accounting for varying endurance times \cite{frey2010endurance}, we use one $F$, $R$ and $r$ for the whole character as a design choice for usability and simple interactivity. 
We note that despite training on one $(F, R, r)$ triplet our policy is able to adapt to new pairs during inference.
This is due to the fact that a single ($F$, $R$, $r$)-triplet alone already corresponds to multiple levels of Residual Capacity over time -- making it easy for the agent to adapt to another combination even though the change of rate of $RC$ may differ.
As such our method can be viewed as a framework for generating a range of motor skills from a single motion clip \cite{lee2021learningFamily}, where the skills are parameterized by fatigue.
%
%
%
\section{Model Representation} \label{sec:model}
%
%
\paragraph{States and Actions}
We evaluate our framework using the 28 DoF humanoid character from \citet{peng2021amp}, as well as their state-space $\mathbf{s}_t$ representation in the IsaacGym implementation with the inclusion of the character's local root rotation \cite{peng2022ase}, as well as the fatigued motor units $M_F$ in the observation space, totalling a state space of 133 dimensions.
The character's local coordinate frame is defined as in \cite{peng2021amp, peng2022ase}.
Each action $\mathbf{a}_t := (\mathbf{u}_t, \mathbf{\beta}_t)$ specifies the target rotations $\mathbf{u}_t$ and stiffness/damping multiplier $\mathbf{\beta}_t)$ for the PD-controllers at each of the character's joints. 
This results in a 29D action space -- one action per DoF, as well as the stiffness/damping multiplier. $M_F$ is randomized during expert training to make the policy agnostic to it before it adapts to it in the fine-tuning stage. 
%
%
\paragraph{Network Architecture}
The policy $\pi(\mathbf{a}_t | \mathbf{s}_t)$ is represented as a neural network, with the action distribution modeled as a Gaussian, where the state-dependent mean $\mu(\mathbf{s}_t)$ and the diagonal covariance matrix $\Sigma$ are specified by the network output
$\pi(\mathbf{a}_t | \mathbf{s}_t) = \mathcal{N}(\mu(\mathbf{s}_t), \Sigma)$.
%
%
The mean is specified by an MLP consisting of two fully-connected hidden layers of 1024 and 512 Rectified Linear Units (ReLU) \cite{nair2010rectified}, followed by a linear output layer. 
The values of the covariance matrix $\Sigma = \text{diag}(\sigma_1, \sigma_2, \dots)$ are manually specified~\cite{peng2021amp} and fixed over the course of training. 
The value function $V(\mathbf{s}_t)$ and the discriminator $D(\mathbf{s}_t, \mathbf{s}_{t+1})$ are modeled by separate networks with a similar architecture as the policy. 
%
%
%
\section{Evaluation and Results} \label{sec:results}
We evaluate our method on five diverse movement skills -- backflip, cartwheel, hopping and locomotion from the CMU dataset [\citeauthor{cmu}] provided in the Isaac Gym \cite{makoviychuk2021isaac} environment, as well as the 360 tornado kick from the SFU dataset [\citeauthor{sfu}].
The experiments below evaluate the following aspects of our method:
First, compared to constant torque limits, our state and time-varying torque limits push the policy network towards movement strategies and patterns not present in the input data $\mathcal{M}$ to overcome the loss of strength.
Second, the found recovery strategies resemble human-like strategies for resting. 
%
%
All experiments are carried out using the high-performance GPU-based physics simulator Isaac Gym \cite{makoviychuk2021isaac}. 
During training 4096 environments are simulated in parallel on a single NVIDIA V100 GPU with a simulation frequency of 120Hz, while the policy operates at 30 Hz.
All neural networks are trained using PyTorch \cite{paszke2019pytorch}. Gradient updates are performed via Proximal Policy Optimization \cite{schulman2017proximal} with a learning rate of $5 \times 10^{-5}$. 
We use an episode length of 300 during pre-training and an episode length of 1000 during fatigue transfer to learn the accumulation of fatigue. 
The gradient penalty coefficient $w_{\text{gp}}$ in Eq.~(\ref{eq:disc_obj}) is set to 0.2 for all but locomotion, for which it is set to 5 \cite{makoviychuk2021isaac}. 
Additional hyper-parameter settings and implementation details can be found in the supplementary document. 
We use the ``Humanoid AMP'' \cite{peng2021amp} character provided in Isaac Gym with 28 internal DoFs and its corresponding rigid body and joint properties. 
Stiffness and damping parameters were set to custom values in accordance to the realistic proportions of a real-life human male. 
$M_A, M_F$ and $M_R$ are randomized at every environment reset.
%
%
\subsection{Fatigue Training} 
\label{sec:fatigue_training}
\begin{figure}
	\includegraphics[trim={19cm 5cm 17cm 7cm},clip,width=0.24\linewidth]{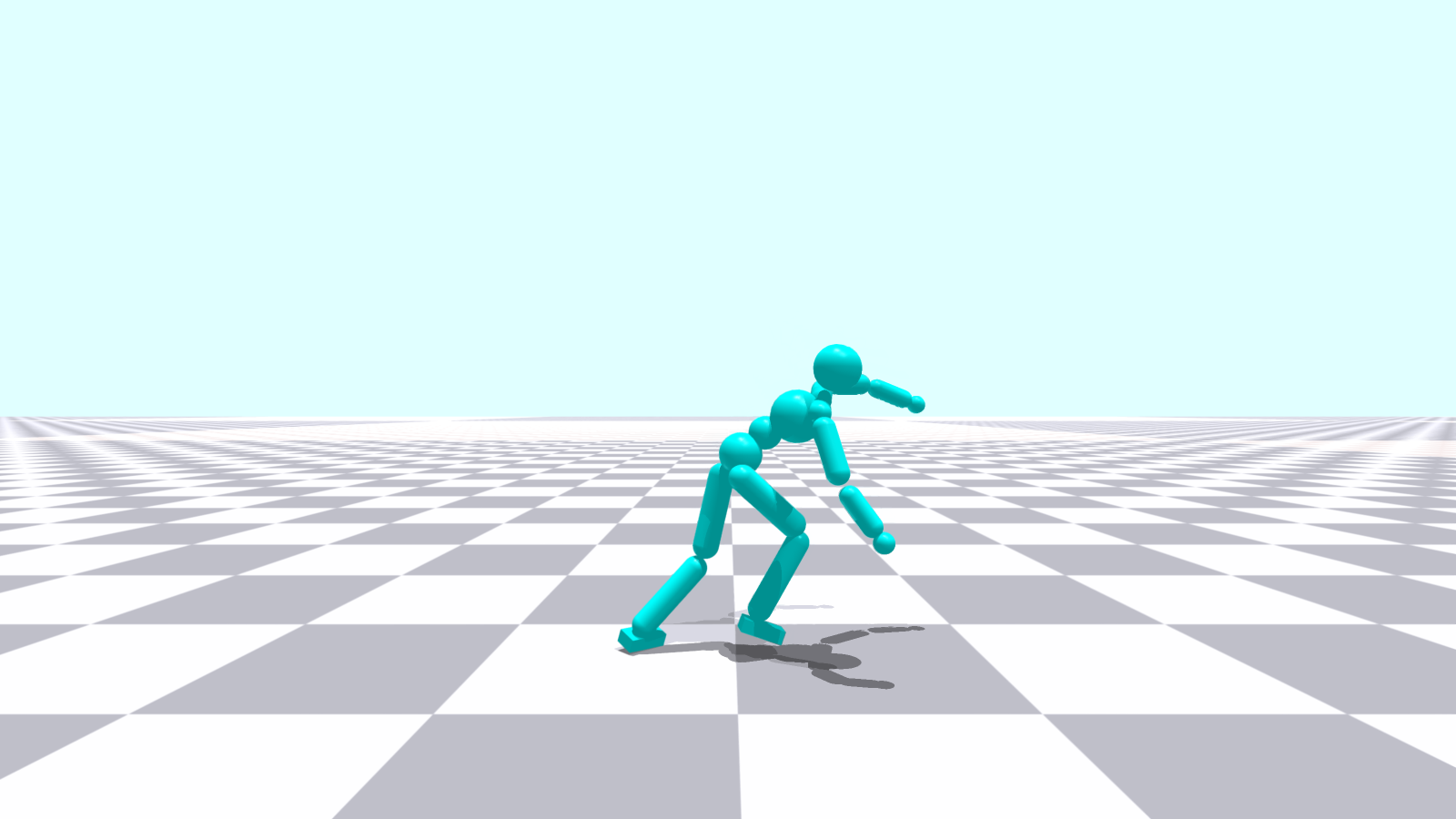}
	\includegraphics[trim={18cm 5cm 18cm 7cm},clip,width=0.24\linewidth]{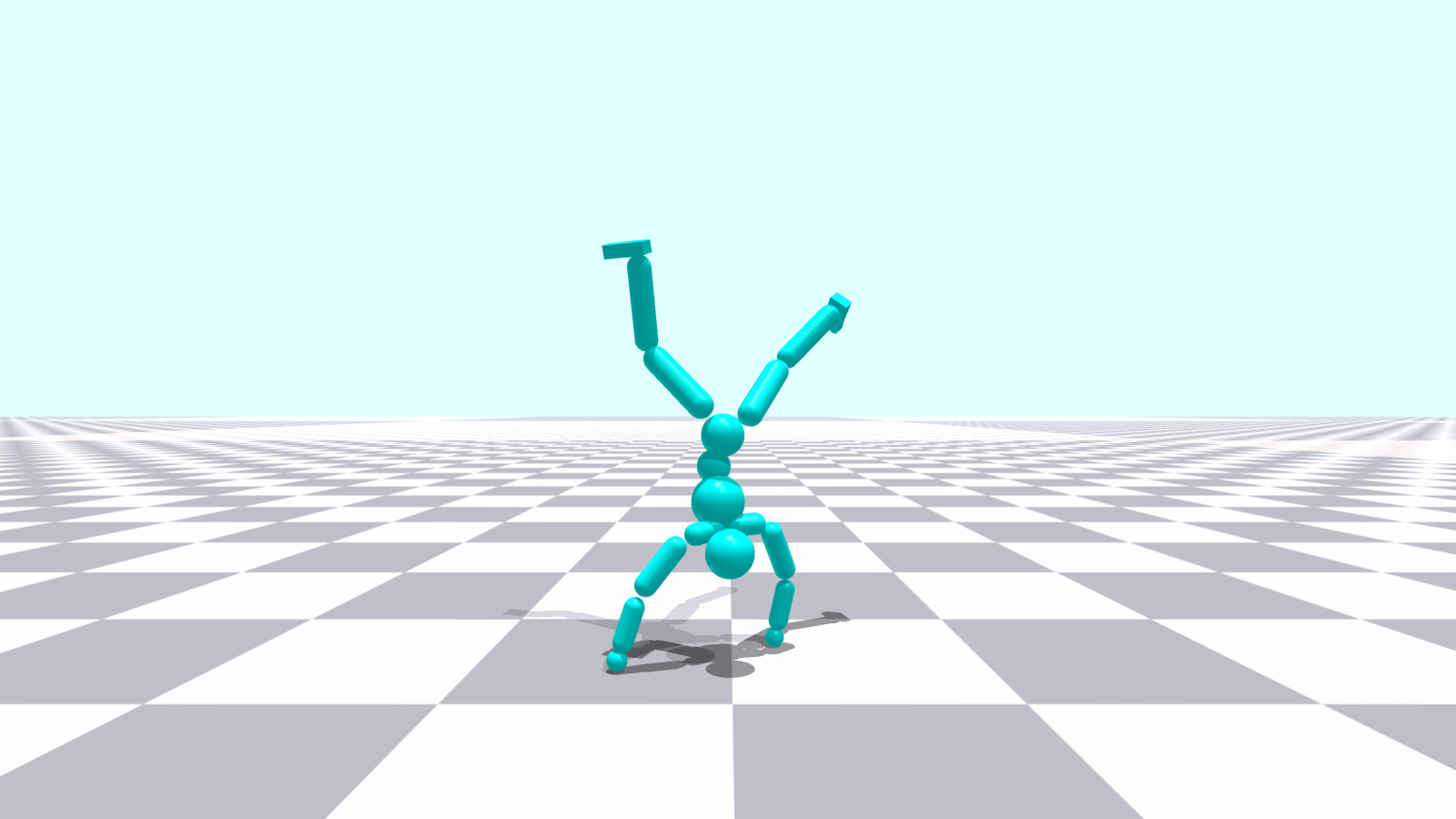}
	\includegraphics[trim={18cm 5cm 18cm 7cm},clip,width=0.24\linewidth]{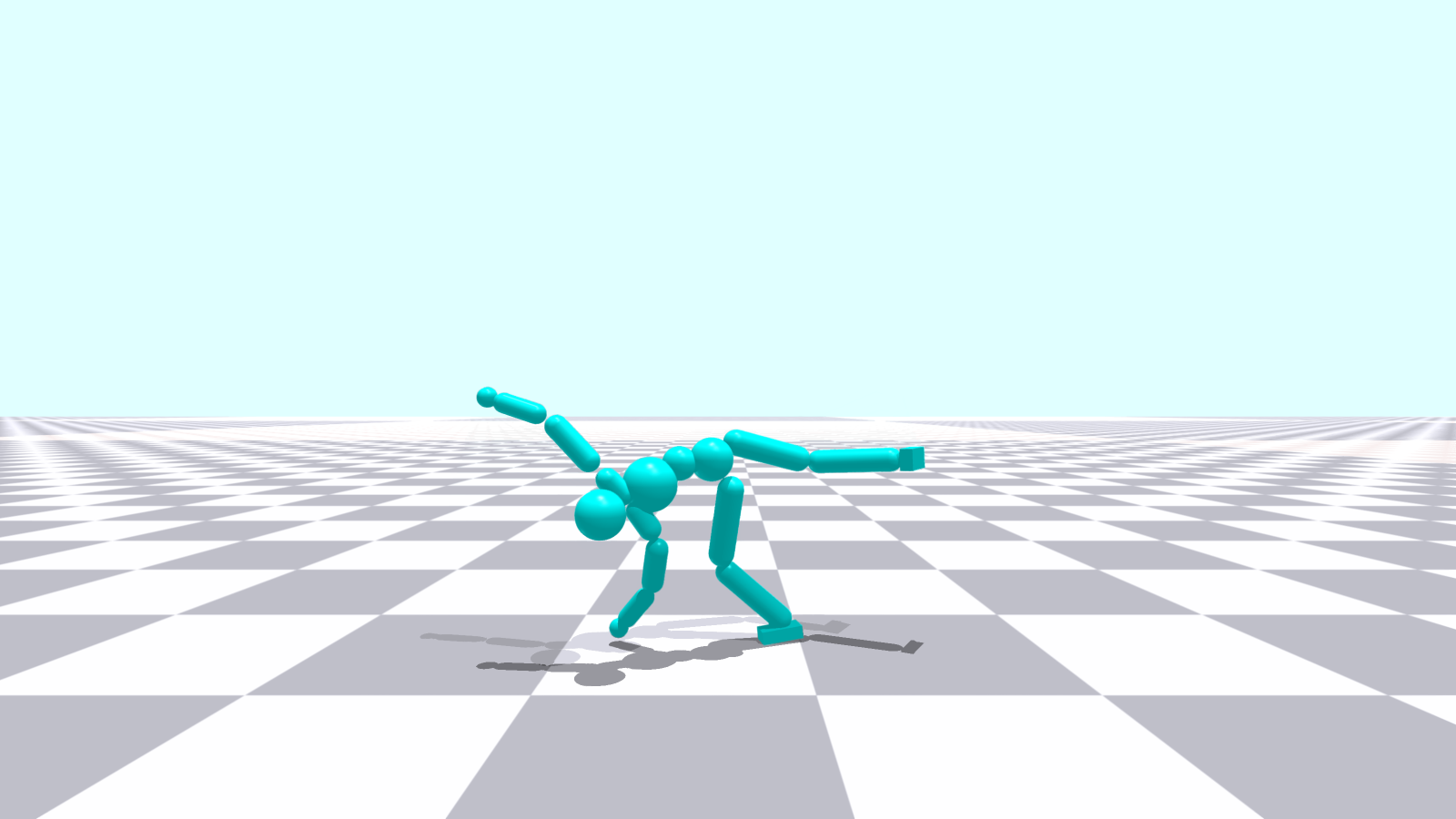}
	\includegraphics[trim={18cm 5cm 18cm 7cm},clip,width=0.24\linewidth]{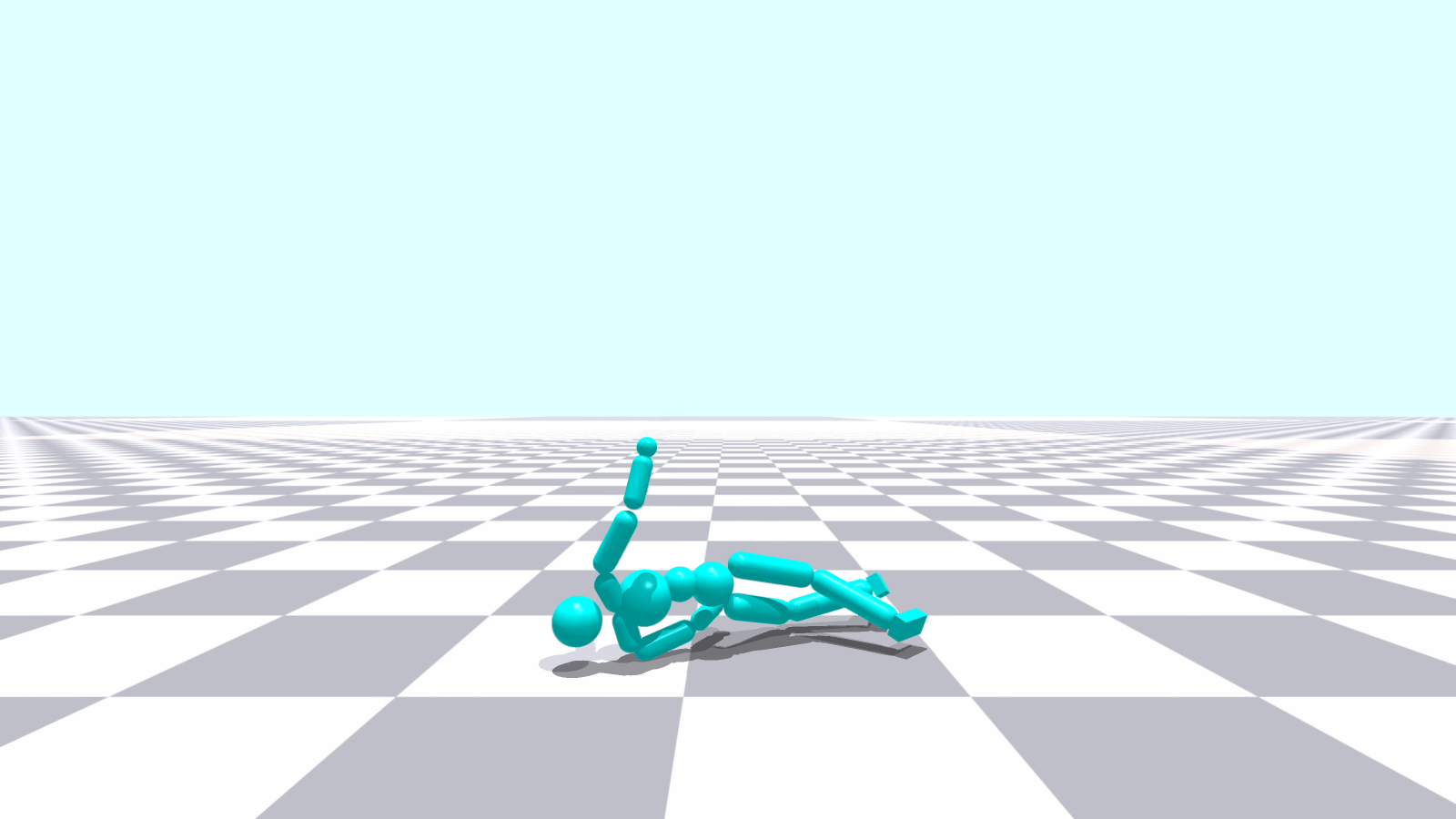}
	\vspace{-12pt}
    \caption
	{
	\textbf{Pre-trained policy with fatigued torque limits during inference without transfer learning} results in task failure, since the policy has not learned to deal with using lower torques and forces.
	}
	\label{fig:fatigue_fail}
\end{figure}

We first note that simply taking a pre-trained policy and adjusting the torque limits during inference only lets the character fall into a termination state and not learn any realistic recovery strategies because the character never learned to deal with less torque as can be seen in Fig.~\ref{fig:fatigue_fail}.
%
%
Thus, we employ a simple transfer learning procedure where we train the expert policies for several iterations until stable behaviours arise.
We disable fatigue behaviour during this pre-training phase. As the observation still contains $M_F$ values for the expert policy training, we ensure that the expert policies are agnostic to the $M_F$ values by randomizing the $M_F$ values at every environment step during this phase, as a strategy for domain randomization \cite{tobin2017domain}.
More specifically, we train these expert policies for 2000 iterations for running and 4000 for others.
For transfer learning, we apply 2000 iterations of additional training for each motion, using the corresponding expert policy as the warm-starting point.
We input $F=1, R=0.01, r=1$ for all transfer learning iterations.
The reference torque estimates for computing $TL$ values are given by the expert policies.
During the transfer learning phase, we randomize the fatigue state at uniform at each reset of the training episode, as to capture as much variability of the fatigue state as possible while observing the input $(F, R, r)$-triplet.
We found that we can train on a single $(F, R, r)$-triplet but test on a variety of combinations (Fig.~\ref{fig:teaser} and \ref{fig:backflip_levels}) if fast enough gradual loss of strength due to fatigue can be observed during training time, as well as some form of increase in strength during recovery periods. 
%
%
\par
%
%
\subsection{Fatigue Movements and Recovery Strategies}
\label{sec:recovery}
\begin{figure}
    \includegraphics[width=0.49\linewidth]{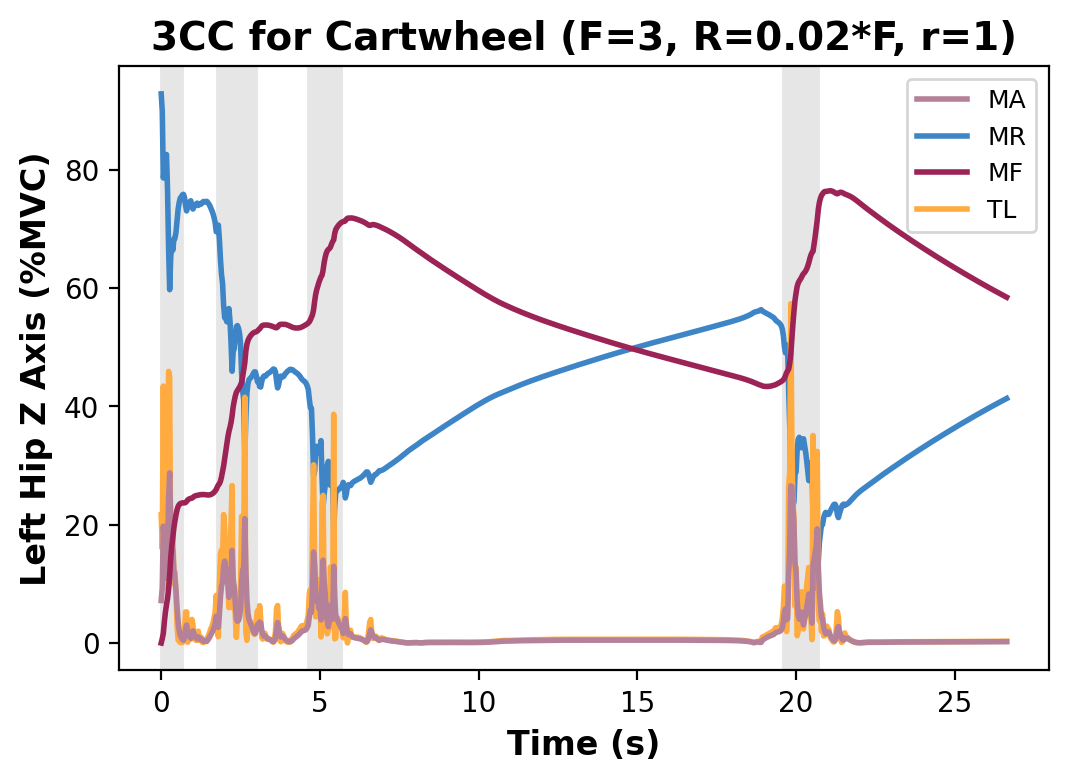}
    \includegraphics[width=0.49\linewidth]{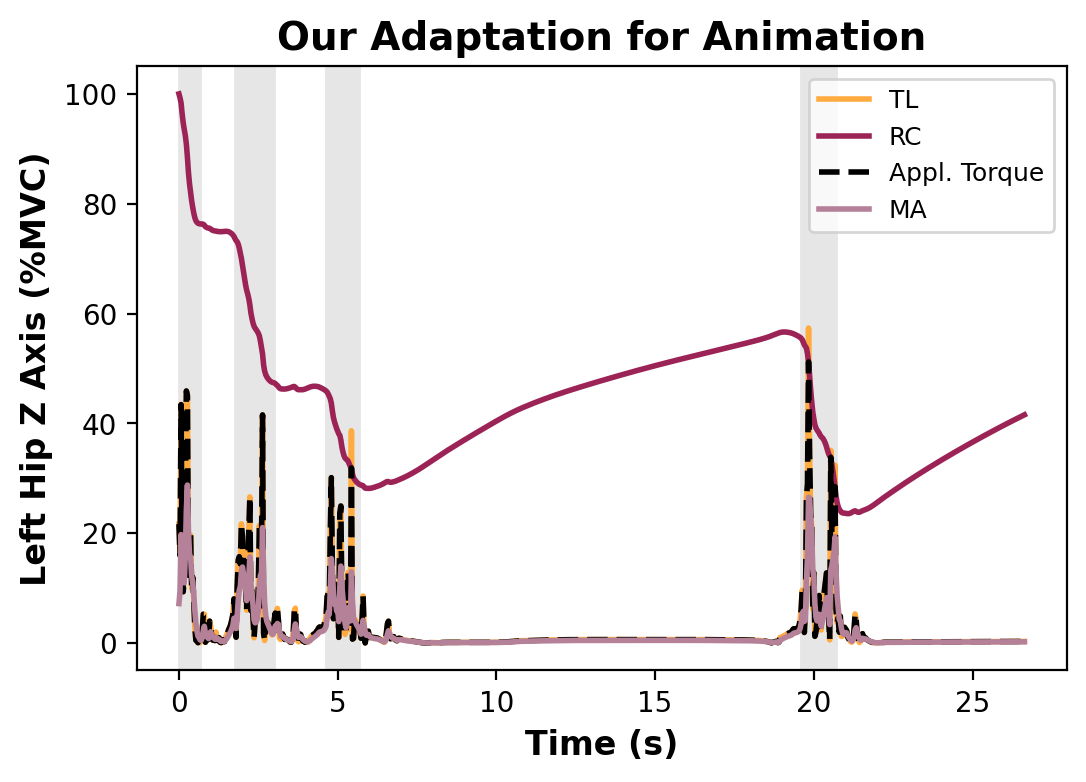}
 \vspace{-12pt}
	\caption
	{\textbf{Results of the 3CC model (\emph{left}) and its adaptation (\emph{right}) for the cartwheel.} The graphs correspond to the cartwheel motion depicted in Fig. \ref{fig:recovery}. The gray areas indicate a successful cartwheel, whereas the areas between 6s and 19s, as well as after 20s correspond to emerging waiting behaviors to rest from the tiring actions. 
	}
	\label{fig:3cc_cartwheel}
\end{figure}

\begin{figure}
    \centering
    \includegraphics[width=\linewidth]{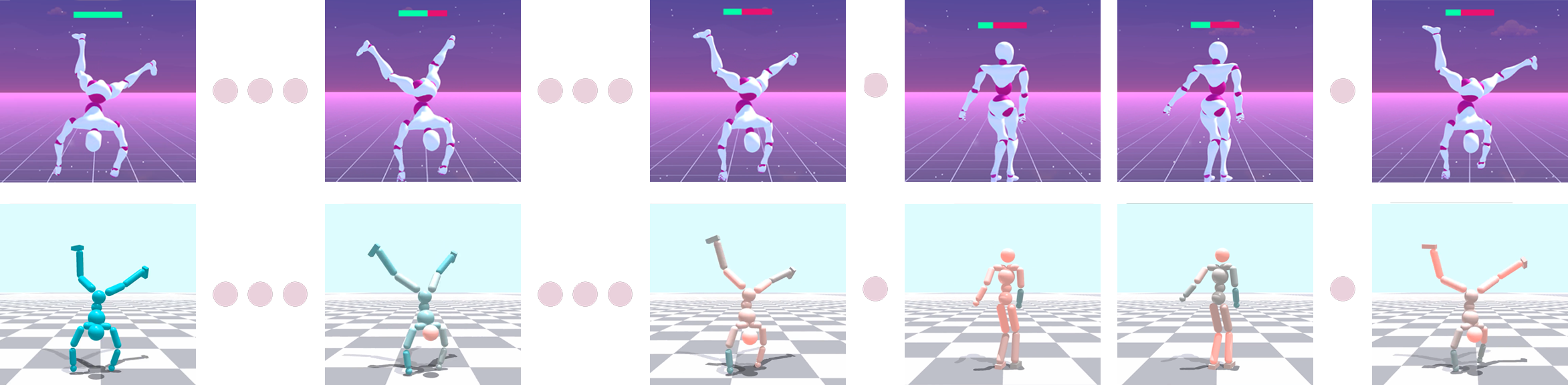}
    \vspace{-15pt}
    \caption{\textbf{Recovery behaviour of a cartwheel motion.} After three cartwheels the agent becomes tired, rests, and does another cartwheel when enough stamina is regained. 
    }
    \label{fig:recovery}
\end{figure}

The loss of strength modeled by our method leads to new movement patterns for the rest recovery strategies, as well as the divergence from the input motion the model was trained on. 
We found the following behaviours, which compensate for the loss of strength: 
\textbf{Waiting} by standing or doing a couple of steps as observed in the cartwheel (Fig.~\ref{fig:recovery}), tornado kick (Fig.~\ref{fig:teaser}) and backflip (Fig.~\ref{fig:backflip_levels}).
Fig.~\ref{fig:teaser}, \ref{fig:3cc_cartwheel}, \ref{fig:recovery} and \ref{fig:backflip_levels} show how the agent regains strength during such rest periods; 
a \textbf{change of performance}, e.g. decrease of height of jumps, especially observable in the hopping and tornado kick motions (Fig.~\ref{fig:performance}); 
a \textbf{change of motion style}, e.g. with increased tucking and knee-bending behaviours in dynamic motions, such as backflip (Fig.~\ref{fig:style}); and 
\textbf{compensation of forces} with movements requiring a lot of momentum such as the cartwheel or backflip by trembling or requiring more suspension (Fig.~\ref{fig:compensation}). 
Additionally, a \textbf{reduction of number of repetitions} (Fig.~\ref{fig:num_reps}) \textbf{and speed} (Fig.~\ref{fig:speed}) can be observed. 
%
%
Fig.~\ref{fig:3cc_cartwheel} shows a comparison between the original 3CC model resulting from the cartwheel motion (\emph{left}), as well as our modification for animation (\emph{right}). Note how the fatigue $M_F$ decreases during the rest period between 6 and 19 seconds (\emph{left}), and increases with each cartwheel. A cartwheel is indicated by the three spikes in $TL$ and $M_A$ in the beginning, as well as the spike at 20 seconds -- in correspondence to the 4 cartwheels in Fig.~\ref{fig:recovery}. For animation, we make use of the residual capacity $RC = 100\% - M_F$ as a strength multiplier for the constant torque bounds. The applied torques (dashed blue line) become similar to $M_A$ when being cut off during fatigue and stay equal to $TL$ during non-fatiguing periods.
\vspace{-5pt}
\paragraph{360 Tornado Kick} During fatigue the character learns to kick with lower foot height and a reduced distance between the legs, which can be best observed in Fig.~\ref{fig:teaser} and Fig.~\ref{fig:performance}. As the motion requires a lot of strength and flexibility in the leg region, the agent learns to recover 
by lowering the jump, as well as waiting between jumps. The most affected joints by fatigue are the kicking knee, as well as the arm joints required to gain the momentum for the jump (Fig.~\ref{fig:joints}).
\vspace{-5pt}
\paragraph{Backflip} The backflips become lower when fatigued with increase in tucking (Fig.~\ref{fig:style}) and compensation of forces (Fig.~\ref{fig:compensation} \emph{bottom}). After rest, the agent is able to do backflips with better forms again but cannot do as many anymore before it needs to rest again (Fig.~\ref{fig:backflip_levels}). 
\vspace{-5pt}
\paragraph{Cartwheel} We find that the character learns to stand or walk in order to rest from the fatigue after a repetition of cartwheels (Fig.~\ref{fig:recovery}). Furthermore, with each repetition, before the character is able to rest fully, the leg height becomes lower.
\vspace{-5pt}
\paragraph{Hopping} The most apparent changes in the hopping motion during fatigue are the jump height/length and frequency (Fig.~\ref{fig:performance} \emph{top}). 
\paragraph{Locomotion} The most apparent change over time is the reduction of speed as well as the change of stride length (Fig.~\ref{fig:speed}).
%
%
\subsection{Learning Diverse Fitness Levels in One Policy}

With the fatigue model driving the behaviour of the policy via the fluctuation of $M_F$, the policy is capable of handling a variety of fatigue states it encounters during deployment.
Here, we highlight a key advantage of using the 3CC fatigue model by demonstrating the capability to model different fitness levels using the same character and simply manipulating the parameters of the 3CC model at deployment.
More specifically, the policy outputs a full spectrum from a high-stamina to a low-stamina character behaviours as a response to intuitive parameter adjustments at runtime.
Figures~\ref{fig:teaser}, \ref{fig:fatigue_levels} and \ref{fig:backflip_levels} juxtapose three different scenarios rendered from deploying the same policy in the same initial state: a non-fatigable character for qualitative baseline (top), a high-stamina character (middle), and a low-stamina character. We further emphasize how varying $(F, R, r)$ parameters over time can be used for the 3CC for more fine-grained control. 
Simply put, we achieve the capability to capture the diversity of fitness levels with the same reference motion, character specification, and policy, which is similar to a \emph{mixture of experts} policy where experts corresponding to different fitness levels emerge depending on the $(F,R,r)$ input without any need for fatigued reference motions.
Additionally, we show that our method can also be used to analyze which joints are being most affected and to what degree by fatigue (Fig.~\ref{fig:joints}).
%
%
\subsection{Ablations and Comparisons}
\label{sec:comparison}
\begin{figure}
    \centering
	\includegraphics[width=0.6\linewidth]{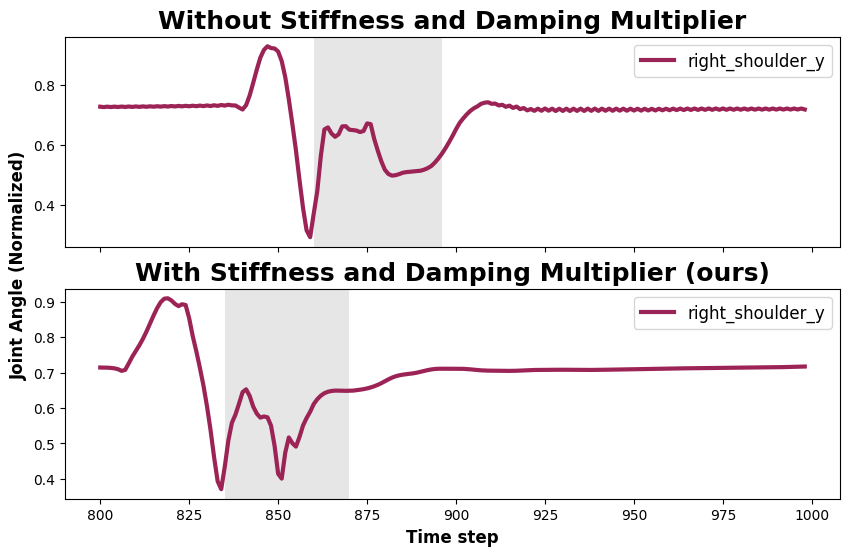}
    \vspace{-15pt}
	\caption
	{
	\textbf{Ablation} without stiffness and damping multiplier (\emph{top}) and with \emph{(bottom, ours)} for the shoulder joint during the backflip motion. While the agent is standing and supposed to rest, the shoulder starts to jitter without our torque coefficient $\beta$, while with it the motion is smooth and still (\emph{bottom}).
	}
	\label{fig:abl_coeff}
\end{figure}

\begin{figure}
    \includegraphics[width=0.07\linewidth]{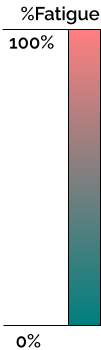}
	\includegraphics[trim={15cm 0cm 17cm 0cm},clip,width=0.17\linewidth]{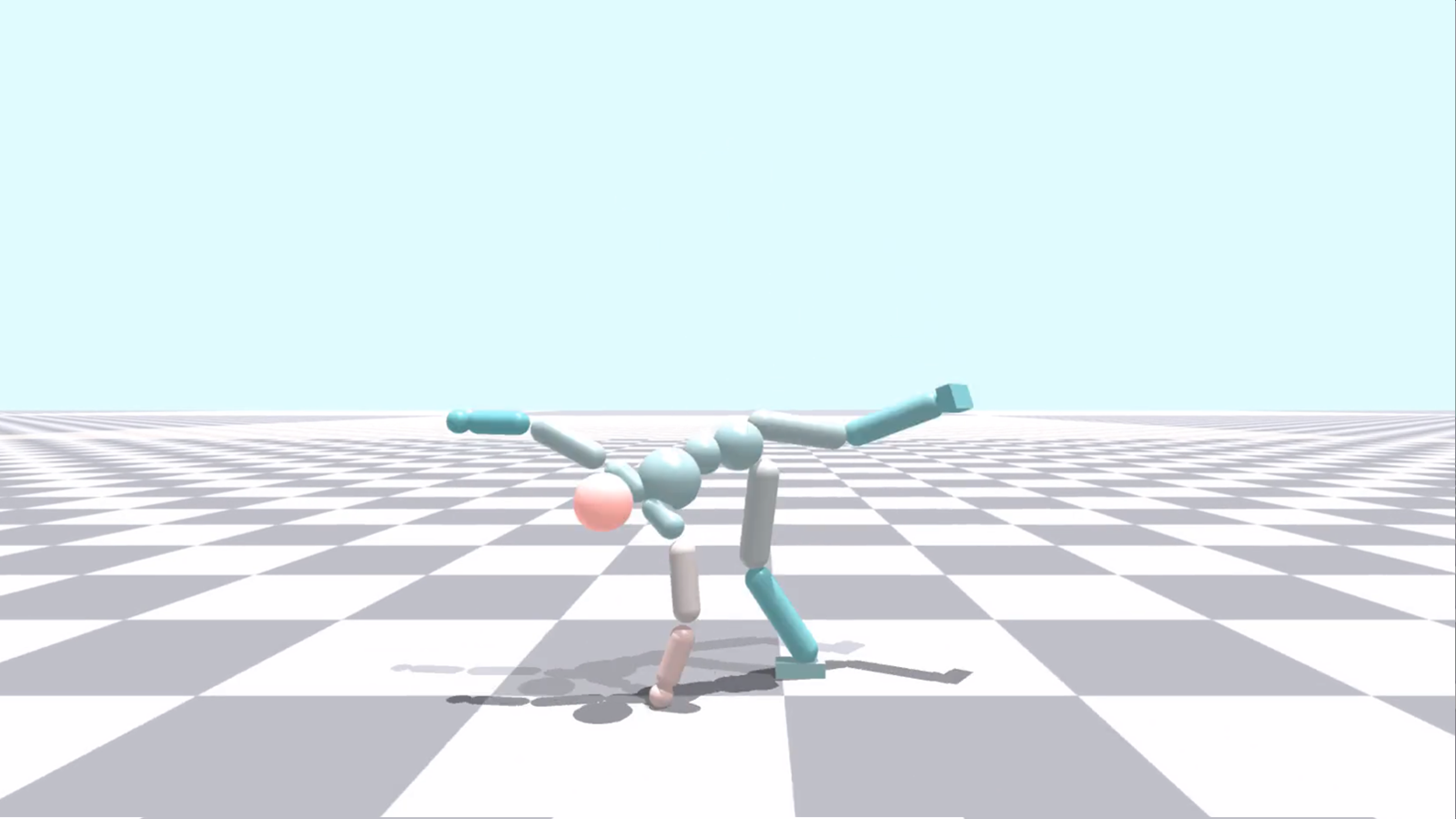}
	\includegraphics[trim={12cm 0 20cm 0cm},clip,width=0.17\linewidth]{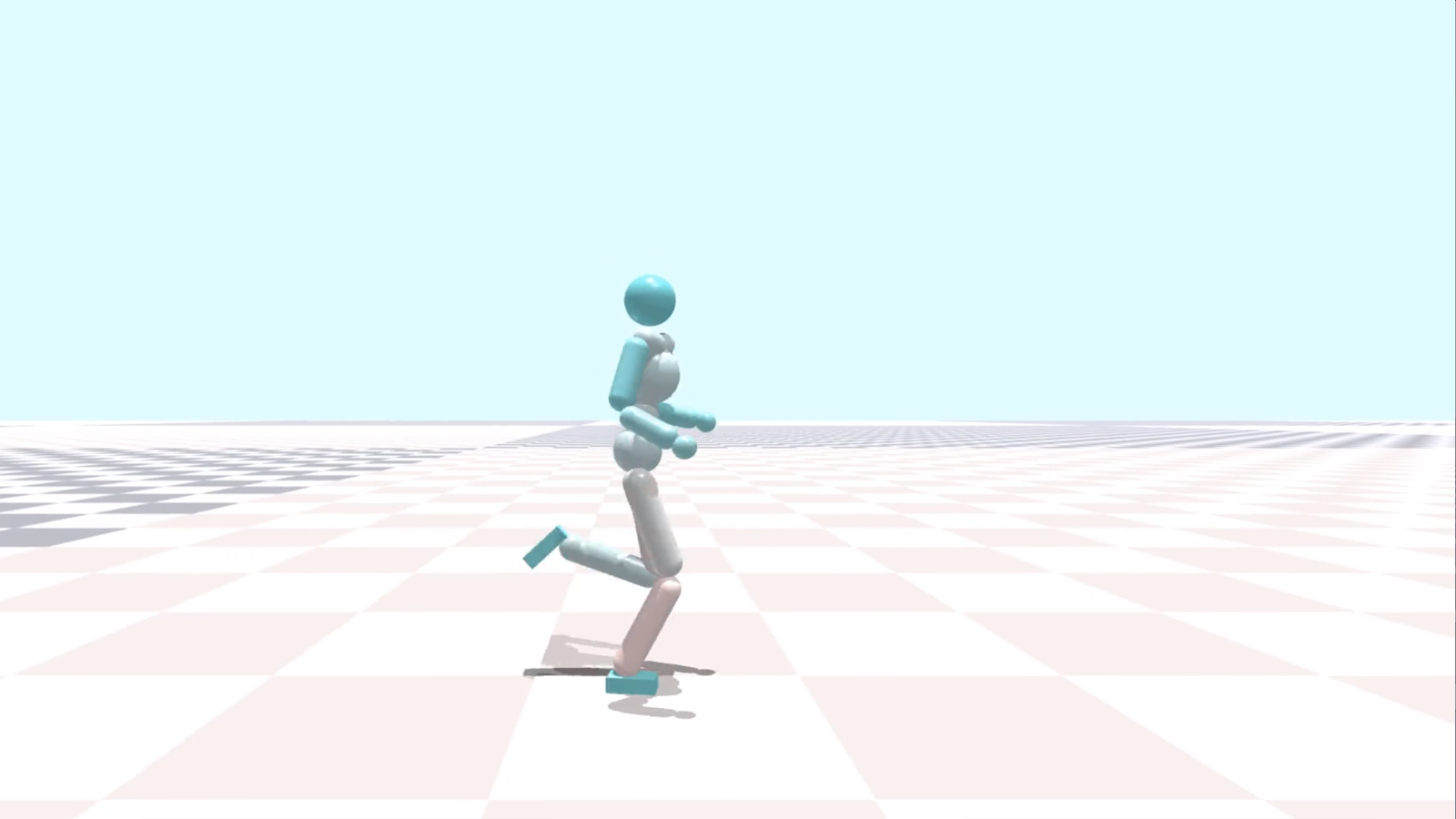}
	\includegraphics[trim={18cm 0 14cm 0cm},clip,width=0.17\linewidth]{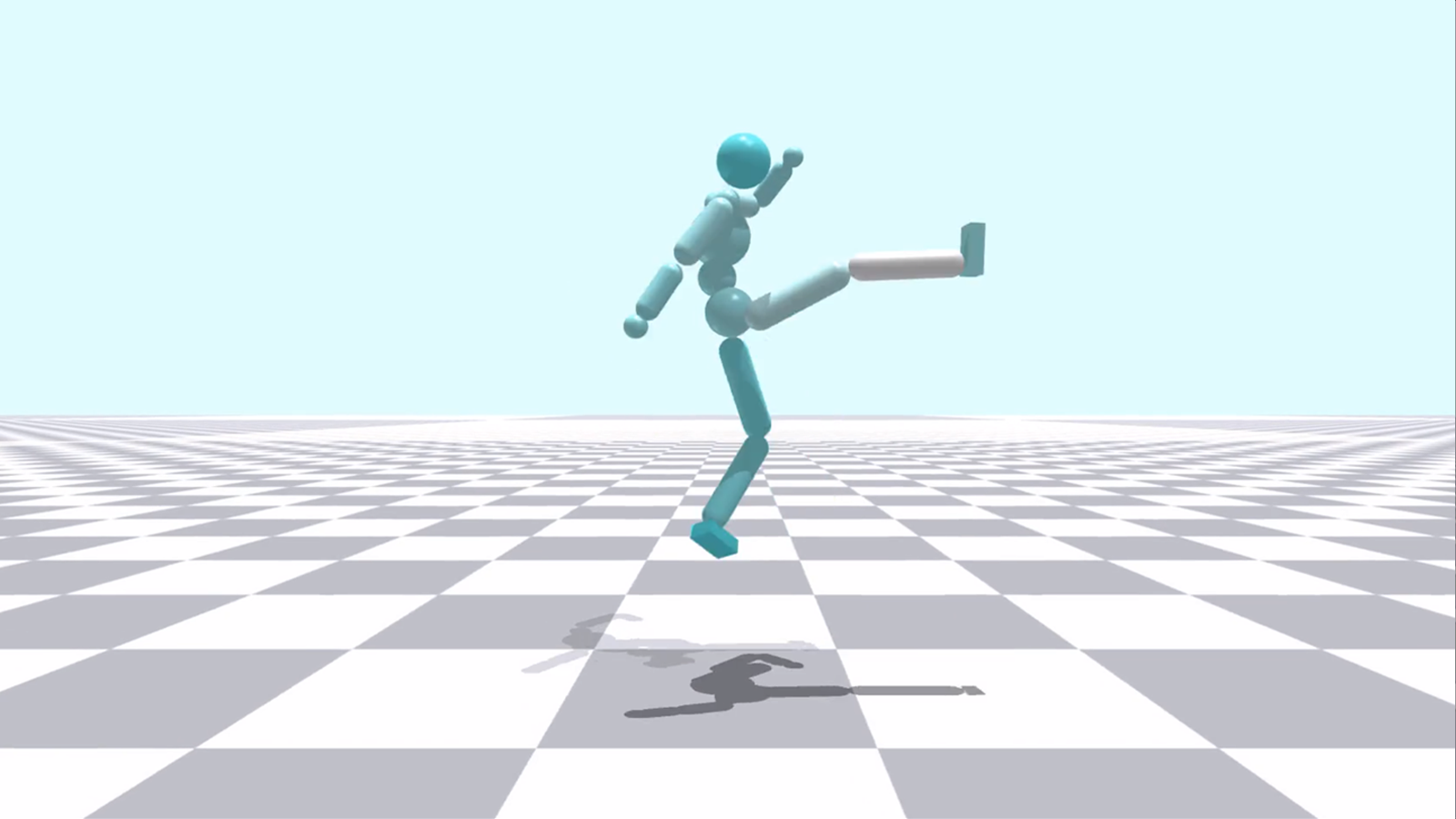}
	\includegraphics[trim={14cm 0 18cm 0cm},clip,width=0.17\linewidth]{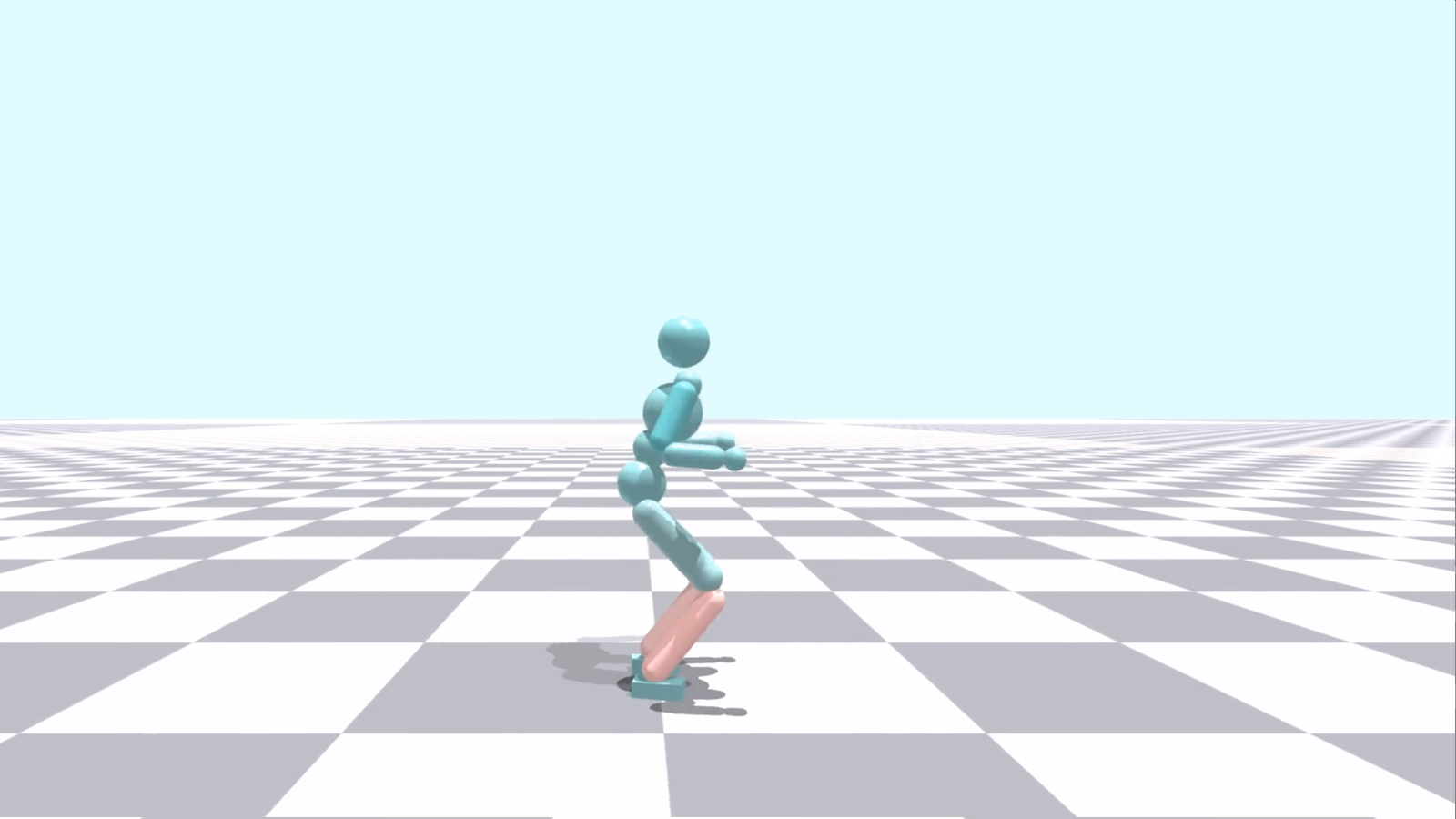}
	\includegraphics[trim={16cm 0cm 16cm 0cm},clip,width=0.17\linewidth]{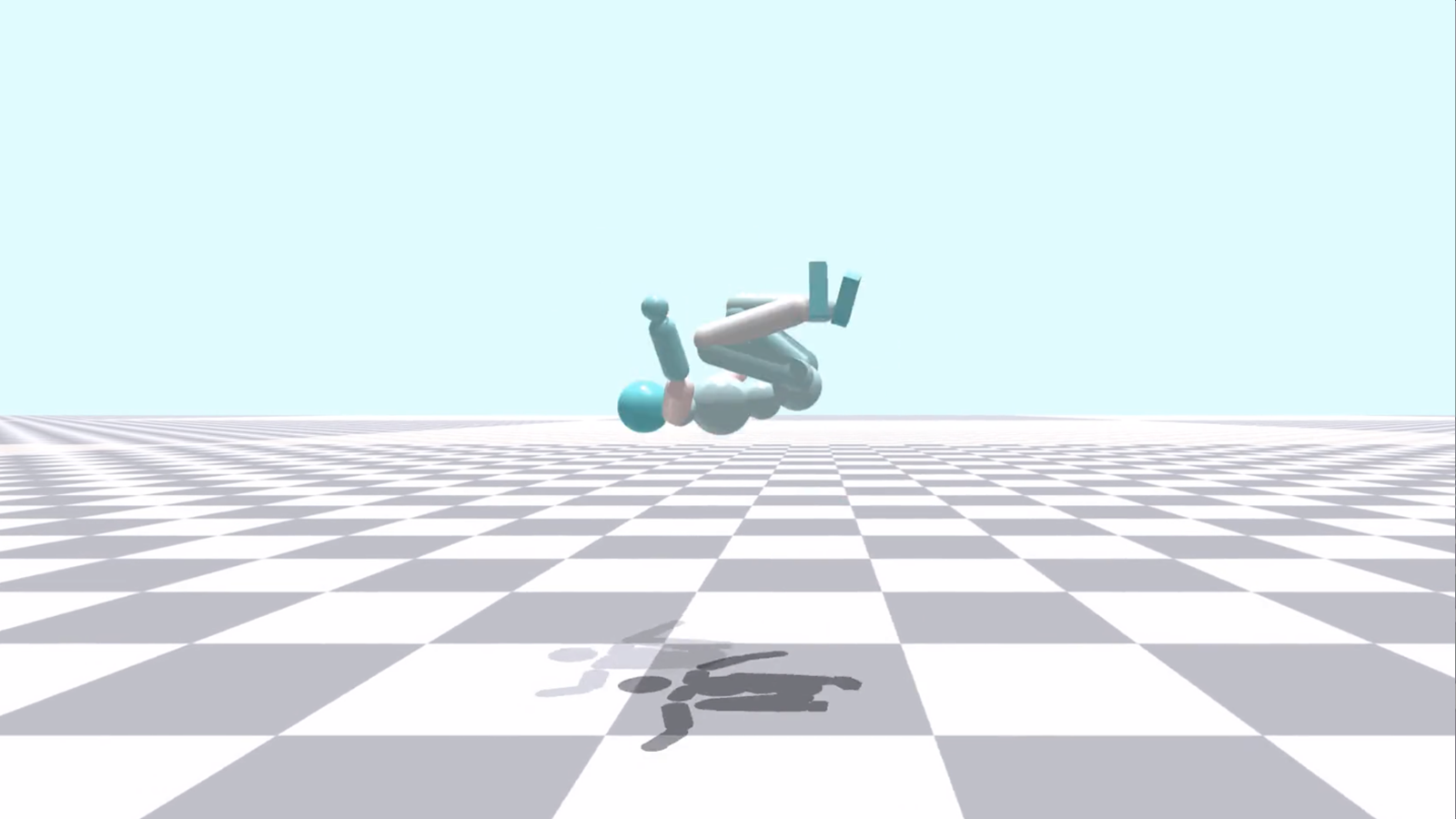}
    \vspace{-12pt}
	\caption
	{
	\textbf{Most affected joints from fatigue.} \emph{From left to right:} Cartwheel, Locomotion, 360 Tornado Kick, Hopping, and Backflip.
	}
	\label{fig:joints}
\end{figure}

\begin{figure}
	\includegraphics[trim={0 3.3cm 0cm 0},clip,width=0.49\linewidth]{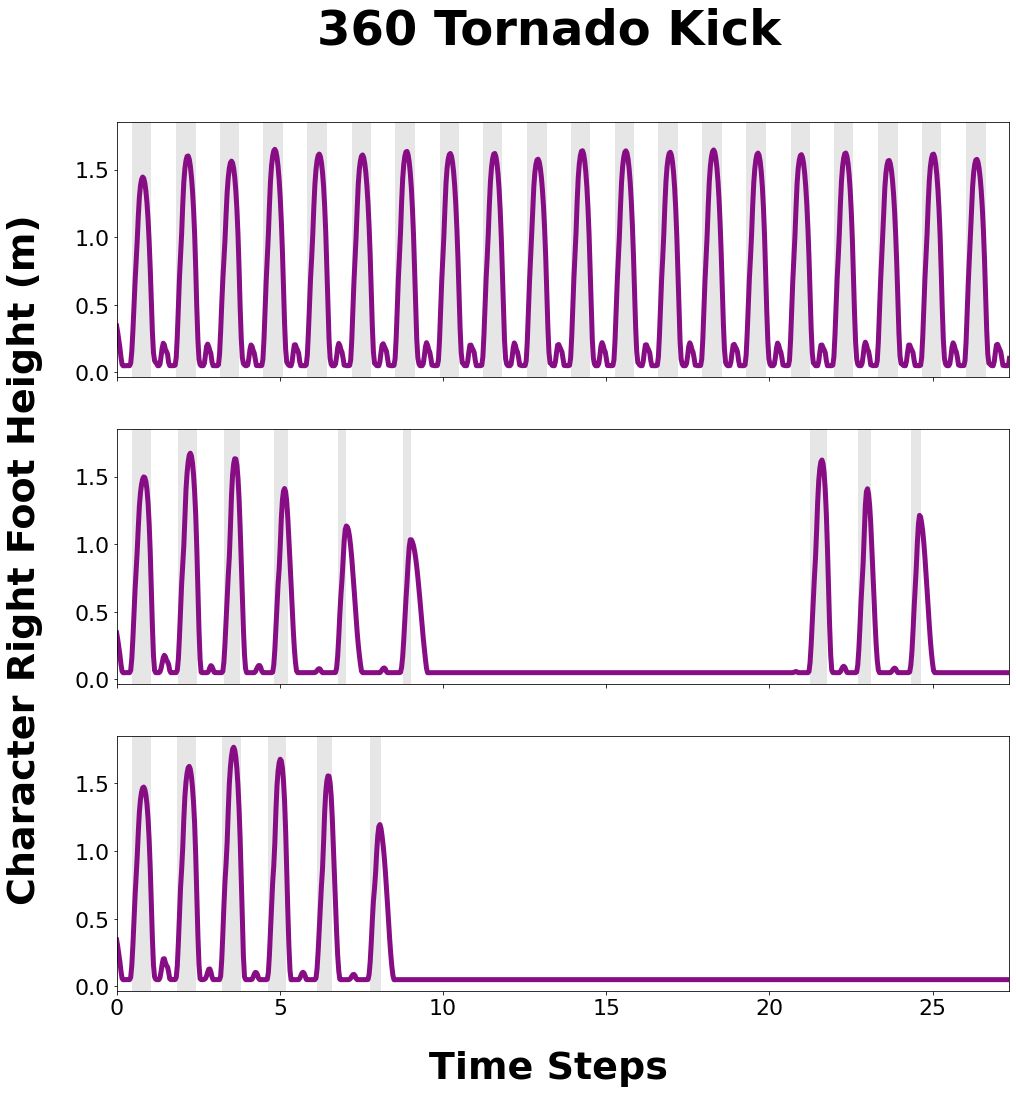}
	\includegraphics[trim={0 3.3cm 0cm 0},clip,width=0.49\linewidth]{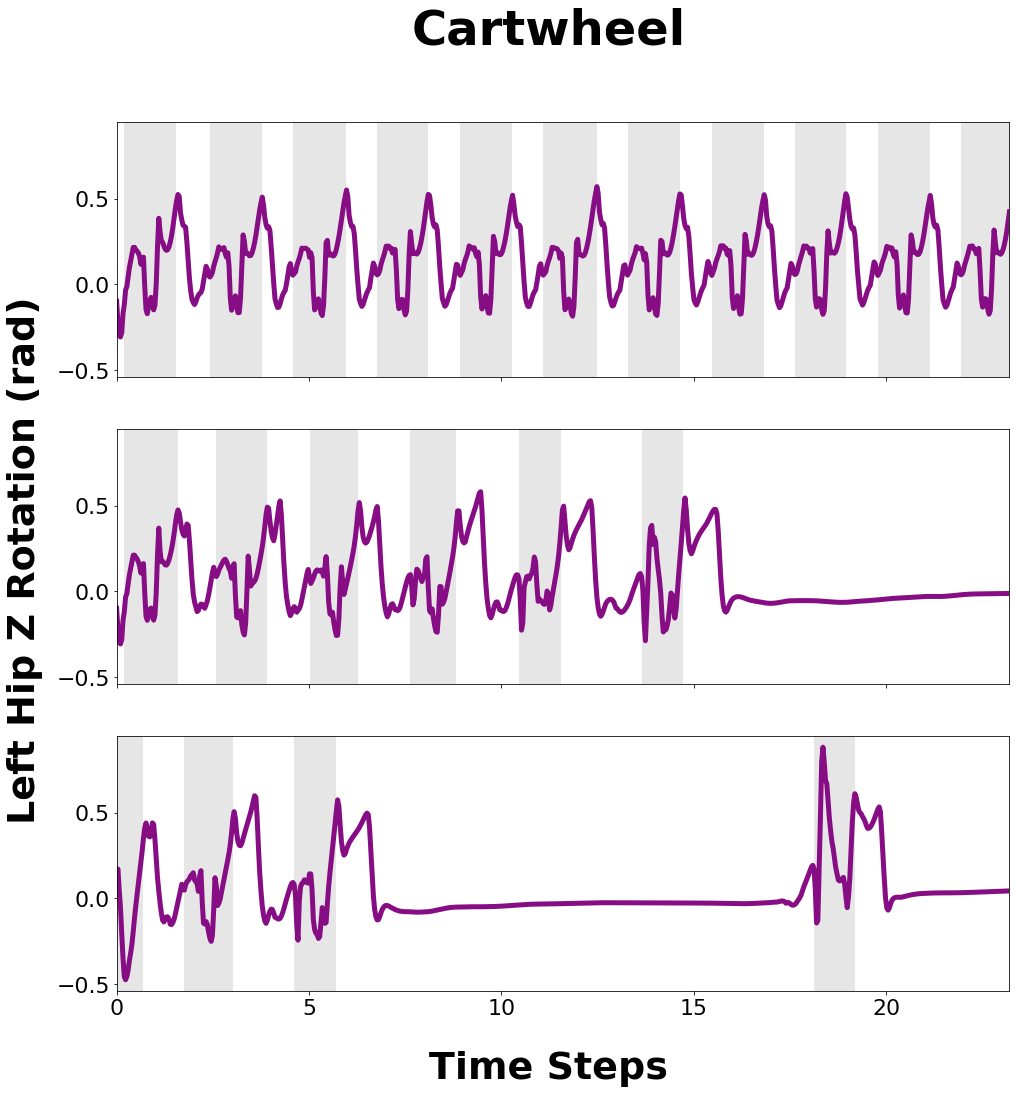}
	\includegraphics[trim={0 0 0cm 3.3cm},clip,width=0.49\linewidth]{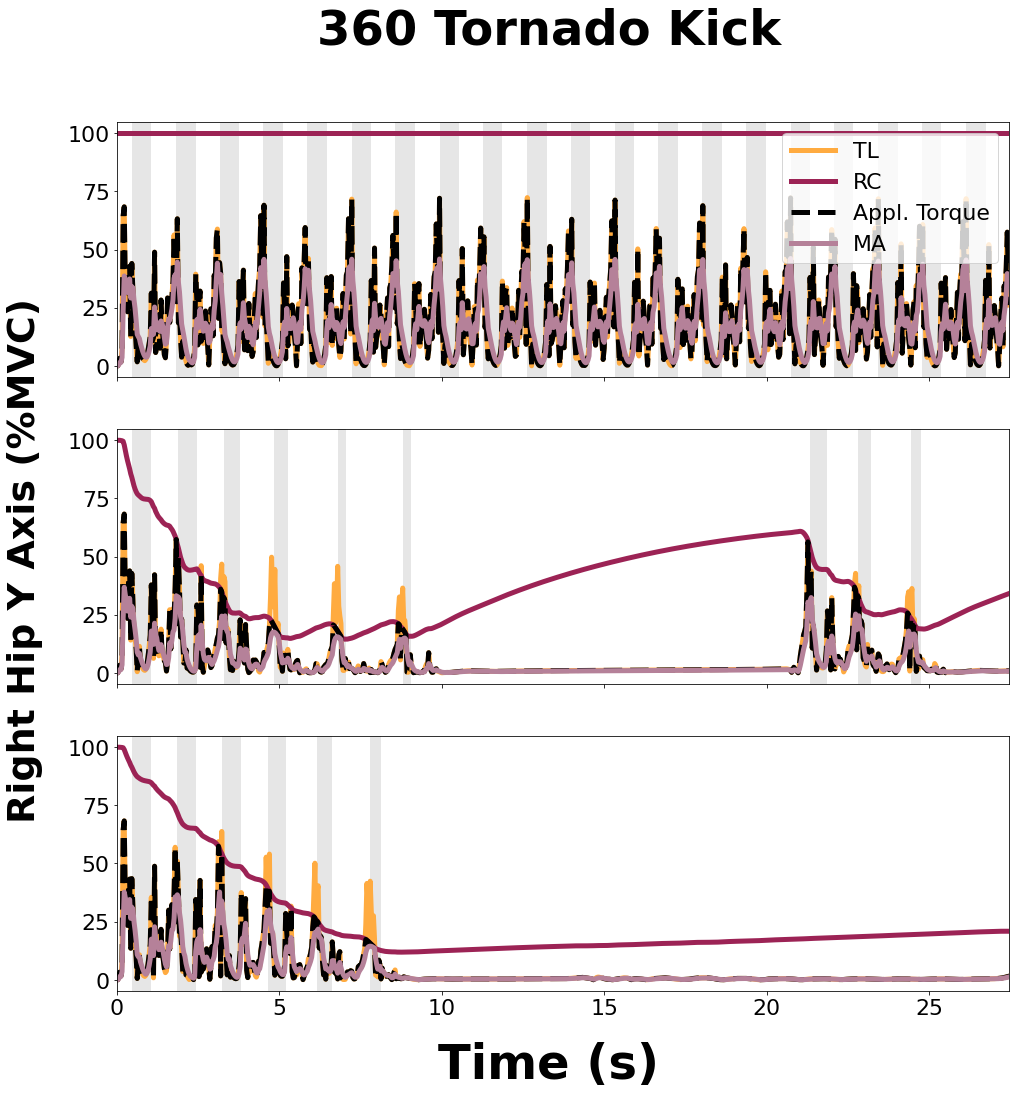}
	\includegraphics[trim={0 0 0cm 3.3cm},clip,width=0.49\linewidth]{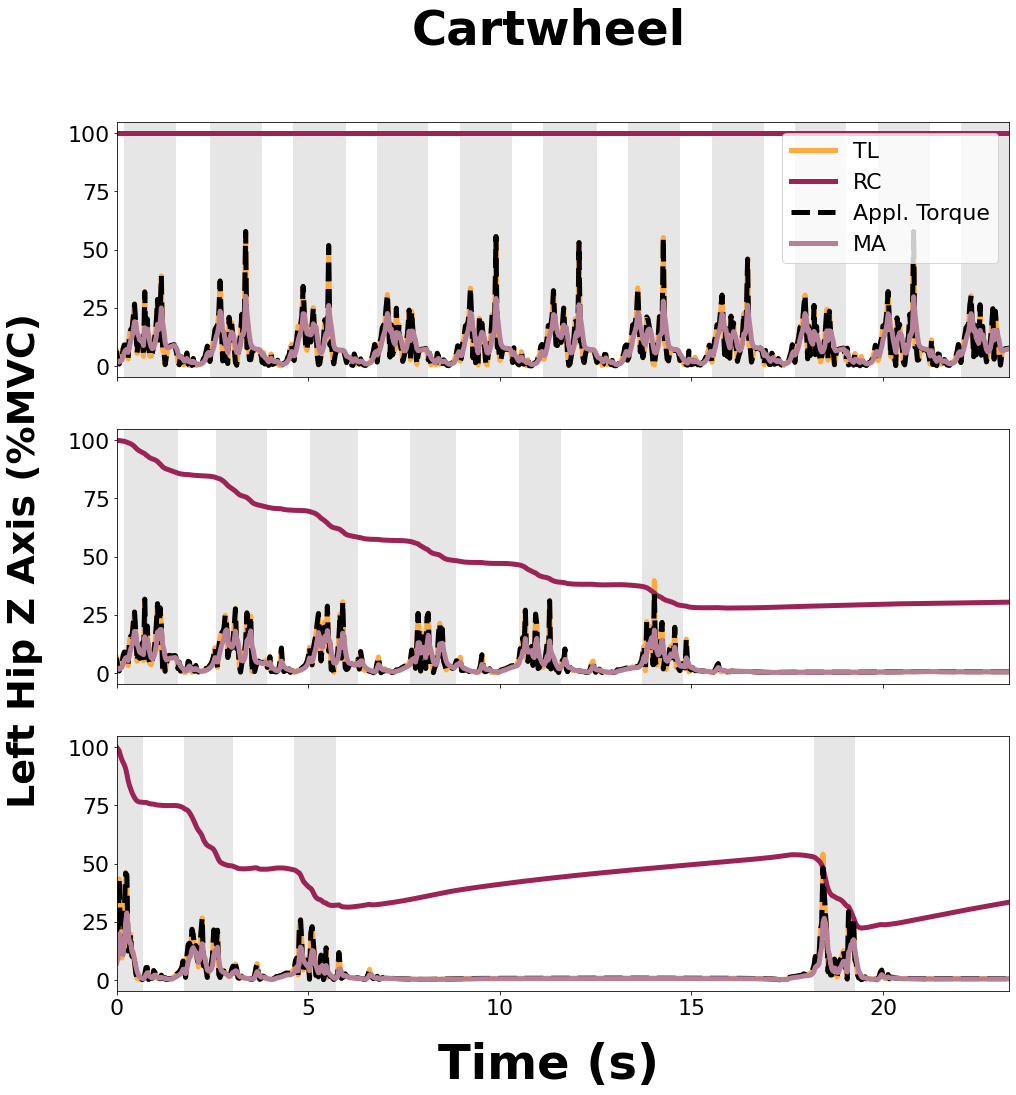}
    \vspace{-12pt}
	\caption{
     \textbf{Movement patterns (\emph{top}) and $\mathbf{\%MVC}$ (\emph{bottom}) for a prominent DoF}. The models show the following $(F, R, r)$ settings (from top to bottom): \textbf{Tornado Kick} -- (0, 0, 0), (2.0, 0.05, 1.0), (1.0, 0.01, 1.0); \textbf{Cartwheel} -- (0, 0, 0), (1.0, 0.01, 1.0), (3.0, 0.02, 1.0)
    The graphs at the bottom show the results for our modification of the 3CC model for animation.
     }
	\label{fig:fatigue_levels}
\end{figure}

\begin{figure}
    \centering
    \includegraphics[trim={5cm 1cm 5cm 6cm},clip,width=0.23\linewidth]{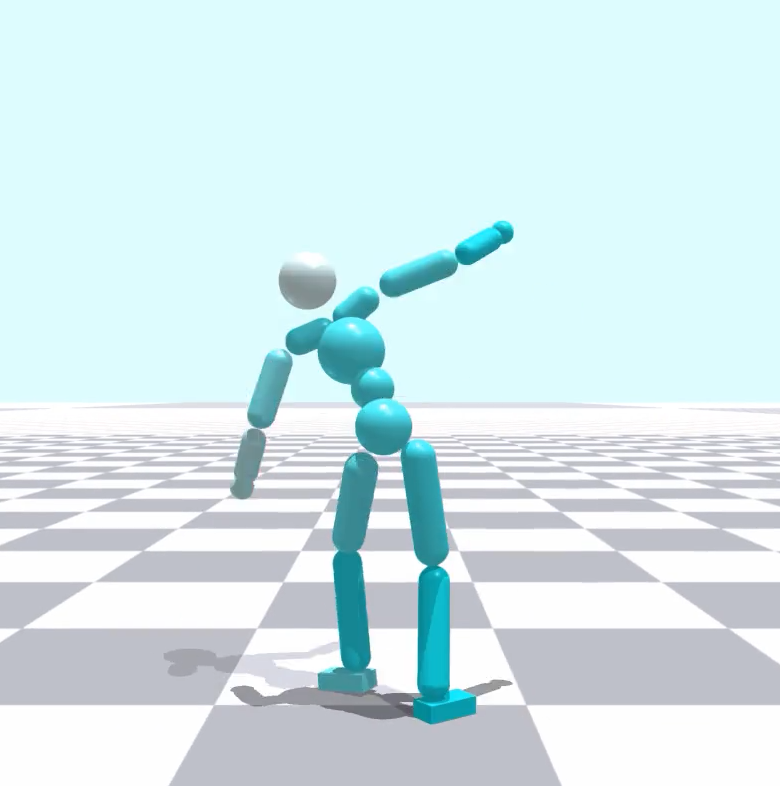}
    \includegraphics[trim={5cm 1cm 5cm 6cm},clip,width=0.23\linewidth]{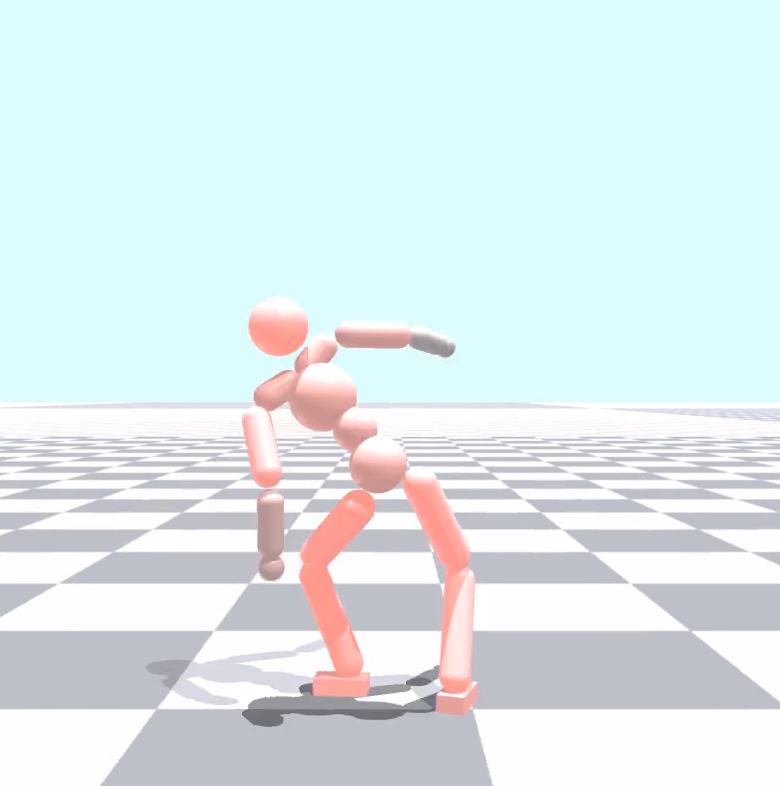}
    \includegraphics[trim={5cm 1cm 5cm 6cm},clip,width=0.23\linewidth]{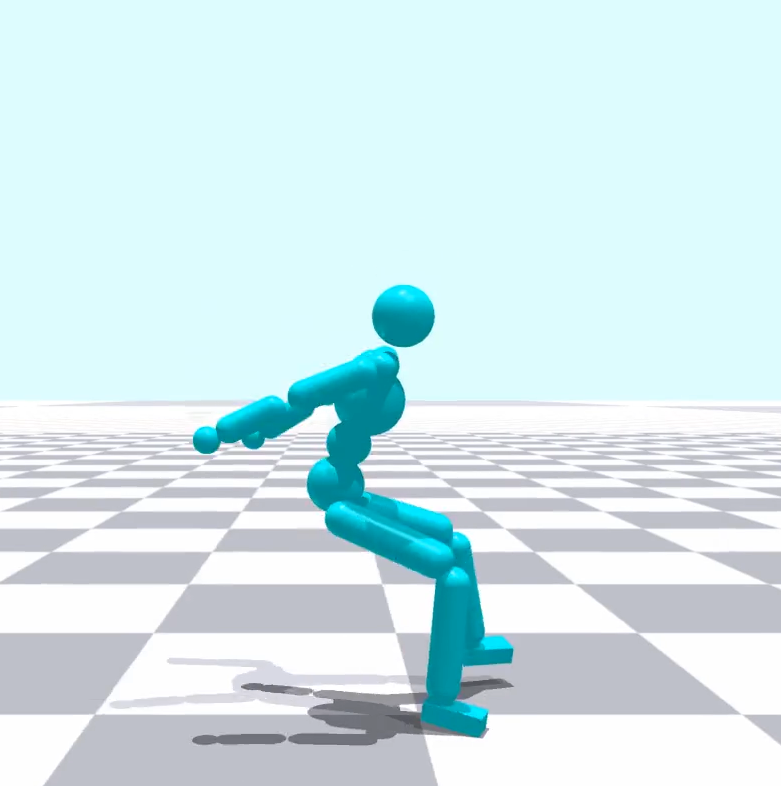}
    \includegraphics[trim={5cm 1cm 5cm 6cm},clip,width=0.23\linewidth]{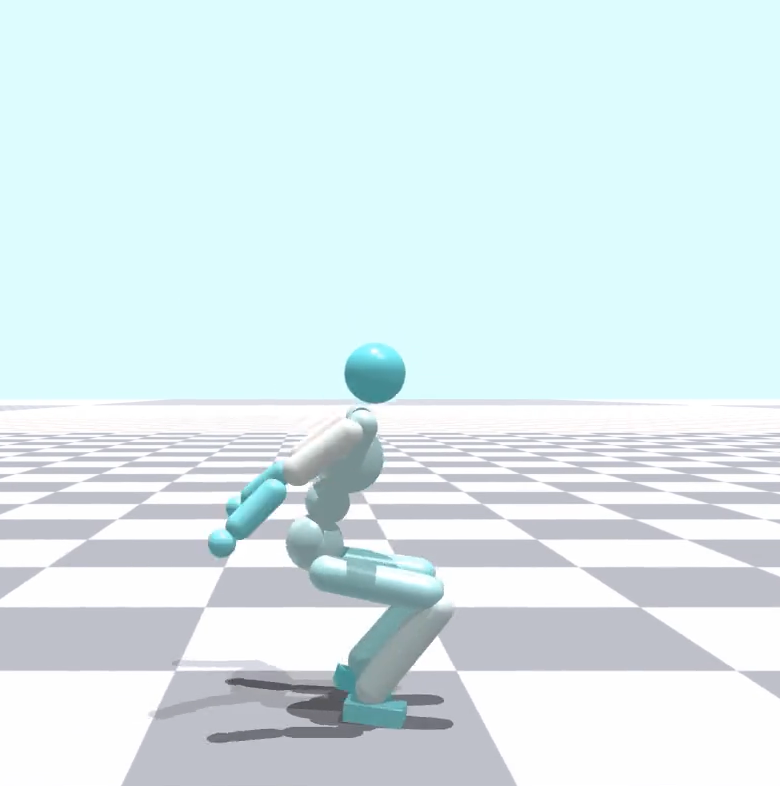}
	\caption
	{
	\textbf{Compensation of forces.} The character compensates forces for movements requiring a lot of momentum such as the cartwheel or backflip by trembling or requiring more suspension.
	}
	\label{fig:compensation}
\end{figure}

\begin{figure}
	\begin{subfigure}{\linewidth}
	\centering
	\includegraphics[trim={0cm 8cm 0cm 5cm},clip,width=0.48\linewidth]{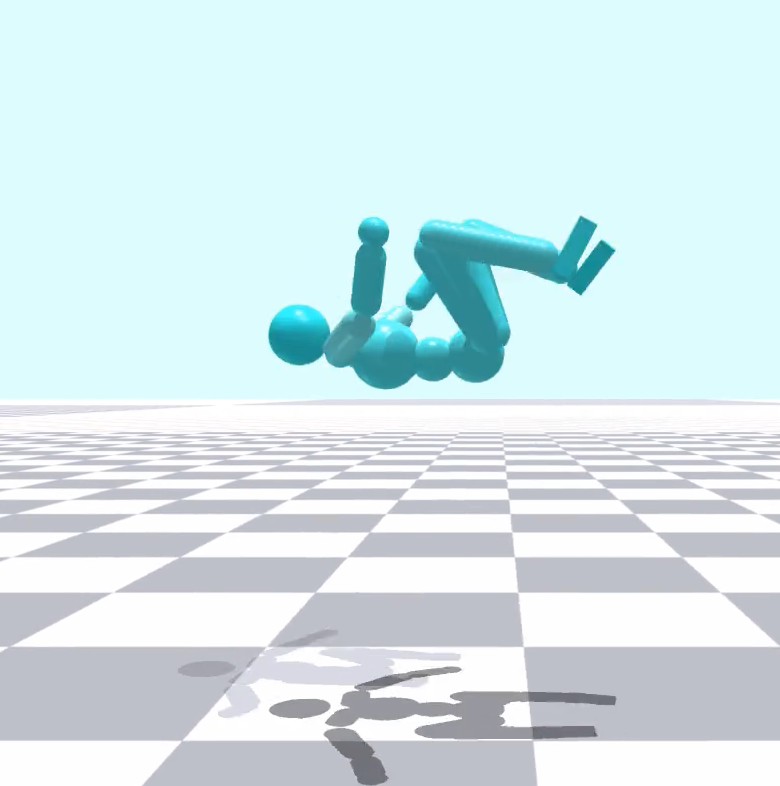}
	\includegraphics[trim={0cm 8cm 0cm 5cm},clip,width=0.48\linewidth]{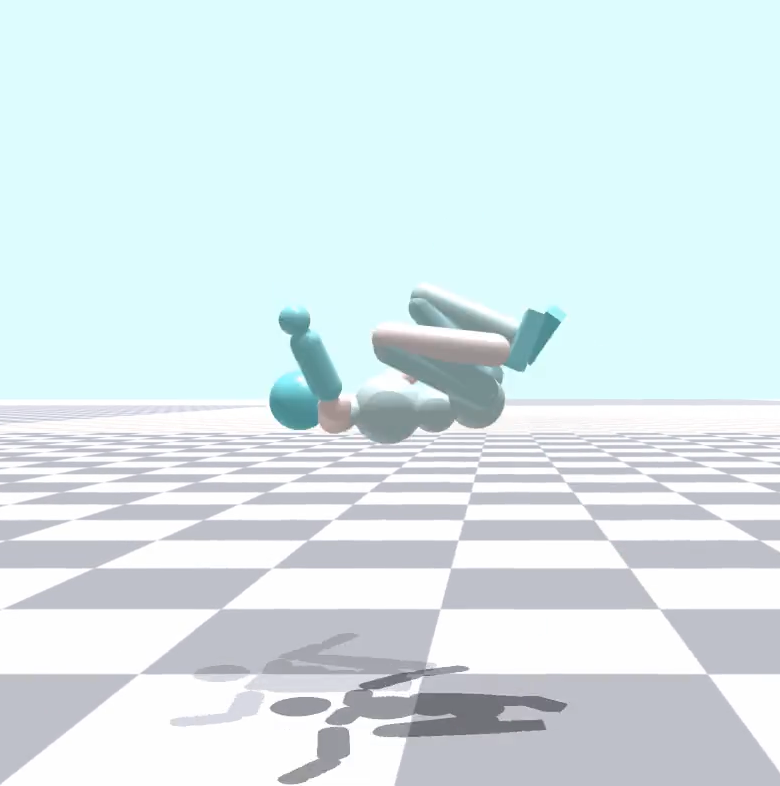}     
	\end{subfigure}
    \vspace{-12pt}
	\caption
	{
\textbf{Change of motion style.} Left: Non-fatigued backflip; Right: Fatigued backflip. The height of the fatigued backflip is lower and the body is less stretched compared to the non-fatigued backflip.
 }
	\label{fig:style}
\end{figure}

\begin{figure*}
    \begin{subfigure}{0.98\linewidth}
    \includegraphics[trim={0 0 30 0},clip,width=\linewidth]{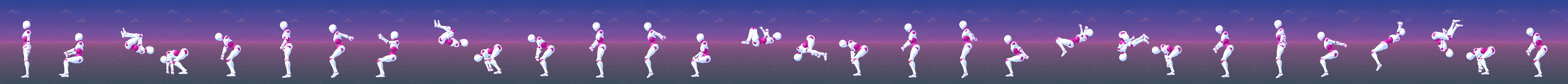}
    \end{subfigure}
    \begin{subfigure}{0.98\linewidth}
    \includegraphics[trim={0 0 30 0},clip,width=\linewidth]{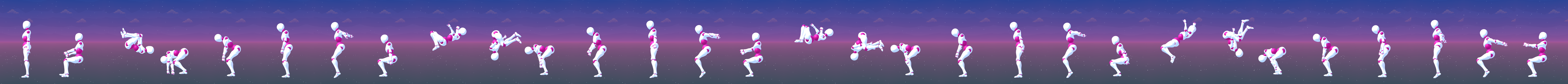}    
    \end{subfigure}
    \begin{subfigure}{0.98\linewidth}
    \includegraphics[trim={0 0 30 0},clip,width=\linewidth]{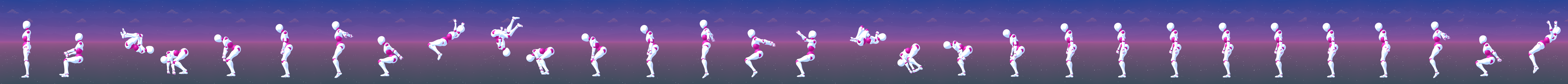}    
    \end{subfigure}
	\caption
	{
	\textbf{Backflip at different ($\mathbf{F,R,r}$)-settings during inference for our model.}
 \emph{Top:} Non-fatigued backflip at $(0, 0, 0)$.
    \emph{Middle:} Moderately fatigued backflips at $(1.0, 0.06, 1.0)$. 
    \emph{Bottom:} High fatigued backflips at $(2.0, 0.07, 1.0)$. 
    }
	\label{fig:backflip_levels}
\end{figure*}

\begin{figure}
    \begin{subfigure}{\linewidth}
    \centering
    \includegraphics[width=\linewidth]{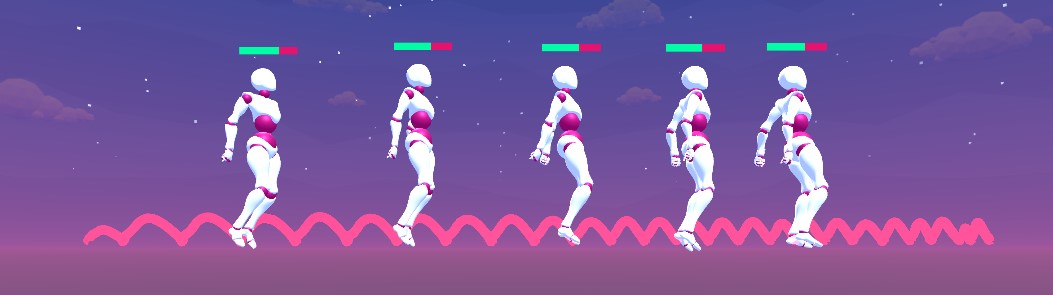}        
    \end{subfigure}
    \begin{subfigure}{\linewidth}
        \centering
        \includegraphics[width=\linewidth]{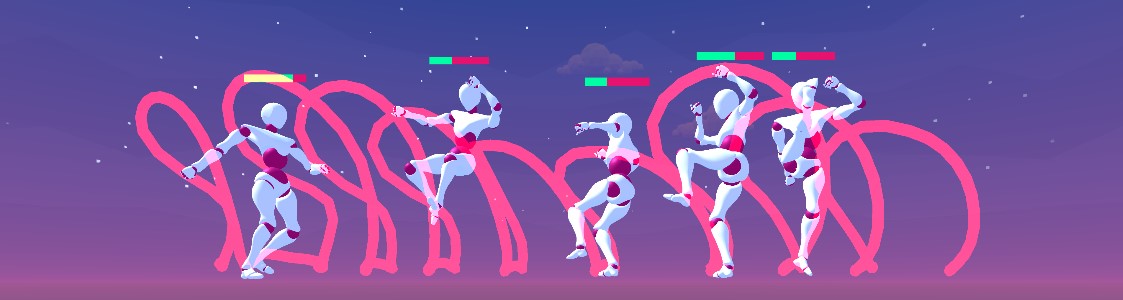}
    \end{subfigure}
    \vspace{-12pt}
	\caption
	{
	\textbf{Change of motion performance.} Hopping (\emph{top}): The distance between each hop decreases with the accumulation of fatigue.
 Tornado kick (\emph{bottom}): The range of the kick and the height of the jump both decrease with the accumulation of fatigue. The character then stands and waits for a significantly long time to recover before performing another kick. 
	}
	\label{fig:performance}
\end{figure}

\begin{figure}
    \begin{subfigure}{\linewidth}
        \centering
        \includegraphics[width=\linewidth, clip, trim={0 60cm 0 60cm}]{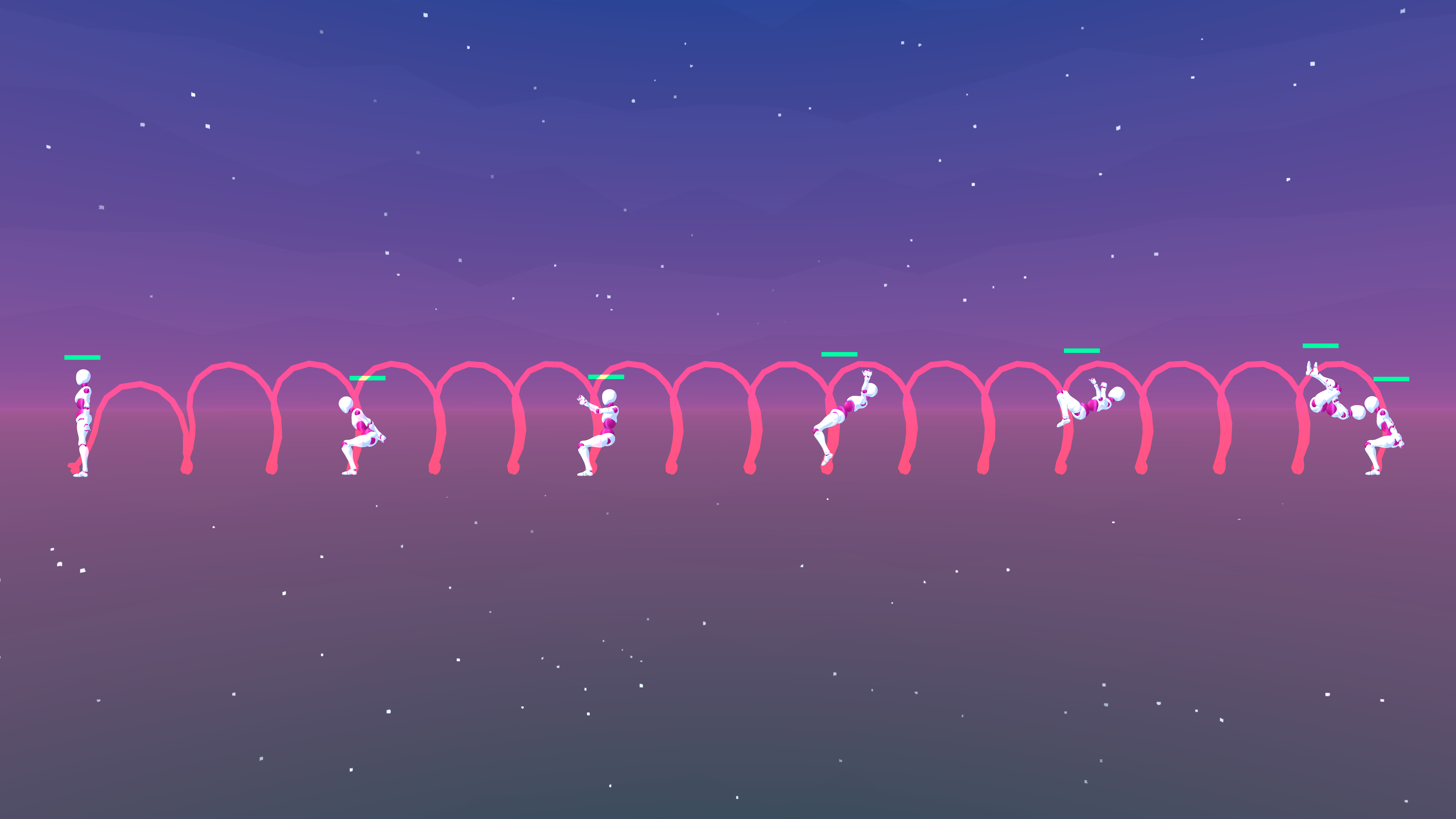}
    \end{subfigure}
    \begin{subfigure}{\linewidth}
        \centering
        \includegraphics[width=\linewidth, clip, trim={0 60cm 0 60cm}]{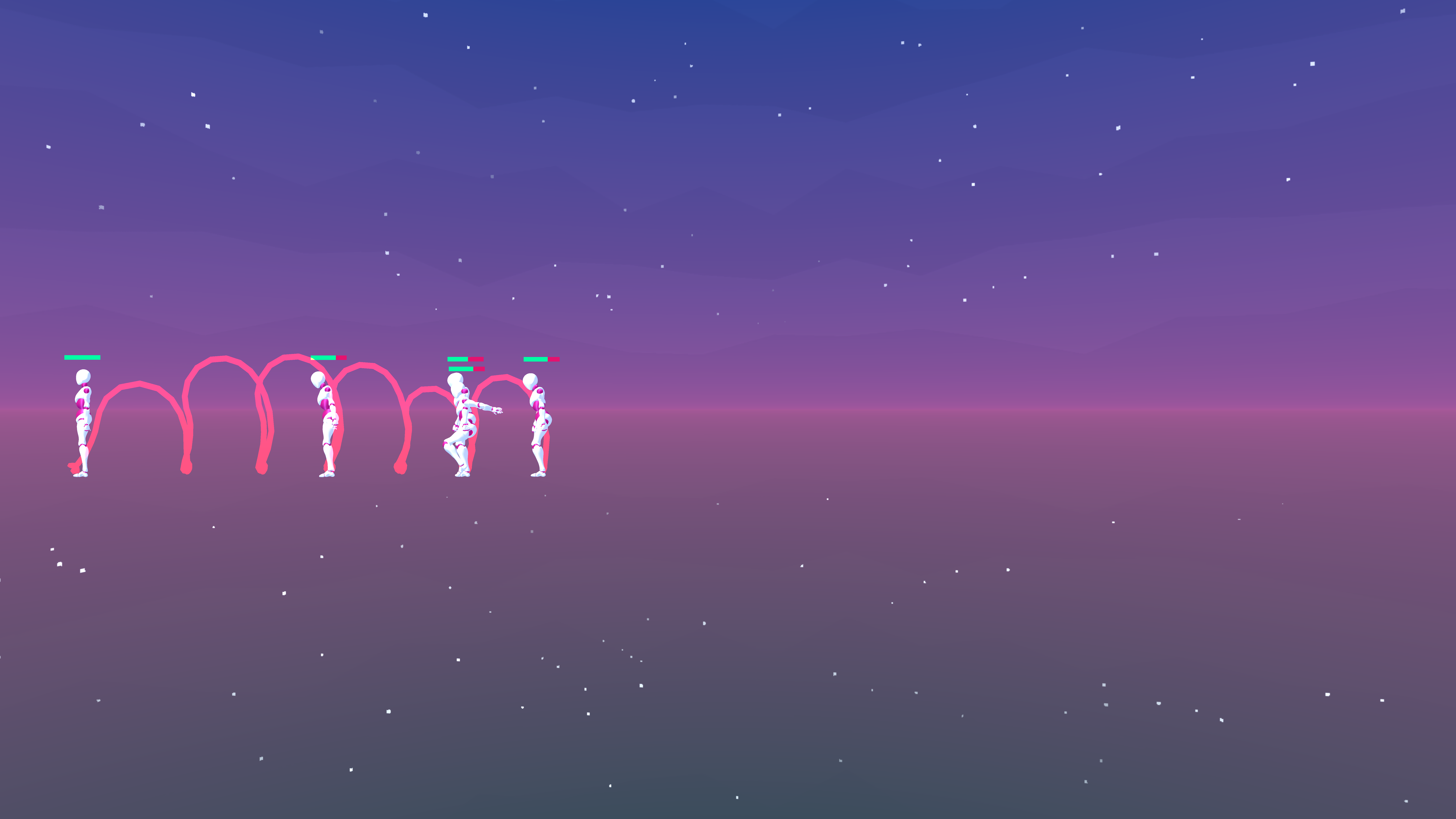}  
    \end{subfigure}
    \begin{subfigure}{\linewidth}
        \centering
        \includegraphics[width=\linewidth, clip, trim={0 60cm 0 60cm}]{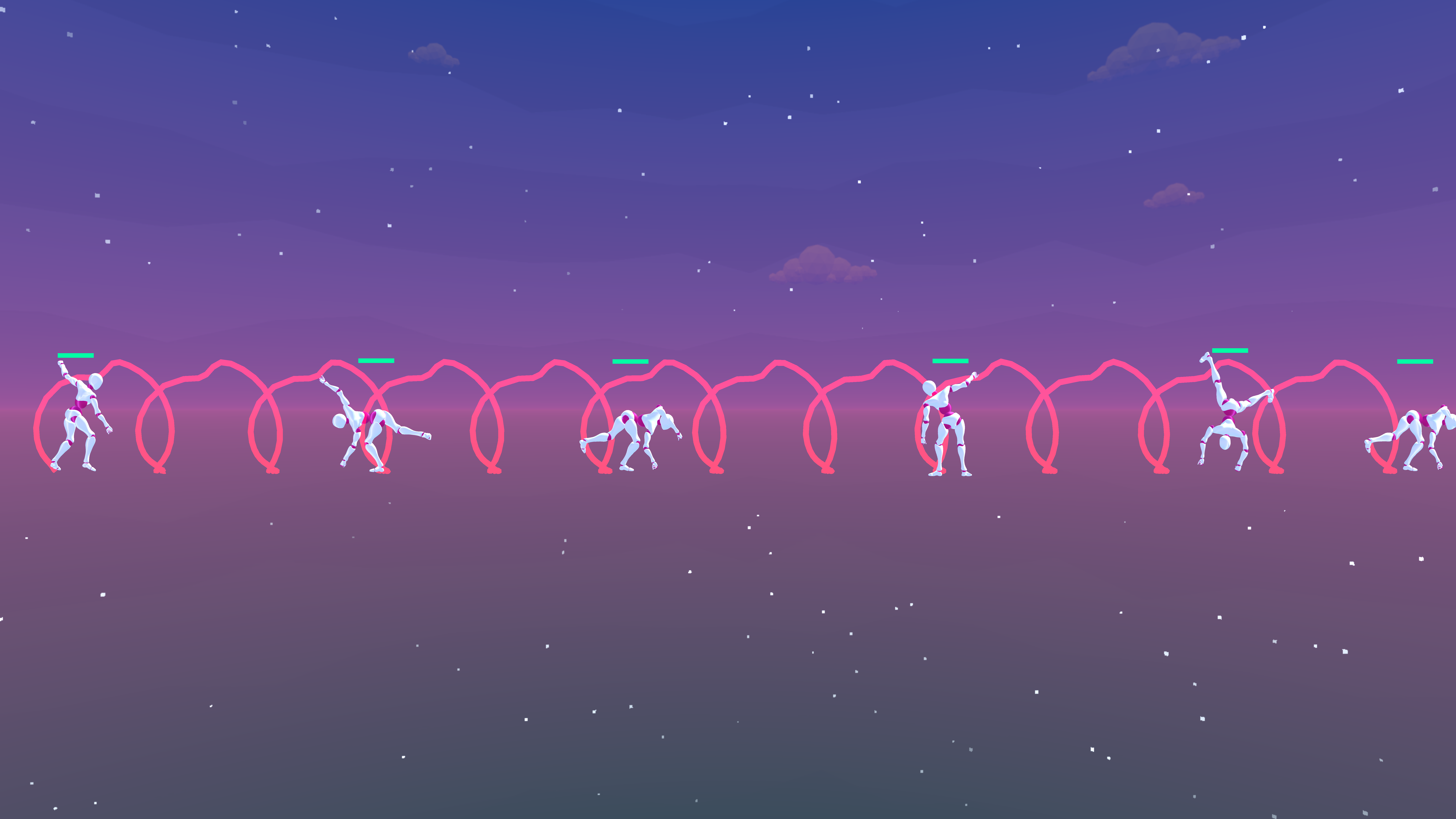}
    \end{subfigure}
    \begin{subfigure}{\linewidth}
        \centering
        \includegraphics[width=\linewidth, clip, trim={0 60cm 0 60cm}]{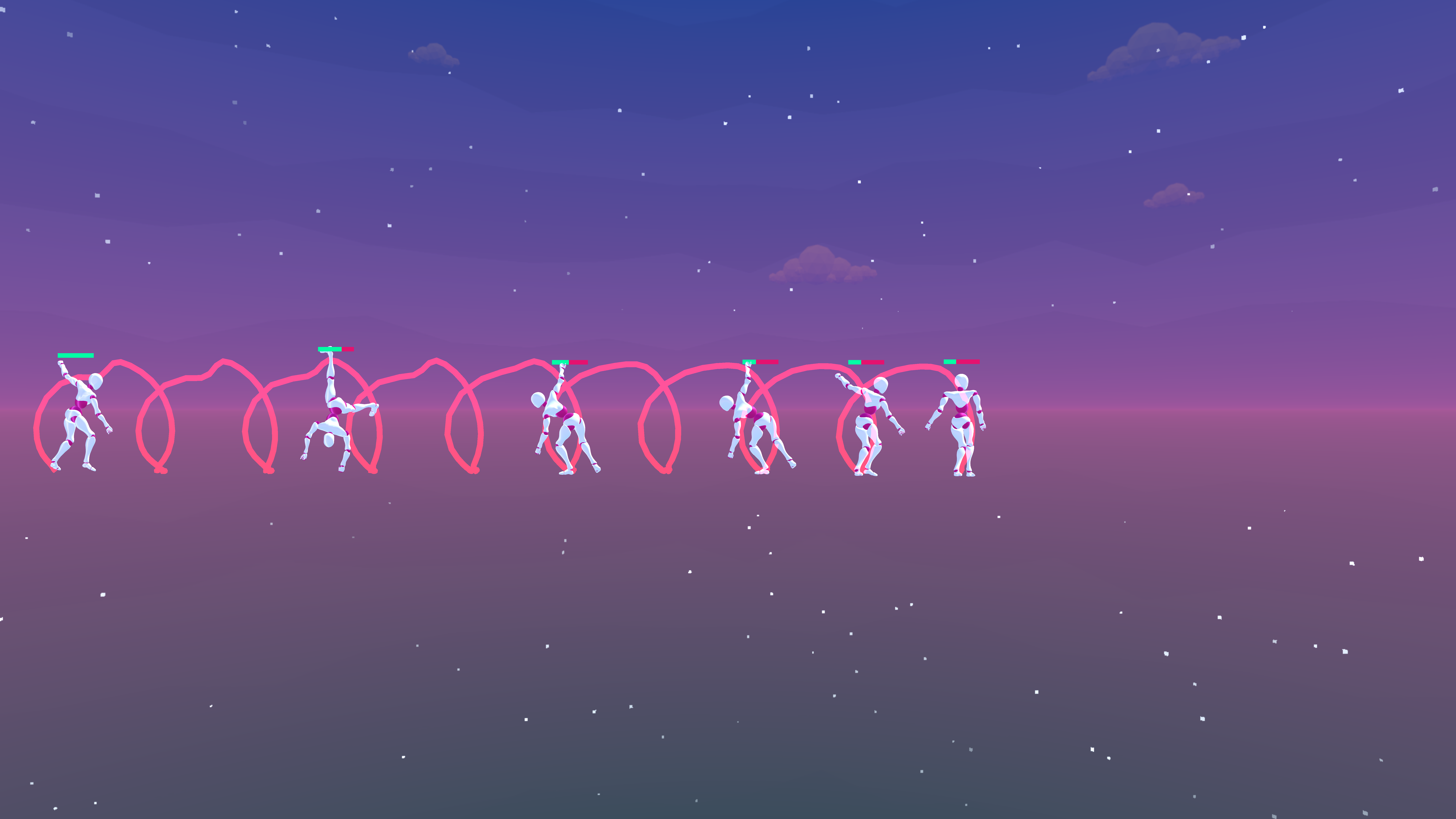}       
    \end{subfigure}
    \caption{\textbf{Reduction of number of repetitions.} The number of repetitions indicated by the number of loops drops between fatigued and non-fatigued motions during the same amount of time. From top to bottom: non-fatigued backflips, fatigued backflips, non-fatigued cartwheels, fatigued cartwheels.}
    \label{fig:num_reps}
\end{figure}

\begin{figure}  
    \begin{subfigure}{\linewidth}
        \centering
    	\includegraphics[width=\linewidth]{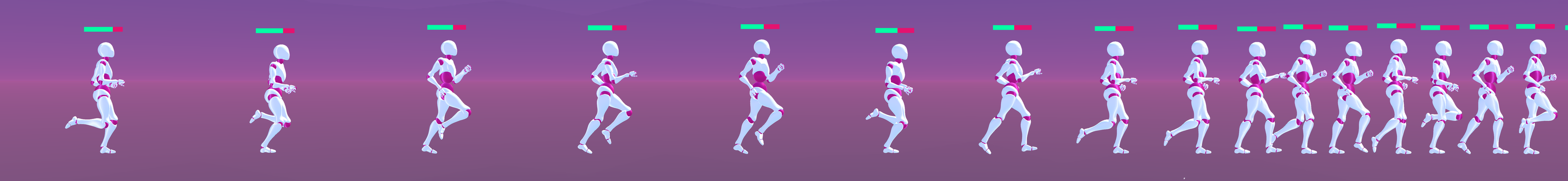}
    \end{subfigure}   

	\caption
	{
	\textbf{Change of speed.} The character becomes slower as it fatigues.
 	}
	\label{fig:speed}
\end{figure}

\begin{figure}
    \includegraphics[width=0.8\linewidth]{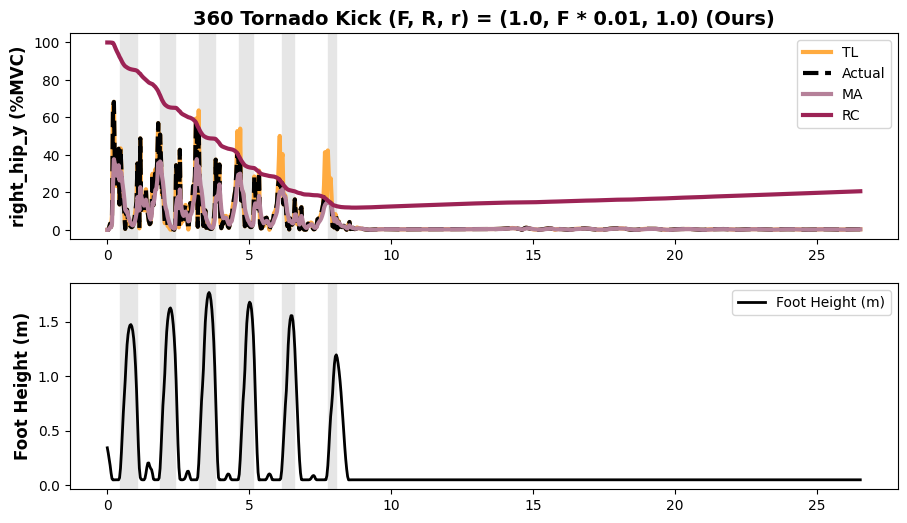}
	\includegraphics[clip, trim={0 0 0 0cm}, width=0.8\linewidth]{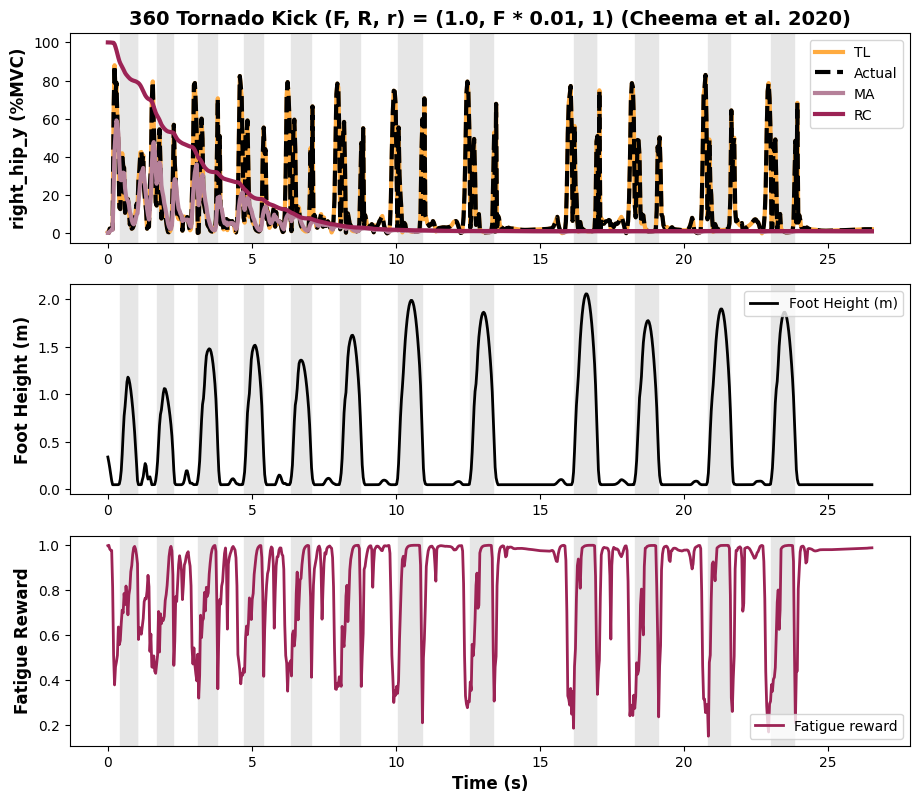}
	\caption{\textbf{Reward-based fatigue transfer learning} \cite{cheema2020predicting}. The grey areas indicate a successful tornado kick. \emph{From top to bottom}: \%MVC for the foot height for the fatigued training setting (1, 0.01, 1) for our method, as well as the corresponding movement pattern. \%MVC for \citet{cheema2020predicting}, as well as the corresponding movement pattern and the fatigue reward \emph{(bottom)}. Note how the movement patterns and applied torque levels do not correctly correspond to the level of fatigue.
 }
	\label{fig:wushu_reward}
\end{figure}

We ablate our method with and without the torque coefficient $\beta$ (Fig.~\ref{fig:abl_coeff}) showing that it improves upon motion smoothness when limiting torques. Not observing fatigue leads additionally to low fidelity motions as can be seen in the supplementary video. To validate our method we show that our model is able to switch from a running clip to a walking clip without motion blending solely based on the fatigue (supp. video). We further compare our method against two baselines: 1) Against a GAIL-baseline based on AMP \cite{peng2021amp}; 2) Against the reward-based fatigue model by \citet{cheema2020predicting}. For the former we use the implementation in Isaac Gym \cite{makoviychuk2021isaac}, whereas for the latter we fine-tune  our pre-trained model with their fatigue-based reward without any torque limitation.
%
%
\begin{table}[]
    \centering
    \caption{\textbf{Comparison between our method and AMP.} Distance comparison against motion capture data between our trained model and \citet{peng2021amp}. The distance describes the average Euclidean distance [and variance] of the normalized joint angles between a generated frame and the closest frame in motion file $\mathcal{M}$. The distance is averaged over an episode.}
    \vspace{-10pt}
    \label{tab:mocap}
    \begin{tabular}{@{}c|c@{ | }c@{ | }c@{ | }c@{ | }c@{}}
         & \textbf{Backflip} & \textbf{Cartwh.} & \textbf{Hopping} & \textbf{Locomot.} & \textbf{360 Kick} \\
         \hline
        \textbf{AMP} & 0.18[0.06] & 0.24[0.07] & 0.16[0.03] & 0.23[0.05] & 0.24[0.08]\\
        \hline 
        \textbf{Ours} & 0.44[0.36] & 0.25[0.09] & 0.32[0.33] & 0.36[0.28] & 0.32[0.17]
    \end{tabular}
    \vspace{-10pt}
\end{table}
\paragraph{GAIL without Fatigue Control} We observe in Tab.~\ref{tab:mocap} that our model is able to synthesize novel fatigued behaviours and recovery strategies not present in dataset, whereas \citet{peng2021amp, peng2022ase} are solely able to imitate existing motion capture disregarding any variance over time due to cumulative fatigue (see also supplemental video). In contrast to that our method allows for interactive control of fatigue and varying degrees of fitness levels over time by changing $(F, R, r)$ values during inference (see Fig.~\ref{fig:teaser}). Previous methods \cite{peng2021amp, peng2022ase} are not able to provide such interactive control and varying degrees of fitness levels in one policy. 
%
%
\paragraph{Fatigue-based Reward}
The closest method~\cite{cheema2020predicting} to our's uses a reward based on the difference between $M_A$ and $TL$. To compare our torque-limit-based method to their reward-based method, we use their reward for fatigue fine-tuning of our GAIL-baseline. The reward is added to $r_t$ in Eq.~(\ref{eq:gail_reward}). 
The results can be observed in Fig.~\ref{fig:wushu_reward}. We show that despite their method being based on the 3CC model and resulting in accurate fatigue analysis, it is not able to synthesize correct fatigued behaviours over time, especially when combined with a GAIL-based reward. Fig.~\ref{fig:wushu_reward} \emph{(forth subplot from the top)} shows that while $RC$ is high, the agent does a lot of fatigued kicks with lower heights. These however become increasingly higher as the $RC$ lowers, since then the agent obtains most of its reward from the imitation reward. The agent furthermore then seems to exploit the imitation learning policy by doing higher jumps than that are in the dataset since the agent can ``rest" during the fall of the increased air time. Using the opposite of this reward instead leads to imitation learning in the beginning and faster motions later on, which require even more energy. This effect holds true for several hyperparameter combinations for their reward. While \citet{cheema2020predicting} have used their method to merely enhance the naturalness of pointing movements during rest periods, as well as to analyse and predict ergonomic differences of user interface configurations, we found that combining their method with an imitation learning policy fails to actually synthesize expected fatigue behaviour and requires careful tuning of reward parameters.
Additionally, we noticed that a reward-based method does not give an intuitive control over the 3CC-parameters as it is the case for our torque-limit based model which is described in Sec.~\ref{sec:fatigue_training}. 
Fig.~\ref{fig:wushu_reward} also shows that a method solely based on reward does not correspond to the actual strength capabilities that should be left with a given $(F,R,r)$ setting, as it is not a hard constraint. Setting $R=0.2F$ means that at maximum fatigue $20\%$ strength should be left. However, a reward-based method does not account for such physiological restrictions, while our joint actuation torques actually correspond to the level of residual capacity.
%
%
\vspace{-5pt}
\section{Conclusion} \label{sec:conclusion}
We presented the first full-body approach for physics-based 3D humanoid motion synthesis with fatigue. 
Our experiments demonstrate the emergence of realistic fatigued movements and recovery behaviours for interactive athletic animations that are difficult to produce with previous techniques (especially without muscoskeletal simulation). 
Our torque-based fatigue simulation system can be efficiently used for real-time interactive 3D virtual character animation, opening up new possibilities for games and animation tools. 
Additionally, we show in the supplementary material that due to the morphology-agnostic nature of the 3CC-model our method can be applied to characters of any physiology, as long as a measure of $\%MVC$ as a function of either torque or force can be provided. It can further be easily extended with additional task-rewards such as target goals or orientations.
We believe this work opens many exciting pathways for biomechanical simulations and physics-based character animation, particularly in terms of discovering automatic emergent behaviours for intelligent agents.
%


\begin{acks}
Noshaba Cheema was supported from the EIC Pathfinder grant Carousel+ (101017779) and an IMPRS fellowship. Rui Xu was supported from the BMWi grant KAI (19A21022C). Nam Hee Kim received support from the FCAI Open Application grant and an NSERC Postgraduate Scholarship (567794-2022). The authors from MPII were supported by the ERC Consolidator Grant 4DRepLy (770784). 
We thank Erik Herrmann and Han Du for their retargeting framework and for providing invaluable feedback, alongside Jaakko Lehtinen. We also wish to express our appreciation to Ahmed `The Music Jnn' Hadrich for the sound design; as well as to Laura Morales, Seth Izen and Tobias Cheema for their additional support with the supplementary video and figures.
\end{acks}

\bibliographystyle{ACM-Reference-Format}
\bibliography{paper}
\clearpage

\clearpage

\title{-Supplementary Document-}

\appendix

\section{Goal-oriented Tasks}
\begin{figure}
    \centering
    \includegraphics[width=\linewidth]{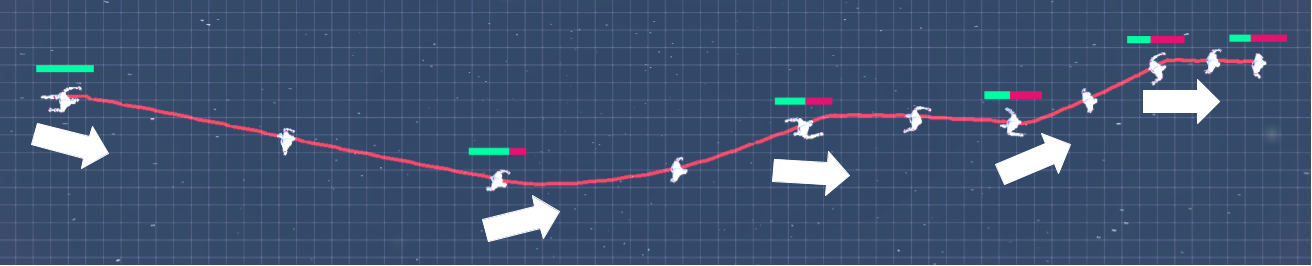}
    \caption{\textbf{Goal-oriented Walking with Cumulative Fatigue.} The character is shown at a 5s interval, while the target direction is changed every 10s as indicated by the arrows. One can observe that the character walks slower while the fatigue accumulates.}
    \label{fig:walk_goal}
\end{figure}

We conduct a goal-oriented simulation task in which a character moves toward a goal while experiencing growing fatigue. To accomplish this, we employed two distinct environments: 1) The humanoid character utilizing the GAIL-based pre-trained expert, 2) a four-legged spider, specifically the Isaac Gym `Ant' with 8 degrees of freedon (DOFs), representing a character with a distinct morphology that is challenging to replicate through motion capture.
\paragraph{GAIL-Humanoid}
We expand upon the state space $\mathbf{s}_t$ outlined in Section 5 of the main document by incorporating a 2D-direction vector representing the character's intended heading. The reward function in Eq. 8 is then augmented by adding the following heading reward:
\begin{equation*}
    r_{dir_t} = \left(\left<\frac{\Gamma_t}{\|\Gamma_t\|},\frac{\hat{\Gamma}}{\|\hat{\Gamma}\|}\right> + 1\right) \cdot 0.5
\end{equation*}
With $\Gamma_t$ describing the current local root x,y-orientation and $\hat{\Gamma}$ the target direction. $\hat{\Gamma}$ is randomized at every environment reset. During inference we set $\hat{\Gamma}$ every 10s, which can be seen in Fig.\ref{fig:walk_goal}. 

\begin{figure}
    \includegraphics[trim={18cm 0cm 18cm 20cm},clip,width=0.19\linewidth]{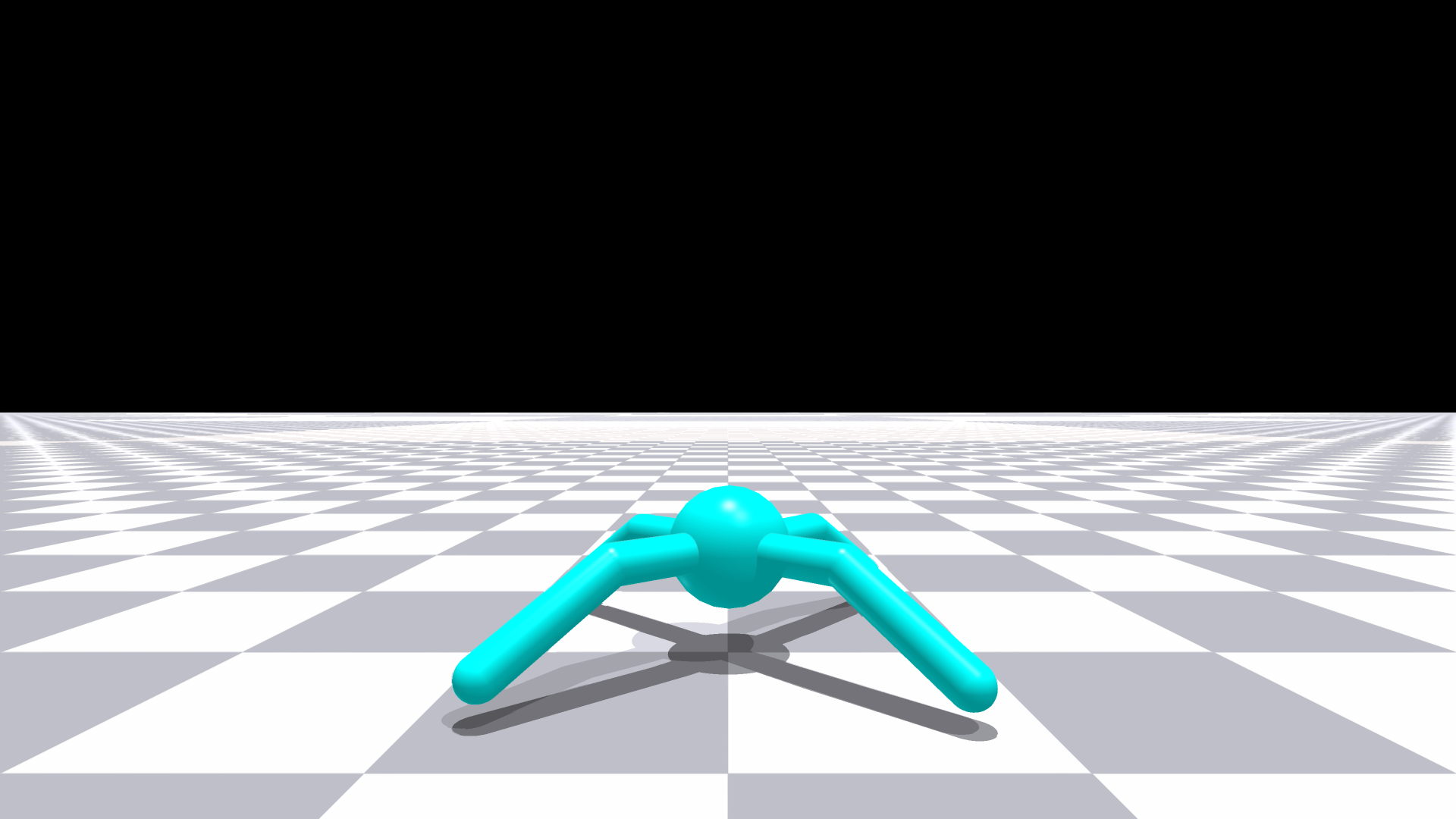}
	\includegraphics[trim={18cm 0cm 18cm 20cm},clip,width=0.19\linewidth]{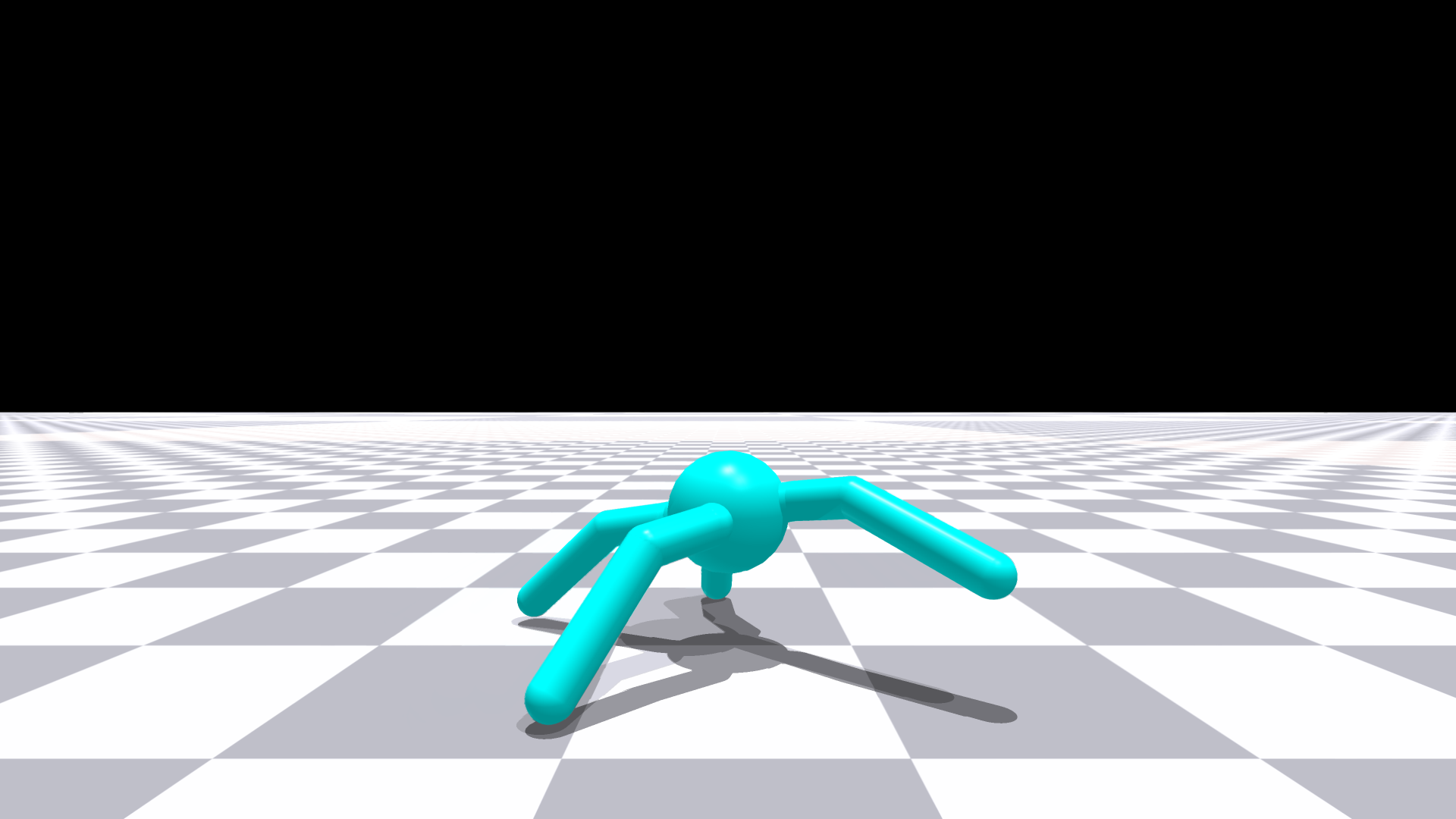}
	\includegraphics[trim={18cm 0cm 18cm 20cm},clip,width=0.19\linewidth]{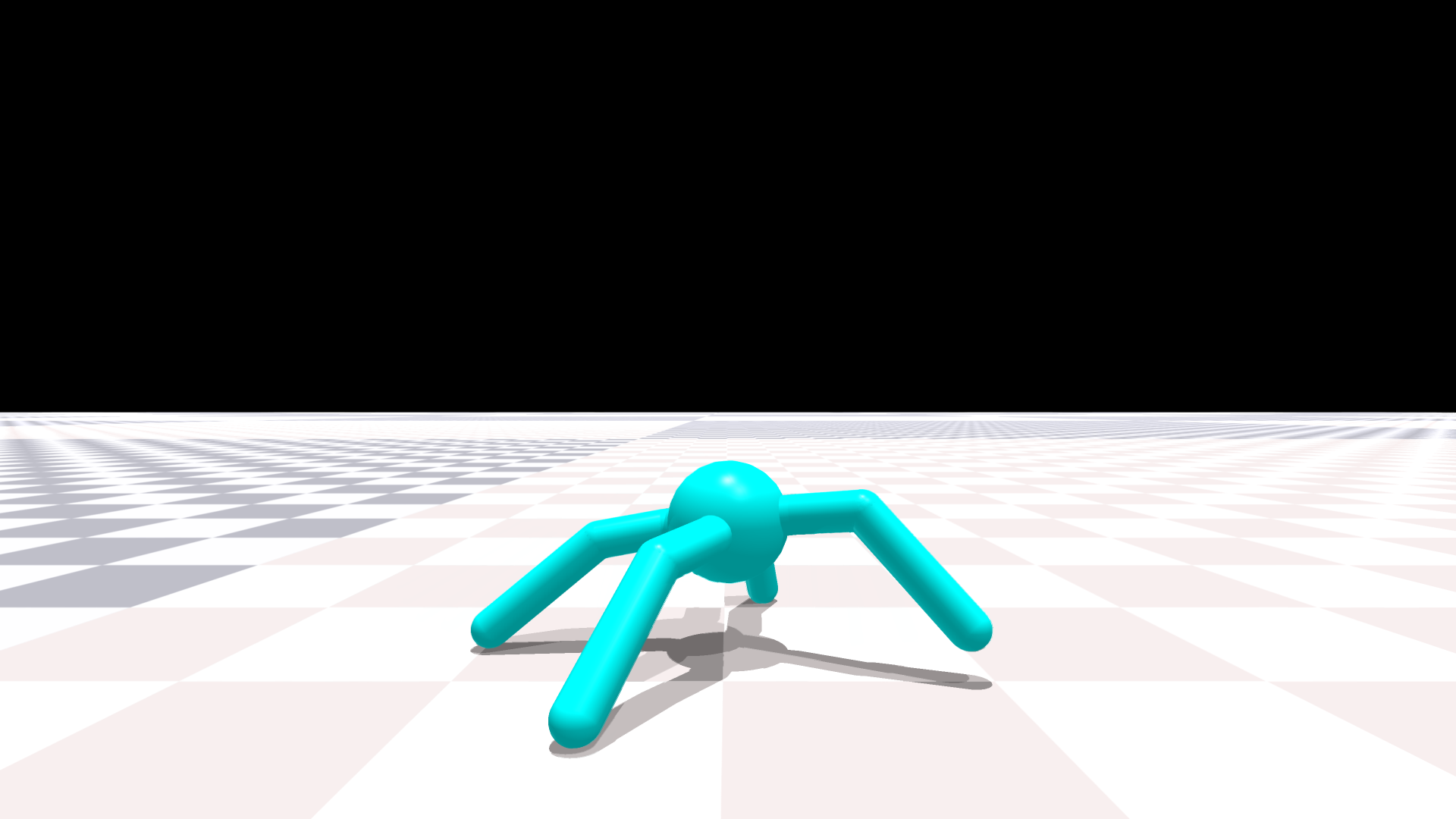}
	\includegraphics[trim={18cm 0cm 18cm 20cm},clip,width=0.19\linewidth]{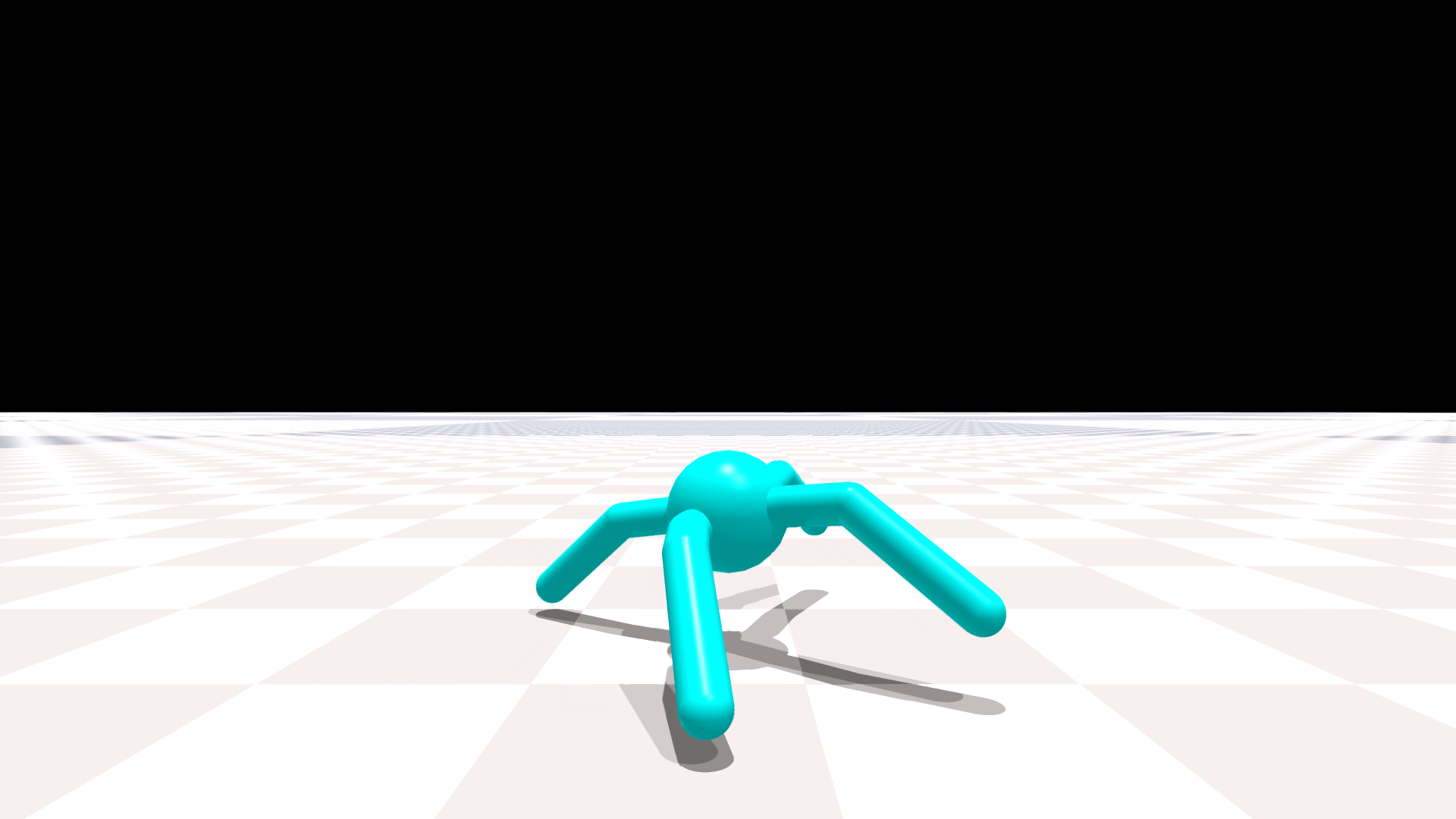}
    \includegraphics[trim={18cm 0cm 18cm 20cm},clip,width=0.19\linewidth]{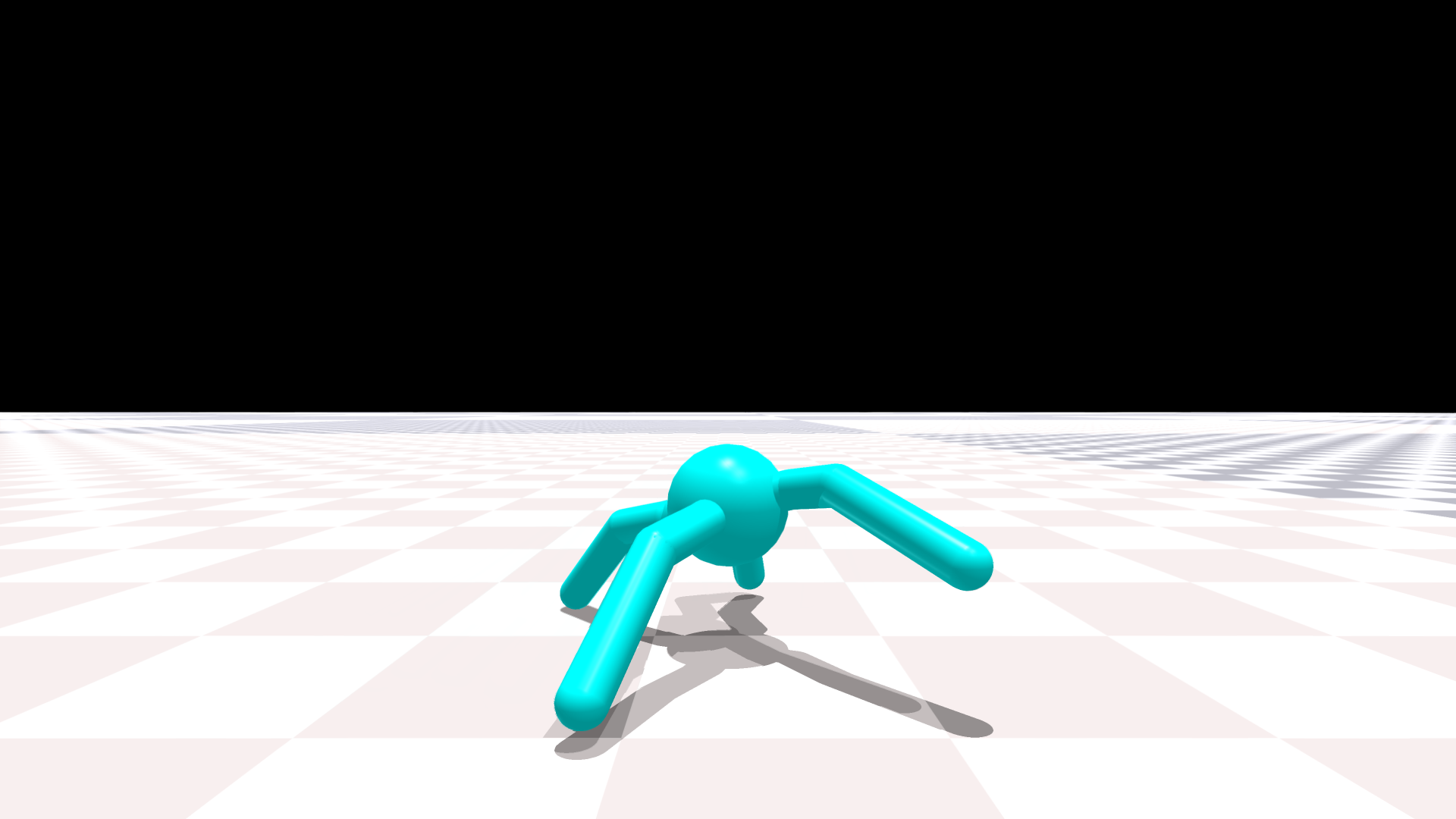}
    \includegraphics[trim={18cm 0cm 18cm 20cm},clip,width=0.19\linewidth]{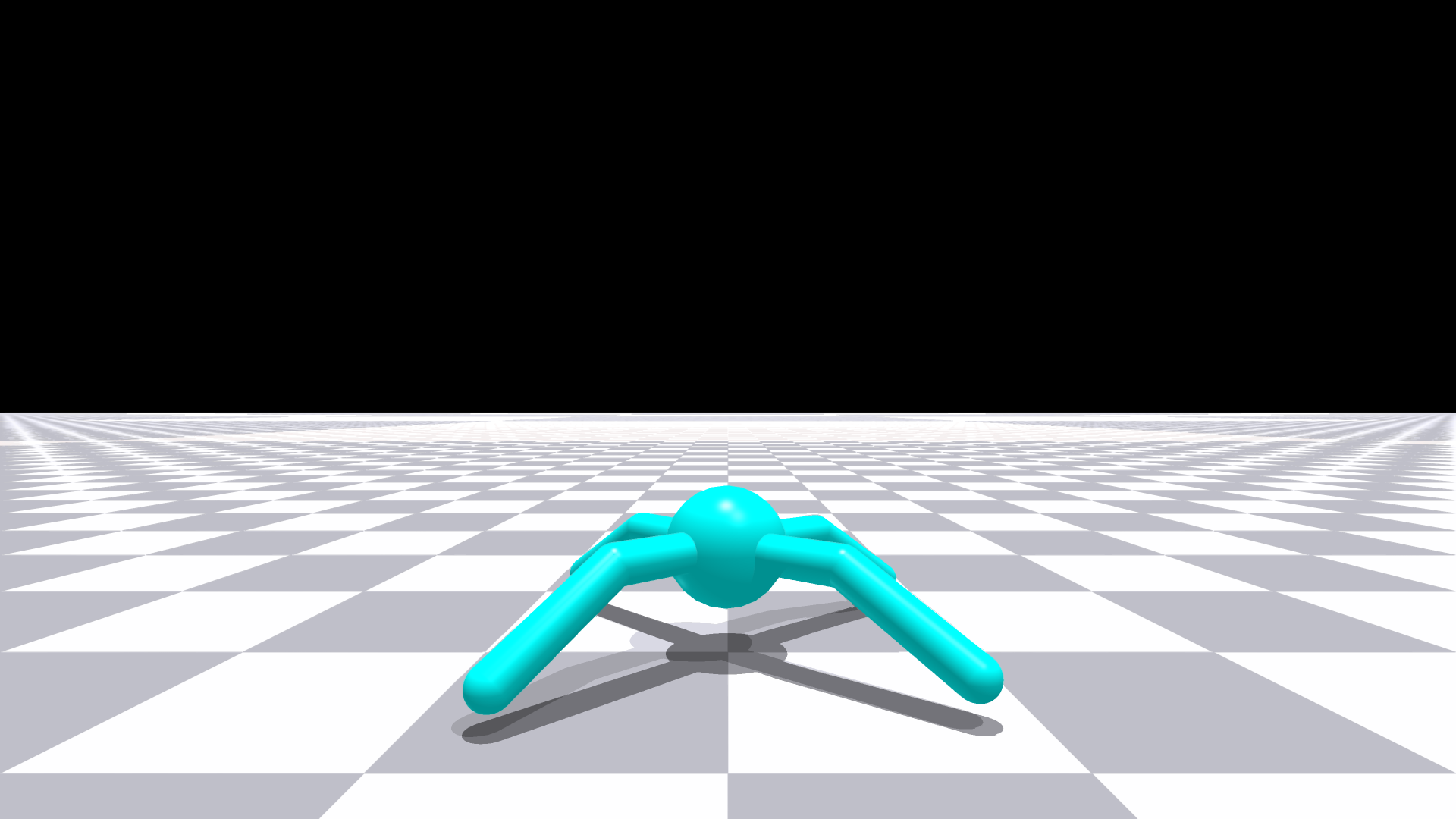}
	\includegraphics[trim={18cm 0cm 18cm 20cm},clip,width=0.19\linewidth]{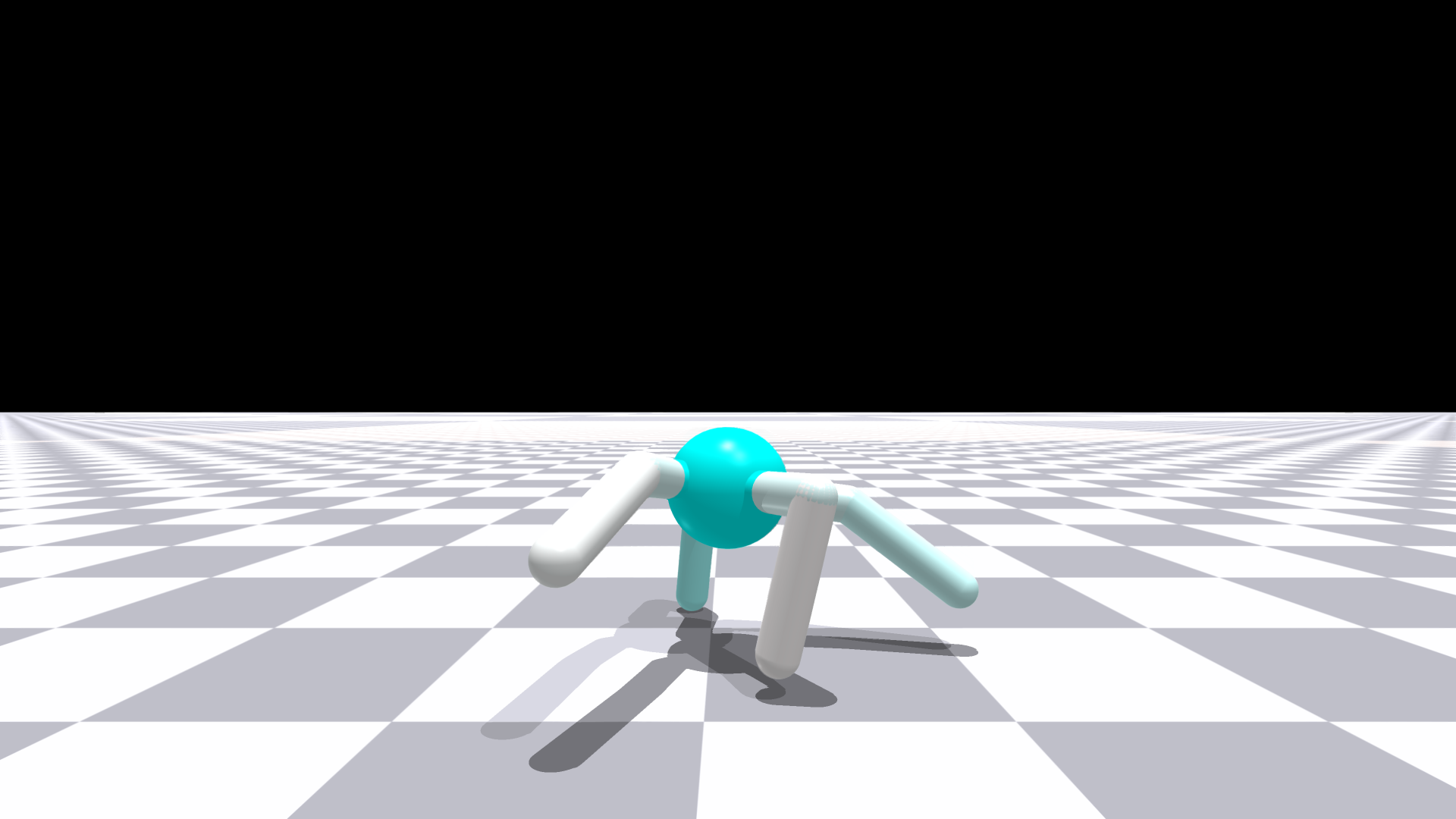}
	\includegraphics[trim={18cm 0cm 18cm 20cm},clip,width=0.19\linewidth]{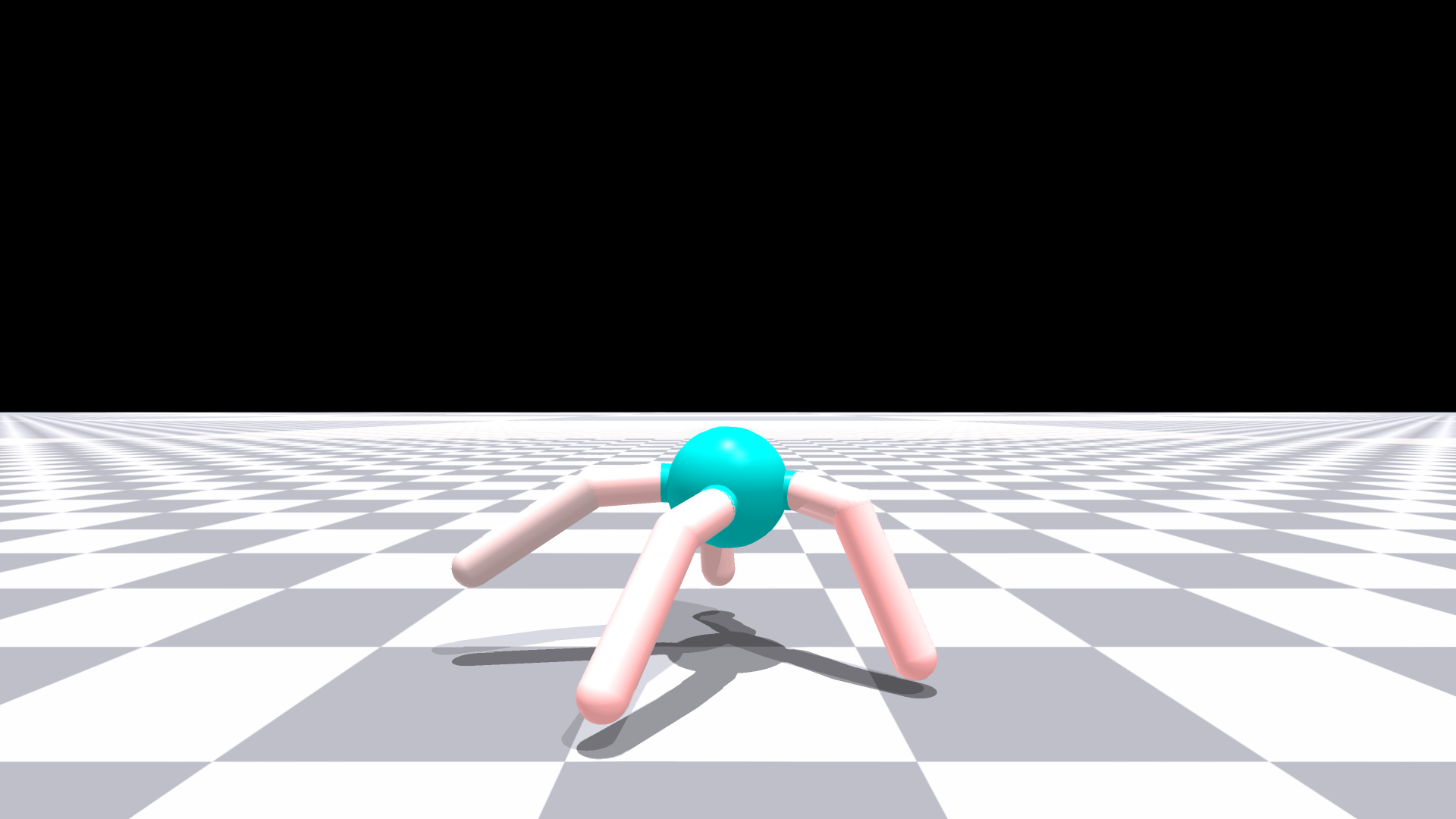}
	\includegraphics[trim={18cm 0cm 18cm 20cm},clip,width=0.19\linewidth]{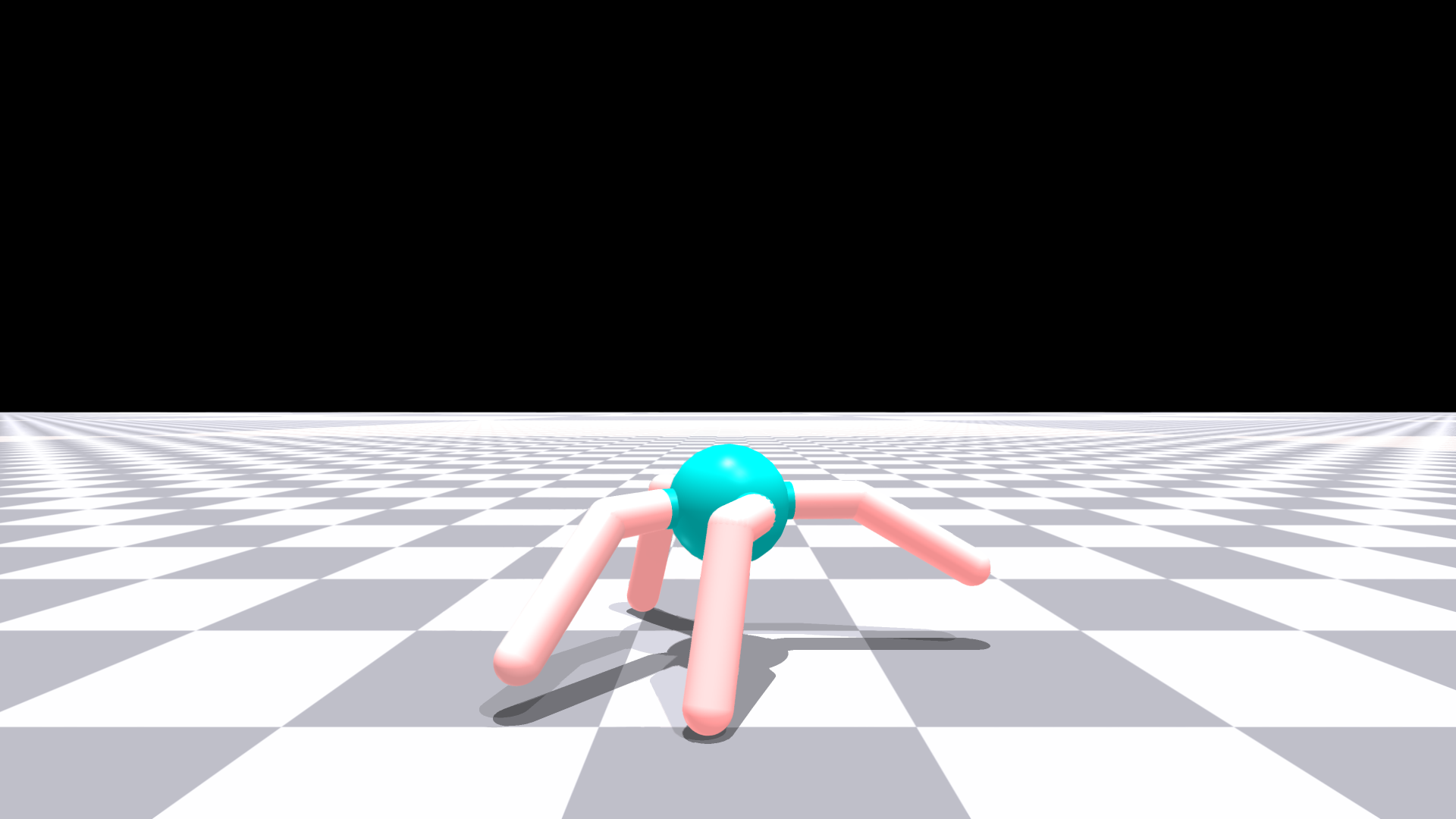}
    \includegraphics[trim={18cm 0cm 18cm 20cm},clip,width=0.19\linewidth]{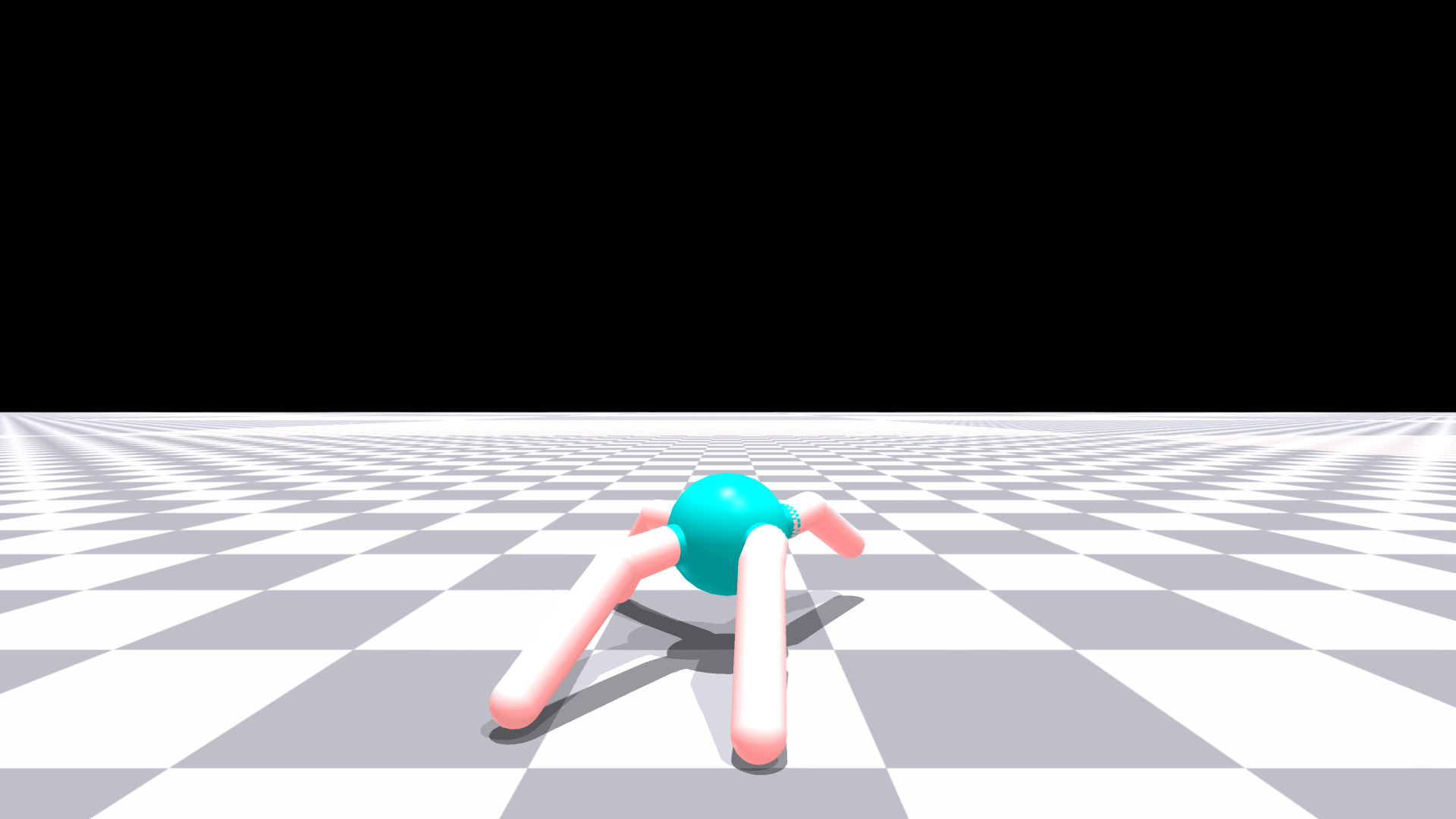}
    \caption
	{
	\textbf{Goal-oriented IsaacGym `Ant'.} \emph{Top:} Without fatigue. \emph{Bottom:} With fatigued torque bounds. Each picture was captured with a 100-frame interval between them. In a fatigued state the Ant keeps its body low and walks slowly as can be seen in the supplementary video.
	\label{fig:ant}
   }
\end{figure}

\begin{figure}
    \centering
    \includegraphics[width=\linewidth]{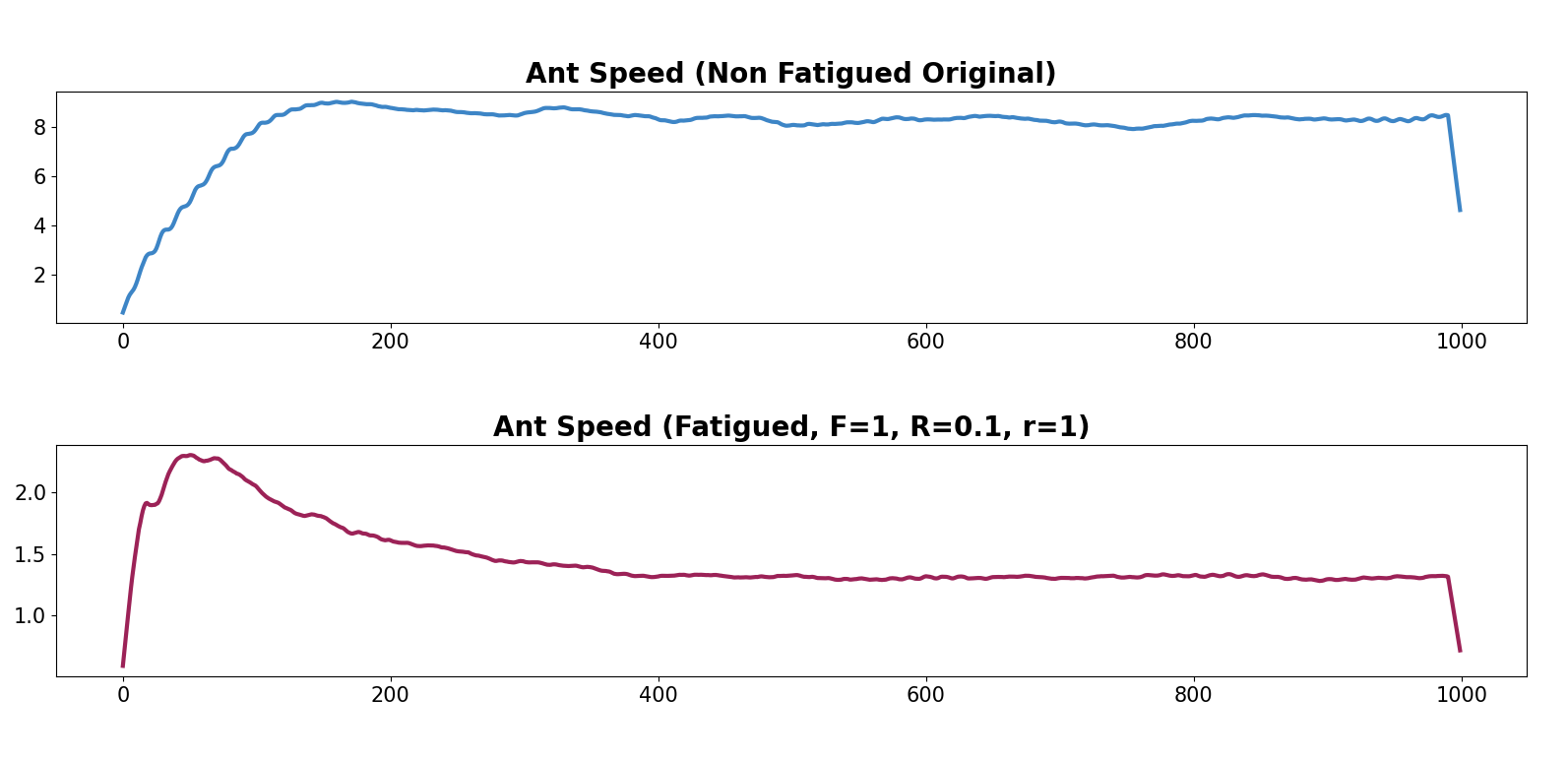}
    \caption{\textbf{`Ant'-Speed} averaged over 64 environments. \emph{Top:} Non-fatigued Ants. \emph{Bottom:} Ants with cumulative fatigue.}
    \label{fig:ant_speed}
\end{figure}

\paragraph{Four-legged Spider}
For this experiment we extend the goal-oriented IsaacGym `Ant'-environment with the fatigue-based torque-limits detailed in Section 4.2, and the addition of $M_F$ into state space $\mathbf{s}_t$. Given that we are dealing with a fictional character, we presume a maximum torque limit of 100 Nm for each DOF. The goal of the Ant is to walk towards a specific target. Cumulative fatigue leads to movements that look more natural for the Ant, where it walks slowly on all fours. This can be observed in last two sub-figures in the second row of Fig.\ref{fig:ant}. The original implementation leads instead to speeding and jumping on its hind legs. The results can be better viewed in the supplementary video. In Fig.\ref{fig:ant_speed} we further show the average speed difference between the two across 64 environments.

%
%
\section{Limitations and Discussion} \label{sec:discussion}
While our method for the first time demonstrates plausible fatigue behaviours and emerging resting strategies in the context of full-body character animation, it still exhibits some limitations. 
Our proposed method is based on the assumption that perceived fatigue can be directly deduced from biomechanical information. However, in reality perceived fatigue can be attributed to a multitude of various factors, such as physiolocal and psychological changes or environmental factors, with fatigue and rest being perceived differently from different individuals \cite{Xia2008ARecovery, jang2017modeling, frey2012knee, hincapie2014consumed, borg1982psychophysical}. In addition, the 3CC model has primarily been only validated on simple isometric tasks.
Furthermore, while our method is able to generate unseen motions, they may still result in motions of lower quality compared to methods that completely rely on imitation learning, since we try to generate motions outside of the training distribution. One could improve this by including additional methods to explore the action space further, such as intrinsic motivation or curiosity \cite{yin2021discovering, pathak2017curiosity}.
For more naturalness, one could also simulate the motor units $MUs$ in the 3CC model as muscle tendons. However, a key-advantage of the 3CC model for animation is that the $MUs$ can be used as a percentage of maximum voluntary contraction defined as percentage of force or torque, which has also been shown to produce accurate fatigue measures in bio-mechanics literature \cite{frey2010endurance, frey2012knee}.
Beyond the general accuracy of fatigue modeling, future directions could include object- or agent-agent interactions \cite{bae2023pmp, hassan2023synthesizing, zhang2023simulating}, as well as the general exploring of motion modeling outside data-distributions similar to \citet{lee2021learningFamily}.
%
%
\par 
Despite these limitations, we believe this work will help pave the way towards developing widely reusable control models for physics-based character animation and biomechanical simulations. 
As such, our work is of importance for biomechanics and animation researchers and practitioners alike, as it can be employed in various applications such as ergonomics analysis and physical skill training in VR/AR environments, as well as fatigue and stamina animation. 
In this regard, Digital Human Modeling (DHM) \cite{maurya2019digital, demirel2007applications} has been widely used in industrial simulation of workers to perform tasks. 
In order to achieve biomechanically correct ergonomics assessment, it is necessary to include an accurate fatigue model of simulated workers. 
In the scenario of virtual physical skill training, such as virtual sports training, an accurate fatigue model can prevent the users from performing unhealthy movements or getting hurt \cite{Fieraru_2021_CVPR}.
Moreover, our interactive fatigue and rest controls make it possible to easily adapt our method to different fatigue settings during inference in real-time, allowing our approach to be employed in interactive applications such as games or animation tools as can be seen in Fig.~\ref{fig:application} in this document and Fig.~1 in the main document. 
The morphology-agnostic nature of the 3CC-model further allows for the extension to characters of different physiology, and easy extension towards goal-oriented learning.

\begin{figure}
    \centering
    \includegraphics[width=\linewidth]{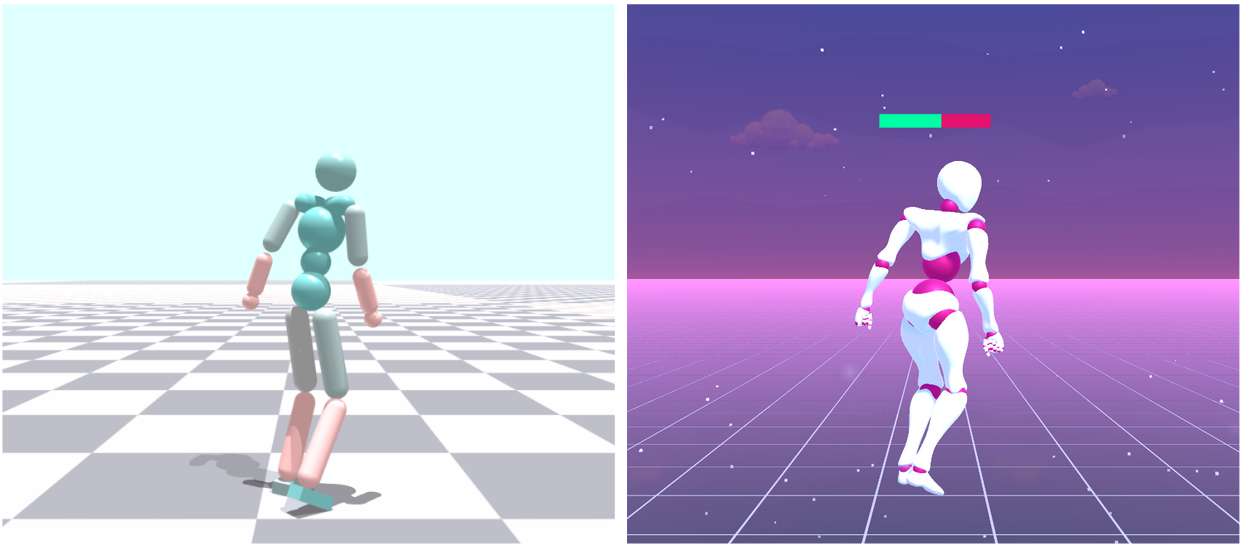}
    \caption{\emph{Left:} \textbf{Ragdoll simulation model.} \emph{Right:} \textbf{Re-targeted visualization model.} Our framework is used to learn fatigued movements for a 28 degrees-of-freedom humanoid character. Pink tints indicates fatigued areas (\emph{left}). The color of the rigid body corresponds to the fatigue of its parent joint (e.g. lower leg corresponds to the knee). The health bar indicates the current average residual capacity (\emph{right}).}
    \label{fig:character}
\end{figure}
\begin{figure}
	\includegraphics[width=\linewidth]{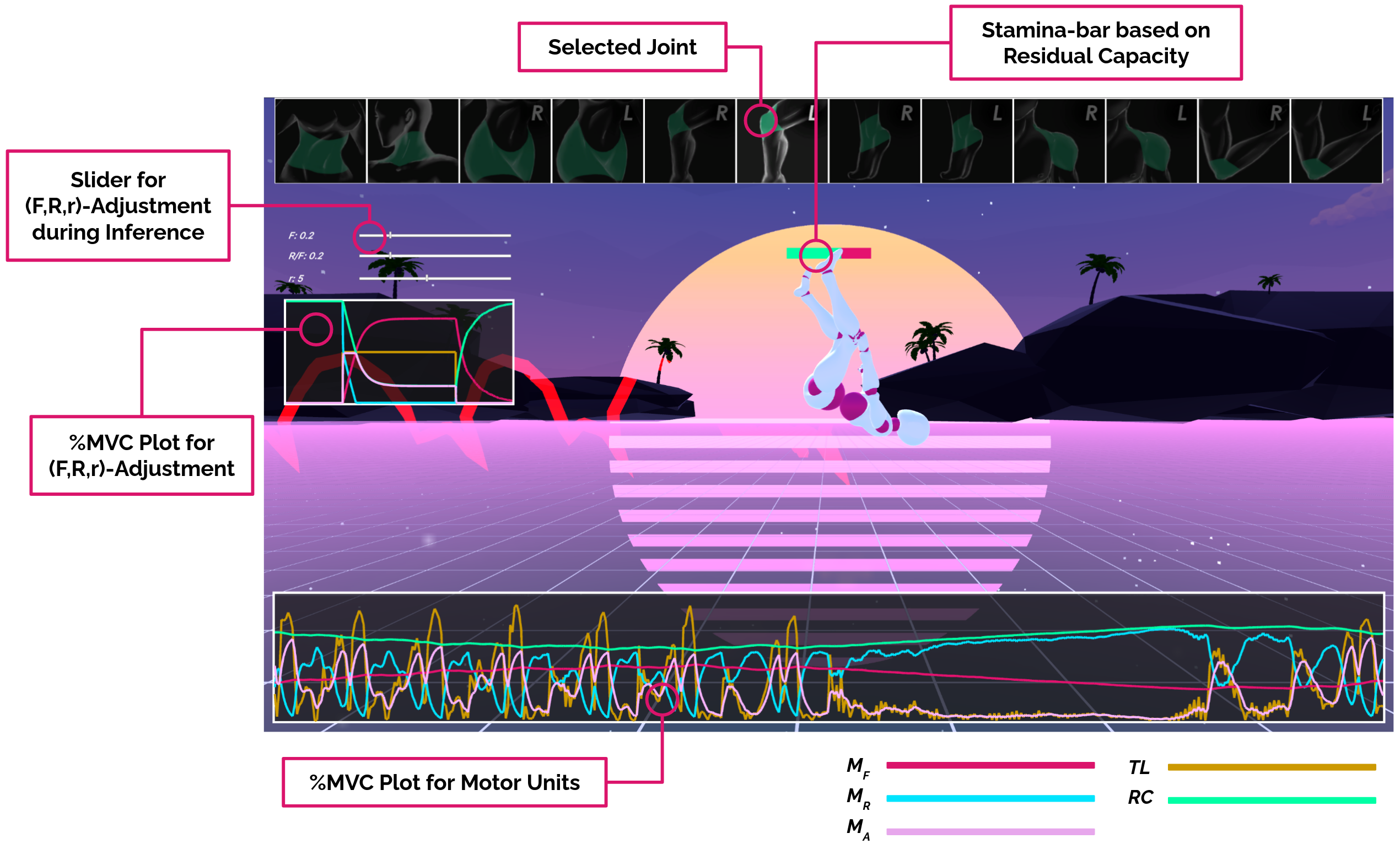}
	\caption
	{
	\textbf{UI for fatigue modeling.} Fatigue synthesis and analysis application based on our method with sliders for setting various ($F$, $R$, $r$)-combinations during inference, and indication of fatigue levels for each joint.
	}
	\label{fig:application}
\end{figure}
\begin{figure}
	\includegraphics[width=0.4\linewidth]{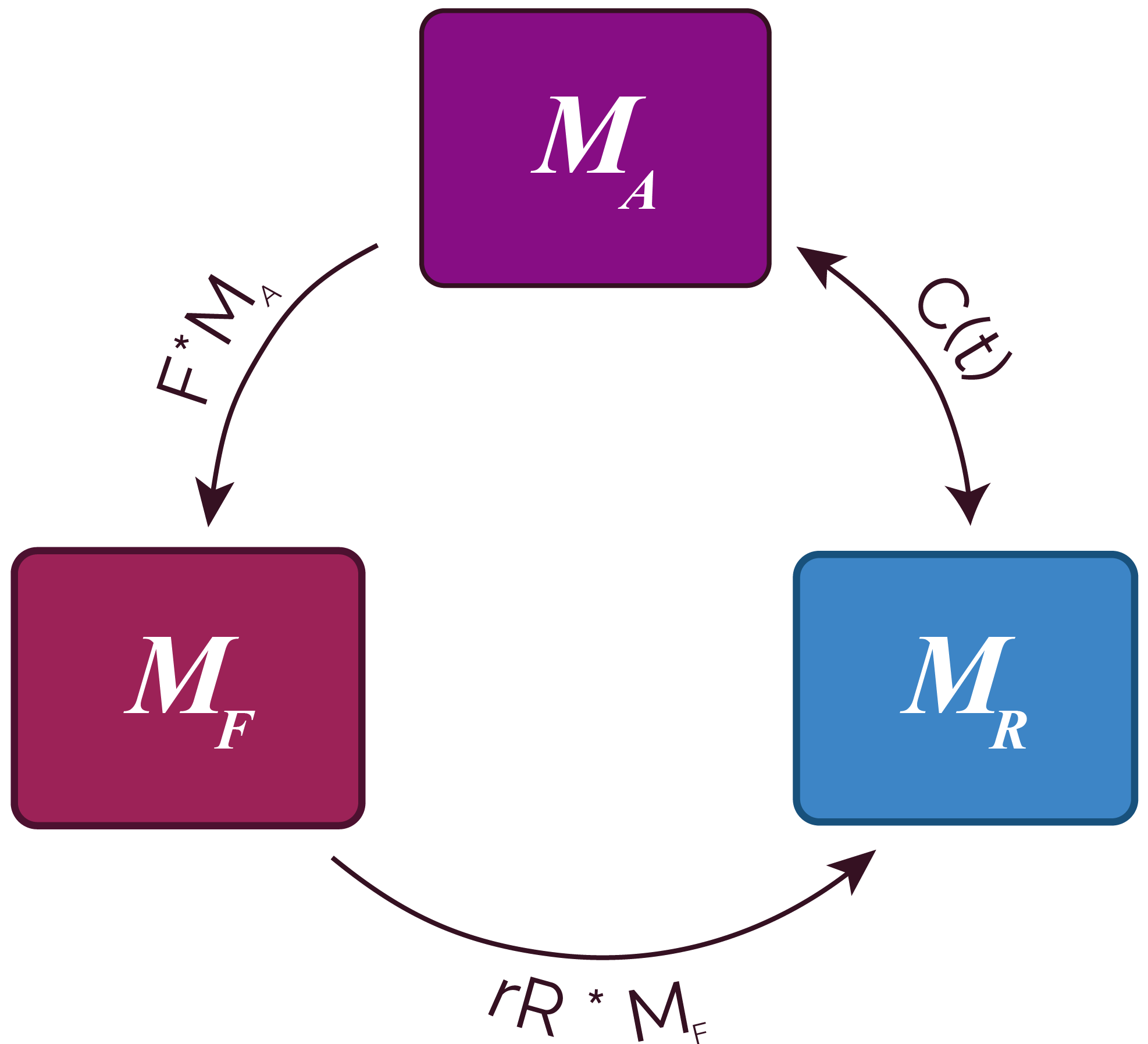}
	\caption
	{
	\textbf{Overview of the Three-Compartment Controller (3CC) model.} 
	$M_A$, $M_R$ and $M_F$ denote the percentage of active, rested and fatigued MUs, respectively. $F$ and $R$ denote the fatigue and recovery coefficients, and $C(t)$ the muscle activation-deactivation drive.
	}
	\label{fig:3cc_model}
\end{figure}
\begin{figure}
	\includegraphics[width=\linewidth]{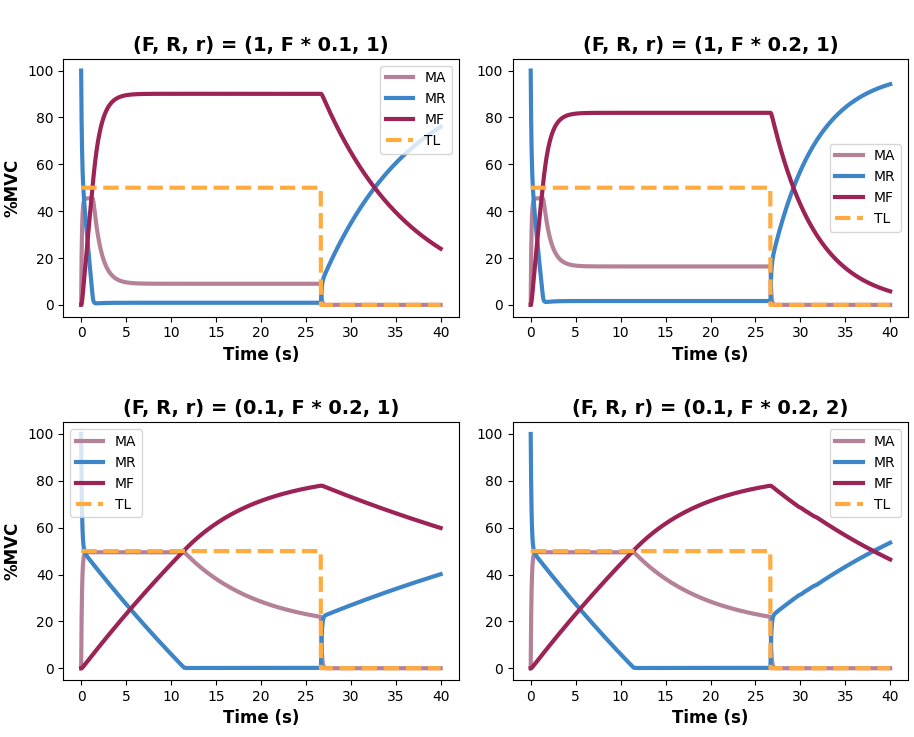}
	\caption
	{
	\textbf{The behaviour of the 3CC model} at 50\% MVC up until $26s$ with different fatigue $F$, rest $R$ and rest recovery $r$ rates. Note how the target load $TL$ cannot be held any longer due to fatigue at ${\sim}3s$ for $F=1$ (top two figures), and ${\sim}12s$ for $F=0.1$ (bottom two figures) as $M_A$ decays and is less than $TL$. A change in $F$ denotes the rate of fatigue. $R$ denotes the limit of fatigued MUs $M_F$: $R = F \cdot 0.2$ for example, indicates that 20\% $MVC$ can be held indefinitely (top right and bottom two figures), as well as overall change of rest. A change in $r$ indicates faster recovery during rest periods, i.e. when $TL = 0$.
	}
	\label{fig:3cc_examples}
\end{figure}
\begin{algorithm}
\caption{Residual Capacity $(RC)$ per DoF}
\label{alg:RC}
\begin{algorithmic}[1]
\State $F \gets$ set the fatigue rate
\State $R \gets$ set the recovery rate
\State $r \gets$ set the rest recovery multiplier
\State $R_r \gets R$ set the rest recovery rate
\State $L_D \gets$ set the muscle force development factor
\State $L_R \gets$ set the muscle force relaxation factor
\State $\Delta t \gets$ interval time between two consecutive simulation steps
\State $\mathcal{T}_{max} \gets$ set the maximum torque
\Function{3CC}{$\mathcal{T}$}
\If{at the beginning of an episode}
    \If{training}
        \State $M_R \gets Uniform.sample(0, 100)$
        \State $M_A \gets Uniform.sample(0, 100 - M_R)$
        \State $M_F \gets 100 - M_R - M_A$
    \Else
        \State $M_R \gets 100$
        \State $M_A \gets 0$
        \State $M_F \gets 0$
    \EndIf
\EndIf
\State $\mathcal{T} \gets clip(\mathcal{T}, -\mathcal{T}_{max}, \mathcal{T}_{max})$  \Comment{Eq.~(11)}
\State $TL \gets \frac{\mathcal{T}}{\mathcal{T}_{max}} \cdot 100$
\If{$M_A < TL$ and $M_R > TL - M_A$}
    \State $C_t \gets L_D \cdot (TL -M_A)$ \Comment{Eq.~(3)}
\EndIf
\If{$M_A < TL$ and $M_R \leq TL - M_A$}
    \State $C_t \gets L_D \cdot M_R$ \Comment{Eq.~(3)}
\EndIf
\If{$M_A \geq TL$}
    \State $C_t \gets L_R \cdot (TL -M_A)$ \Comment{Eq.~(3)}
    \State $R_r \gets r \cdot R$ \Comment{Eq.~(2)}
\EndIf
\State $M_A \gets M_A + \Delta t \cdot (C_t - F \cdot M_A)$ \Comment{Eq.~(1a)}
\State $M_R \gets M_R + \Delta t \cdot (-C_t + R_r \cdot M_F)$ \Comment{Eq.~(1b)}
\State $M_F \gets M_F + \Delta t \cdot (F \cdot M_A - R_r \cdot M_F)$ \Comment{Eq.~(1c)}
\State $RC \gets (100 - M_F)/100$ \Comment{Eq.~(4)}
\State \Return{$RC$}
\EndFunction
\end{algorithmic}
\end{algorithm}
\begin{algorithm}
\caption{Fatigue Transfer Learning}\label{alg:training}
\begin{algorithmic}[1]
\State \textbf{input} $\mathcal{M}:$ reference motion
\State \textbf{input} $\mathcal{T}_{max}$: maximum torque
\State $D \gets$ initialize discriminator with pre-trained model
\State $\pi \gets$ initialize policy with pre-trained model
\State $V \gets$ initialize value function with pre-trained model

\State $\mathcal{B} \gets \emptyset$ initialize replay buffer

\While{not done}
  \For{trajectory $i=1 \dots m$}
    \State $\tau^i \gets \emptyset$ initialize trajectory
    \State $s_0 \gets$ sample start state from pre-defined distribution
    \For{time step $[t=0, \dots, T-1]$}
        \State $a_t \gets$ sample action (target joint position) from $\pi$
        \State $\mathcal{T} \gets PDControl(a_t)$ \Comment{Eq.~(9)}
        \For{each DoF $d$}
            \State $RC_d \gets$ 3CC$(\mathcal{T}_d)$ (see Alg. \ref{alg:RC})
            \State $\mathcal{\widetilde T}_{max_d} \gets RC_d \cdot \mathcal{T}_{max_d}$ \Comment{Eq.~(12)}
            \State $\mathcal{\widetilde T}_d \gets clip(\mathcal{T}_d, -\mathcal{\widetilde T}_{max_d}, \mathcal{\widetilde T}_{max_d})$ \Comment{Eq.~(13)}
        \EndFor
        \State $s_{t+1} \gets$ get next state by applying $\mathcal{\widetilde T}$
        \State $r_t \gets -\log(1 - D(\Phi(\mathbf{s}_t), \Phi(\mathbf{s}_{t+1})))$ \Comment{Eq.~(8)}
        \State record $r_t$ in $\tau^i$
    \EndFor
  \State store $\tau^i$ in $\mathcal{B}$
  \EndFor
  \State update $\pi$, $V$ and $D$ using data from $\mathcal{B}$ and $\mathcal{M}$
\EndWhile

\end{algorithmic}
\end{algorithm}

    

\section{Additional Implementation Details}

For the benefit of the broader reinforcement learning and movement synthesis community, we identify and outline implementation issues and our workarounds used to address them.
To leverage the massive GPU parallelization offered by NVIDIA Isaac simulator, we elected to use the carefully tuned default simulation parameters for the \texttt{HumanoidAMP} environment.
This entails the relatively sparse simulation frequency at 120 Hz and policy query frequency at 30 Hz while using Isaac's native position control mode, which implements stable PD control into the inner loop of the Featherstone articulation program.
The inner loop remains inaccessible as Isaac is at a closed-source preview stage as of this writing.

Given the limited integration between the constraint solver and the simulator, we were unable to directly obtain accurate actuation readings from the simulator, specifically the torque values in Nm applied to the motors at each timestep. As an alternative, we considered using the effort mode control mode, but found that it resulted in an unstable simulation at reasonable frequencies.
As a result, we decided to use position mode control and estimate the torque values using the explicit PD control formula. We used stiffness and damping parameters modulated as appropriate for each controlled degree of freedom. We are confident that this approximation is reasonable, as the 3CC model used in our research is based on ratios instead of absolute units or scales.
For the $k_p$ and $k_d$ parameters at each controlled DoF, we used stiffness and damping parameters modulated as appropriate.
We deem this approximation reasonable as the 3CC model is based on ratios instead of absolute units or scales.
Recall that the modulated stiffness and damping parameter values are written as 
$\tilde k_{p, t} = \beta_t \dot k_p$ 
and 
$\tilde k_{d, t} = \beta_t \dot k_d$, 
respectively.
As such, the intended torque $\mathcal{T}$ and the clipped torque ${\mathcal{\tilde T}_{max}}$ were estimated as
$$
    \tilde{\mathcal{T}} = \tilde k_{p, t} \cdot \left( \hat \theta_t - \theta_t \right) - \tilde k_{d, t} \cdot \dot \theta_t
$$
where 
$\hat \theta_t$ is the PD target angles corresponding to the policy action $a_t$,
$\theta_t$ is the current DoF angles,
and $\dot \theta_t$ is the current DoF angular velocities,
all at timestep $t$.
To implement the torque limiting described in the main paper, in case the estimated intended torque $\mathcal{T}$ is above the the effective ceiling torque ${\mathcal{\tilde T}_{max}}$, we compute the ratio between them, which is then multiplied to the intended torque to implement the clipping, i.e.,
$$
\tilde{\mathcal{T}} = \begin{cases}
\mathcal{T} \cdot \frac{\mathcal{\tilde T}_{max}}{\mathcal{T}} \\
\mathcal{T} \qquad \text{otherwise.}
\end{cases}
$$
We set the simulator's joint stiffness and damping values to $\tilde k_{p, t}$ and $\tilde k_{d, t}$, respectively.

\begin{table}[]
\caption{\textbf{Custom stiffness and damping parameters used for the humanoid character.}}
\begin{tabular}{l | l | l}
\hline
\textbf{Joint} & \textbf{Stiffness} & \textbf{Damping} \\ \hline
Abdomen        & 120                & 12               \\
Neck           & 90                 & 9                \\
Shoulders      & 100                & 10               \\
Elbows         & 110                & 11               \\
Hips           & 320                & 32               \\
Knees          & 370                & 37               \\
Ankles         & 120                & 12              
\end{tabular}
\end{table}

Table \ref{tab:hyperparams} reports additional hyperparameters for training our approach such as learning rate and batch size.
Table \ref{tab:max_torques} reports the found maximum torque bounds. We report bounds for each action and joint. Note that those bounds are found automatically during our proposed pre-training.

\begin{table}[h]
    \centering
    \caption{\textbf{Additional PPO-Hyperparameter Settings.}}
    \label{tab:hyperparams}
    \begin{tabular}{l|c}
    \hline
    \multicolumn{2}{c}{\textit{PPO-Hyperparameters}}\\
    \hline
    \hline
         \textbf{Parameter} & \textbf{Value}\\
         \hline
         Horizon Length & 16\\
         Minibatch Size & 4096\\
         Batch Size & 512\\
         KL Coeff. & 0.008\\
         $\gamma$ Discount & 0.99\\
         GAE($\gamma$) & 0.95\\
         TD($\gamma$) & 0.95\\
         Adam stepsize & $5 \times 10^{-5}$
    \end{tabular}
\end{table}
\begin{table}[h]
  \centering
  \caption{\textbf{Found maximum constant torque bounds for each task during pre-training.} Note that $z$ denotes the up-axis. Symmetric joints present on the left and right side of the body use the same values. Bold values highlight the maximum value across motions, hence the denominators of $TL$ value calculation.}
  \label{tab:max_torques}
  \begin{tabular}{c|*{6}{S[table-auto-round,table-format=-1.0]}}
  \hline
  \multicolumn{6}{c}{\textit{Maximum constant torque bounds found during Training}}\\
  \hline
  \hline
 \textbf{DoF} & \textbf{Run} & \textbf{Hop} & \textbf{Cartwheel} & \textbf{Backflip} & \textbf{Kick}\\
 \hline
 $Abdomen_x$ & 44.86 & 67.24 & 163.27  & \textbf{370.27} & 289.66 \\
$Abdomen_y$ & 37.48 & 196.68 & 234.43  & \textbf{382.34} & 181.37 \\
$Abdomen_z$  & 56.89 & 97.52 & 154.58  & \textbf{240.88} & 234.43 \\
\hline
$Neck_x$  & 13.55 & 25.52 & 58.88 & 21.11  & \textbf{78.61}  \\
$Neck_y$  & 26.97 & \textbf{160.34} & 133.49  & 77.09  & 69.46  \\
$Neck_z$  & 14.96 & 15.98 & 61.56 & \textbf{63.66}  & 46.88  \\
\hline
$Shoulder_x$ & 19.55 & 64.14 & 148.68 & \textbf{177.67} & 143.16 \\
$Shoulder_y$ & 35.13 & 98.05 & 89.79 & \textbf{362.03} & 170.81 \\
$Shoulder_z$ & 12.77 & 55.59 & 116.39 & \textbf{306.93} & 173.37 \\
\hline
$Elbow$ & 21.82 & 106.75 & 80.32 & 91.37 & \textbf{136.1}  \\
\hline
$Hip_x$  & 150.29 & 393.96 & 580.35 & 143.27 & \textbf{598.84}  \\
$Hip_y$  & 202.58 & 420.74 & \textbf{796.42} & 387.99 & 691.47 \\
$Hip_z$  & 78.68 & 100.95 & 247.19 & 135.1 & \textbf{373.14} \\
\hline
$Knee$  & 420.83 & 741.27 & 496 & 340.21 & \textbf{809.59} \\
\hline
$Ankle_x$  & 23.25 & 91.8 & 98.97 & 73.17 & \textbf{138.58} \\
$Ankle_y$  & 227.65 & 206.63 & 119.59 & 407.15 & \textbf{451.8}  \\
$Ankle_z$  & 37.95 & 69.69 & 114.9 & 90.14 & \textbf{201.1}  \\
  \end{tabular}
\end{table}

\end{document}